\documentclass[twocolumn,preprintnumbers,amsmath,amssymb,superscriptaddress]{revtex4-1}
\usepackage{graphicx}
\usepackage{dcolumn}
\usepackage{bm}
\usepackage{natbib}
\usepackage{lineno}
\usepackage{amsmath}
\usepackage{hyperref}
\usepackage{url}

\hypersetup{
  pdfauthor={...},
  pdftitle={...},
  pdfsubject={...},
  urlcolor=blue,
}

\def\apjl{Astrophys. J. Lett.}

\def\mnras{Mon.Not.Roy.As.Soc.}

\def\aap{Astron. \& Astrophys.}
\def\ltsima{$\; \buildrel < \over \sim \;$}
\def\simlt{\lower.5ex\hbox{\ltsima}}
\def\gtsima{$\; \buildrel > \over \sim \;$}
\def\simgt{\lower.5ex\hbox{\gtsima}}
\def\roe{Institute for Astronomy, University of Edinburgh, Royal Observatory, Blackford Hill, Edinburgh EH9 3HJ}
\def\imperial{Imperial Centre for Inference and Cosmology, Imperial College, Blackett Laboratory, Prince Consort Road, London SW7 2AZ, UK}
\def\durham{Institute for Computational Cosmology, Department of Physics, Durham University, South Road, Durham DH1 3LE, UK}
\def\sydney{Sydney Institute for Astronomy, School of Physics, University of Sydney, NSW 2006, Australia}
\def\ljmu{Astrophysics Research Institute, Liverpool John Moores University, IC2, Liverpool Science Park, 146 Brownlow Hill, Liverpool, L3 5RF, UK}
\def\sussex{Astronomy Centre, University of Sussex, Falmer, Brighton BN1 9QH, UK}
\def\swin{Centre for Astrophysics \& Supercomputing, Swinburne University of Technology, P.O. Box 218, Hawthorn, VIC 3122, Australia}
\let\sec=\section

\def\[{\begin{equation}}
\def\]{\end{equation}}

\def\m@th{\mathsurround=0pt }
\def\eqalign#1{\null\,\vcenter{\openup1\jot \m@th
 \ialign{\strut\hfil$\displaystyle{##}$&$\displaystyle{{}##}$\hfil
 \crcr#1\crcr}}\,}

\begin{document}
%
%
\title[GAMA: Clipping the Galaxy Power Spectrum]{Galaxy and Mass Assembly (GAMA):  \\ Redshift Space Distortions from the Clipped Galaxy Field} 
%
\author{F.~Simpson} \thanks{E-mail:~fergus2@icc.ub.edu} 

\affiliation{ICC, University of Barcelona (UB-IEEC), Marti i Franques 1, 08028, Barcelona, Spain.}
\affiliation{\roe}
\author{C.~Blake}
\affiliation{\swin}
\author{J.A.~Peacock}
\affiliation{\roe}
\author{I.K.~Baldry}
\affiliation{\ljmu}
\author{J.~Bland-Hawthorn}
\affiliation{\sydney}
\author{A.F.~Heavens}
\affiliation{\imperial}
\author{C.~Heymans}
\affiliation{\roe}
\author{J.~Loveday}
\affiliation{\sussex}
\author{P.~Norberg}
\affiliation{\durham}

\date{\today}
\newcommand{\ud}{\mathrm{d}}
\newcommand{\fpe}{f_\perp}
\newcommand{\fpa}{f_\parallel}
\newcommand{\om}{\Omega_m}
\newcommand{\dmax}{{\delta^{\rm{max}}}}
\newcommand{\kmax}{{k_{\rm{max}}}}
\newcommand{\Veff}{{V_{\rm{eff}}}}
\newcommand{\eff}{{\rm{eff}}}
\newcommand{\lcdm}{$\Lambda$CDM }
\newcommand{\hmpc}{ \, h \rm{ Mpc}^{-1}}
\newcommand{\hmpcv}{ \, h^3 \rm{ Mpc}^{-3}}
\newcommand{\hinvmpc}{ \, h^{-1} \rm{ Mpc}}
\newcommand{\tripleint}{\int \! \! \int \! \! \int}
\newcommand{\kfund}{k_f}
\newcommand{\fm}{f_m}
\newcommand{\dint}{\int  \!\!  \int}
\newcommand{\bfx}{\mathbf{x}}
\newcommand{\bfr}{\mathbf{r}}
\newcommand{\dclip}{\delta_{c0}}
\newcommand{\dThresh}{\delta_{0}}
\newcommand{\deltaX}{\delta^{(X)}}
\newcommand{\rh}{{\rho}(r)}
\newcommand{\clip}{{\mathcal{C}}}
\newcommand{\Plin}{{P_\mathrm{L}(k)}}
\newcommand{\Plinclip}{{P_\mathrm{cL}(k)}}
\newcommand{\Plinclipz}{{P_\mathrm{cL}^{s}(k)}}
\newcommand{\Alin}{{a_{\mathrm L}}}
\newcommand{\Atwo}{{a_2}}
\newcommand{\Poneloop}{{P_\mathrm{1loop}(k)}}
\newcommand{\Ponetwo}{{P_{\mathrm{c}12}(k)}}
\newcommand{\doa}{{ t_0}} 
\newcommand{\fid}{{\mathrm{fid}}}
\newcommand{\amp}{{ b \sigma_8 }}
\newcommand{\fsig}{ f \sigma_8}
\newcommand{\GAMAfsig}{ 0.29}
\newcommand{\GAMAfsigError}{{ 0.10 }}
\newcommand{\GAMAz}{ {0.18}}

\date{\today}

\begin{abstract}
We present the first cosmological measurement derived from a galaxy density field subject to a `clipping' transformation. By enforcing an upper bound on the galaxy number density field in the Galaxy and Mass Assembly survey (GAMA), contributions from the nonlinear processes of virialisation and galaxy bias are greatly reduced. This leads to a galaxy power spectrum which is easier to model, without calibration from numerical simulations. 

We develop a theoretical model for the power spectrum of a clipped field in redshift space, which is exact for the case of anisotropic Gaussian fields.  Clipping is found to extend the applicability of the conventional Kaiser prescription by more than a factor of three in wavenumber, or a factor of thirty in terms of the number of Fourier modes. 
By modelling the galaxy power spectrum on scales $k < 0.3 \hmpc$ and density fluctuations $\delta_g < 4$ 
we measure the normalised growth rate $f\sigma_8(z= \GAMAz) = \GAMAfsig \pm \GAMAfsigError$.
\end{abstract}

\maketitle

\section{Introduction}

The spatial distribution of galaxies encodes a wealth of information relating to the composition and evolution of the Universe. The apparent positions of galaxies in redshift space offers a glimpse into both the density and velocity perturbations associated with dark matter. These in turn are influenced by a number of phenomena in fundamental physics, such as the mass of the neutrino and the nature of gravity.  However two key factors have thus far restricted our view: (a) the advanced stages of gravitational collapse are highly unpredictable and (b) the uncertainty associated with galaxy bias, defined as the manner in which the galaxy distribution reflects the dark matter distribution. In Fourier space, conventional analyses impose a maximum wavenumber beyond which the data points are considered unpredictable and are simply discarded. For example, despite utilising numerical simulations to calibrate the non-linear power spectrum, recent studies of redshift space distortions typically truncate the power spectrum at $k < 0.2 \hmpc$  \cite{BlakeWigglezRSD} or exclude galaxy pairs closer than $25 \hinvmpc$ \cite{BOSSRSD2012}. While most of the nonlinear behaviour is successfully disposed of, so too is much of the cosmological information. 

A number of different methods have been proposed to allow smaller clustering scales to be exploited. A phenomenological model has been developed by \citet{2012KLL}, and expanded by \citet{2013LinderRSD}, to model the power spectrum based on fits to numerical simulations. \citet{2014ReidBOSSRSD} extract information from small-scale clustering in BOSS using a model for the halo occupation distribution. While achieving a significant increase in precision, this technique also relies heavily upon calibration from numerical simulations.  Alternatively, various local transformations have been explored as a means of reducing the influence of nonlinearities, such as Gaussianisation \cite{1992MNRAS.254..315W, 2015arXiv151102034M} and the logarithmic transformation  \cite{Neyrinck2009}. In \citet{2013SimpsonClip} it was shown that simply enforcing a maximum density could greatly increase the number of Fourier modes that could be modelled with the standard set of theoretical tools. Recently it has also been found to enhance the observational signature in models of modified gravity which invoke a screening mechanism \cite{2015arXiv150104961L}. This clipping technique serves as the focus of the present work. 

Our inability to model the small scale power spectrum is not the only source of information loss. More fundamentally, the onset of nonlinear gravitational collapse degrades the amount of information held by the power spectrum  \cite{Rimes_Hamilton_2005}; this loss occurs for any non-Gaussian field. To extract some of this missing information, we can either perform additional measurements, such as higher-order statistics, or manipulate the field prior to evaluating the power spectrum. Previous attempts to extract cosmological information from the spatial distribution of galaxies beyond the conventional two point statistics include measuring three-point statistics \cite{2002MNRAS.335..432V, 2013MarinBlake3pt, 2014Hector}, Minkowksi functionals \cite{2014BlakeBerian}, and the shapes of voids \cite{2014SutterVoids}.  Local transformations also appear promising in restoring this information to the power spectrum (see for example \cite{Neyrinck2009, 2011WangLog, 2013Carron, 2014Carron}). \citet{2014Carron} present a transformation that is optimised for extracting information from the power spectrum of a Poisson-sampled field.  Our approach differs slightly: instead of optimising the extraction of all information from the observable galaxy density field $\rho_g$, we seek to selectively extract only the \emph{predictable} information from the field. At the highest values of $\rho_g$ both the nonlinear structure of the dark matter and the environmental impacts on galaxy formation present formidable obstacles in interpreting its value. Conversely, regions closer to the mean density are expected to behave in a more predictable and robust manner. The action of clipping preserves the location of galaxy clusters as useful information while discarding information relating to the precise value of their density contrast. 

Clipping is already known to be a highly effective technique for improving the theoretical modelling of the galaxy bispectrum \cite{SimpsonClip} and power spectrum \cite{2013SimpsonClip}, when applied to fields in real space.  Before we can apply this technique to data from real surveys, it must first be verified in redshift space. It would also be desirable to develop a deeper understanding of why it has been successful. These are two of the goals of this paper. The third is to apply clipping to the GAMA survey in order to obtain a low-redshift measurement of the normalised growth rate $f \sigma_8$.  

In \S \ref{sec:Theory} we review the theoretical background associated with the  two point statistics of clipped fields and consider its extension to anisotropic fields. This theoretical framework is placed into a cosmological context in \S \ref{sec:applications}, where we develop a model for the form of the clipped galaxy power spectrum.  In \S \ref{sec:sims} we apply clipping to simulated dark matter and galaxy fields in redshift space, with the results illustrated in \S \ref{sec:sim_results}.  The GAMA dataset is introduced in \S \ref{sec:data}, while the main results of this work are presented in \S \ref{sec:results} before our concluding remarks in \S \ref{sec:conclusions}. 

\section{Statistical Properties of Clipped Fields} \label{sec:Theory}

Clipping is a local transformation characterised by the application of a saturation value $\delta_0$ to a scalar field $\delta(\bfx)$ such that
\[ \label{eq:clip}
\eqalign{
\delta_c (\bfx)  &=  \delta_0 \qquad (\delta (\bfx) > \delta_0 )  \cr
\delta_c (\bfx)  &=  \delta(\bfx) \quad   (\delta (\bfx) \leq  \delta_0 ),  \cr
}
\]
\noindent  yielding the clipped field $\delta_c (\bfx)$. In this section we explore the statistical properties of $\delta_c (\bfx)$, with particular attention paid to its autocorrelation 
\[ \label{eq:xi}
\xi_c (\bfr) \equiv \langle \left[ \delta_c (\bfx) -  \bar{\delta}_c \right]  \left[ \delta_c (\bfx+\bfr) -  \bar{\delta}_c \right]  \rangle \, ,
\]
and corresponding power spectrum $P_c(\vec{k})$.  We note that the clipping transformation induces a non-zero mean in $\delta_c(x)$, which is why must  be careful to specify the more general form of the autocorrelation function, as defined by (\ref{eq:xi}).  In the analysis of cosmological fields the subtraction of the mean is conventionally omitted from this definition, since the mean is usually zero by construction. 

We begin by reviewing the special case where $\delta(\bfx)$ is an isotropic Gaussian field before generalising to anisotropic and higher-order fields.

\subsection{Isotropic Gaussian Fields}

\noindent  For the case of Gaussian fields we may invoke Price's theorem \cite{price1958useful, gross1994snr} to evaluate the two-point statistics associated with the clipped field $\delta_c (\bfx)$ in terms of the original correlation function $\xi(r)$:
\[  \label{eq:price}
\frac{\partial \xi_c(r)}{\partial \xi(r)} =  \iint \frac{\partial g[\delta_1]}{\partial \delta_1} \frac{\partial g[\delta_2]}{\partial \delta_2} p(\delta_1, \delta_2, r)  \ud \delta_1 \ud \delta_2  \, ,
\]

\noindent  where  $g[\delta]$ is the local transformation defined by (\ref{eq:clip}), $ \xi_c(r)$ is the correlation function of the transformed field, and $p(\delta_{1}, \delta_{2}, r)$ is the joint probability distribution for a Gaussian process
\[ \label{eq:jointprob}
p(\delta_{1}, \delta_{2}, r) = \frac{1}{2 \pi \sigma^2 \sqrt{1 - \rh^2}} \exp \left[  \frac{2 \rh \delta_{1} \delta_{2} - \delta_{1}^2  - \delta_{2}^2}{2 \sigma^2 \left[ 1 - \rh^2 \right]}  \right]
\]


The functional derivative of the clipping transformation is unity below the threshold, and zero above the threshold. This simple behaviour permits an analytic solution of the clipped correlation function  $\xi_c(r)$,  which is given by \cite{2013SimpsonClip}

\[  \label{eq:clipAnalytic}
 \xi_c(r) = f_c^2  \xi(r)  +  \sigma^2 \sum_{n=1}^{\infty} \left[ \frac{\xi(r)}{\sigma^2} \right]^{n+1} \! \! C_n(u_0) \, ,
\]
where $\sigma^2$ is the variance of the field prior to clipping, $u_0$ is the normalised threshold value  $u_0 \equiv  \delta_0/\sqrt{2} \sigma$, $f_c$ is the fraction of the field which lies below the threshold
\[ \label{eq:alin}
f_c = \frac{1}{2} \left[ 1 + \mathrm{erf}\left(u_0\right) \right],
\]
and $C_n(x)$ is the distortion coefficient 
\[ \label{eq:cn}
C_n(x) = \frac{H_{n-1}^{2} \! \left( x \right)}{\pi 2^n (n+1)!}  e^{-2x^2} \, ,
\]
where $H_n(x)$ is the Hermite polynomial of order $n$. Beyond a scale-independent reduction in amplitude, clipping induces a distortion in the shape of the correlation function. However provided the slope of the spectral power is not too steep $(|n| \lesssim 3)$, and the clipping remains weak $(u_0 \gtrsim 1)$, the leading order correction makes only a small contribution to the resultant power. Furthermore, terms at higher values of $n$ decay rapidly. 

\subsection{Anisotropic Gaussian Fields} \label{sec:clipAniso}

For the more general case of anisotropic Gaussian fields we may express the two-point correlation function $\xi(r, \vec{\theta})$ in terms of both the pair separation $r$ and the orientation vector $\vec{\theta}$.  In deriving the expression given by (\ref{eq:clipAnalytic}) we made use of the joint probability distribution for a Gaussian process as given by (\ref{eq:jointprob}), which is not directly applicable to anisotropic fields. However we can proceed by applying this single-parameter transformation separately at each fixed value of $\vec{\theta}$, which in itself constitutes a single-parameter Gaussian process. Since the variance of each subspace is the same, $ \sigma^2 = \xi(r \! = \! 0, \vec{\theta})$ for all orientations, the transformation maintains the same functional form for all values of $\vec{\theta}$. Therefore the expression $(\ref{eq:clipAnalytic})$ can be generalised to anisotropic fields. 

Transforming to Fourier space leaves us with the expression for the clipped power spectrum
\[ \label{eq:clipAnalyticAnisotropic}
 P_c(\vec{k}) =  f_c^2  P(\vec{k}) \, + \sigma^2 \sum_{n=1}^{\infty} C_{n} \left( u_0 \right) \left[ \frac{P(\vec{k})}{\sigma^2} \right]^{*(n+1)} \! \! ,
\]
where the notation $*n$ represents a self-convolution of order $n$. In practice it is computationally more straightforward to evaluate the higher order terms using powers of the correlation function, rather than performing multiple convolutions of the power spectrum. Further details of this calculation, as applied to mock cosmological density fields,  can be found in Appendix \ref{sec:testAnisotropicClipping}.

In line with the definition of the correlation function (\ref{eq:xi}), our definition of the power spectrum in (\ref{eq:clipAnalyticAnisotropic}) is specified in terms of the mean subtracted field. For a clipped Gaussian field the mean is given by
\[
\bar\delta_c = \delta_0(1-f_c) - \frac{\sigma}{\sqrt{2\pi}} \exp(-\delta_0^2/2\sigma^2) \, .
\]
In practice, it is not critical to account for this constant offset since it only contributes to the power spectrum at $k=0$.

\subsection{Second Order Anisotropic Fields} \label{sec:higherOrder}

In order to gain insight into how higher order terms respond to clipping, we repeat the procedure above using the square of a Gaussian random field, $e(x) \equiv \delta^2(x) -  \sigma^2$. The correlation function of this second order field, clipped at $e=\delta_0$, is well approximated by \cite{2013SimpsonClip}
\[
\xi_c(r, e) = \Atwo \xi(r,e) + \mathcal{O} \left( \frac{ \xi^2(r,e) }{\sigma^{2}}  \right) \, ,
\]
\[
\Atwo = \frac{1}{\pi} \left[\sqrt{\pi} \mathrm{erf}(u'_0) - 2 u'_0 e^{-u_0'^{2}} \right]^2  \, ,
\]
\[
u'_0 \equiv \frac{\sqrt{\delta_0 + \sigma^2}}{\sqrt{2} \sigma} \, ,
\]
where $\sigma$ refers to the standard deviation associated with the original Gaussian field, and $\Atwo$ quantifies the amplitude of the power spectrum relative to the original field. Following the same line of reasoning given in the previous subsection, we can generalise this result to the anisotropic case:
\[ \label{eq:secondOrderClip}
P_c(\vec{k}, e) \simeq a_2  P(\vec{k}, e)   \, .
\]
As before, the clipped two-point statistics of a second order field maintain the same shape as the unclipped case, at least for weak transformations. Note that when clipping at equivalent thresholds, $u_0 = u'_0$, the higher order field is subject to a significantly stronger suppression of its two point statistics than for the Gaussian case, and this trend strengthens with yet higher order fields. 

\subsection{Hybrid Fields} \label{sec:hybrid}

Practical applications of clipping will inevitably involve the superposition of a Gaussian field with other components that contaminate the desired signal, particularly where the amplitude of the field is large. In this scenario, clipping can assist in extracting  the power spectrum associated with the original Gaussian field.  

Consider a hybrid field $h(x)$ that is a linear combination of a Gaussian field $\delta(x)$, a higher order field $\delta^2(x)$, and a nuisance field $\delta^{X}$, which characterises some unknown departure from the model:
\[  \label{eq:hybrid_field}
h(x) = \alpha_1 \delta(x)  + \alpha_2 (\delta^2(x) - \sigma^2) + \delta^{X}(x)  \, .
\]
Upon clipping at a given threshold $\delta_0$, and provided the nuisance field is constant where $h(x) < \delta_0$,  we can subtract the mean to remove any residual contribution from $\delta^{X}$
\[  \label{eq:clipped_hybrid_field}
h_c(x) = \alpha_1 \delta_c + \alpha_2 (\delta_c^2 - \sigma^2)   \, ,
\]
where the component fields $ \delta_c$ and  $\delta_c^2$ are now also clipped fields.  This result is important as it shows that we can cleanly remove any trace of our nuisance field $\delta^{X}$. In most practical applications the recovery will be imperfect, as the nuisance field is likely to vary outside of the clipped region. However this is a much better state of affairs than the conventional approach  - be it perturbation theory or a model of galaxy bias - where we assume that the extra terms missing from our model (as specified by $\delta^{X}$) vanish everywhere. With clipping, we can now make the much more reasonable approximation that $\delta^{X}$ only vanishes where $h(x)$ is small. 

The two component fields $ \delta_c$ and  $\delta_c^2$ each experience their own distinct thresholds, which may be found by solving (\ref{eq:clipped_hybrid_field}) with the condition $h_c(x) = \delta_0$.
The power spectrum of $h_c(x)$ is given by
\[  \label{eq:hybridClipPower}
P_c (k) = \alpha_1^2 \Plinclip +  2 \alpha_1  \alpha_2 \Ponetwo + \alpha_2^2 P_{c22}(k) \, .
\]
where  $\Plinclip$  and   $P_{c22}(k)$ denote the power spectra associated with the clipped $\delta(x)$ and $\delta^2(x)$ fields respectively.  The cross spectrum $\Ponetwo$ vanishes in the limit of a high threshold, but becomes increasingly prominent as the threshold is lowered.  In order to estimate  $\Ponetwo$, we may decompose it as
\[
\langle  \delta_c^{(1)}  \delta_c^{(2)} \rangle = \langle r_1 r_2 \rangle -  \langle r_2 \delta^{(1)}  \rangle  - \langle   r_1  \delta^{(2)}  \rangle \, ,
\]
where the residual fields are defined as $r_i \equiv \delta^{(i)} - \delta_c^{(i)} $.
Provided the clipping is weak, $\delta_0 > \sigma$, the residual fields are closely related $r_2 \simeq 2 \doa \frac{ \alpha_2}{\alpha_1} r_1$, where $\doa$ is the threshold experienced by the Gaussian field. This may be re-expressed in the form
\[ \label{eq:P12}
\Ponetwo \simeq  -   \left[ 2 \doa \left( f_c^{-1} - 1\right) P_{L}(k) + \frac{\alpha_1 - \alpha_2}{2 \doa \alpha_2}  P_{22}(k) \right]   .
\]
Since $\Ponetwo$ may be expressed as a linear combination of the first and second order power spectra, we can now rewrite (\ref{eq:hybridClipPower}) in the form
\[   \label{eq:A11A22model}
P_{c}(k) \simeq   \Alin  \Plin   +   \Atwo  P_{22}(k)    \, .
\]
where $\Alin$ and $\Atwo$ are the apparent amplitudes of the original linear and second order spectra. This result helps explain why the simple model used in \citet{2013SimpsonClip} was particularly successful at reproducing the clipped dark matter power spectrum, without explicitly accounting for the cross-spectrum $P_{c12}(k)$. 

Adopting this higher order model, rather than relying on the linear solution from \S \ref{sec:clipAniso},  holds two advantages. First of all weaker clipping thresholds can be used, allowing the power spectrum to maintain a high amplitude. In addition, the inclusion of a higher order term potentially allows the degeneracy between linear bias and $\sigma_8$ to be lifted.  The disadvantage of this approach is the difficulty in estimating $\Atwo$, which could either be calibrated from simulations, or simply treated as an additional free parameter.
 
\section{Galaxy Density Fields} \label{sec:applications}

Galaxy redshift surveys continue to develop an intricate mozaic of the low redshift Universe. By convolving the point-like positions of galaxies with a suitable kernel, a continuous density field can be generated. These galaxy fields are heavily influenced by both nonlinear structure and galaxy bias, which have proved highly challenging to model. In each case, it is the highest density regions which are particularly troublesome, and this motivates the application of clipping. The three dimensional nature of the data ensures that clipping can be applied very efficiently. Selecting a threshold that affects only $\sim 1 \%$ of the field's volume typically leads to a reduction in large scale power by a factor of two. Maps which are two dimensional projections, such as those derived from a photometric survey, could also be subject to clipping but a greater proportion of the area would need to be clipped in order to achieve the same degree of suppression. 

In this section we explore the consequences of clipping a galaxy density field in redshift space. We shall work in the distant-observer approximation such that all line-of-sight displacements may be considered parallel. 

\subsection{Galaxy Bias} \label{sec:galbias}

If we model the fractional overdensity of galaxies $\delta_g \equiv \rho_g/\bar{\rho}_g - 1$ as an arbitrary function of the local dark matter density  \cite{fry1993biasing},
\[
\delta_g = \sum_{k=0}^{\infty} b_k \frac{\delta_m^k}{k!} \, ,
\]
then applying the clipping transformation suppresses higher order $b_k$ terms in the same way that higher order terms in perturbation theory are suppressed. The simplest extension to the linear bias model would be the introduction of $b_2$, which for the case of Gaussian dark matter fluctuations leaves us with a hybrid field as defined in (\ref{eq:hybrid_field}).  With a sufficiently low clipping threshold, the linear bias parameter $b_1$ dominates such that the clipped galaxy power spectrum is highly insensitive to the initial value of $b_2$. This linearisation process was demonstrated explicitly in Figure 4 of \citet{2013SimpsonClip}. Even in the context of more complex models of galaxy bias, such as those induced by tidal fields \cite{2012Bald}, we expect a similar behaviour. Non-linear contributions to the galaxy bias still predominantly arise in regions where $\delta$ is large, and these are the regions suppressed by the clipping transformation. 

There is however a fundamental limit on how much we can shield ourselves from the influence of the highest density regions,  and this stems from the estimation of the mean number density. In defining the fractional overdensity, $\delta_g \equiv \rho_g/\bar{\rho}_g - 1$,  the quantity $\bar{\rho}_g$ necessarily incorporates the abundance of galaxies across the whole volume, prior to clipping. Unlike dark matter perturbations where the total particle number is conserved, no such restriction applies to galaxy bias. For example if baryonic effects reduce the abundance of galaxies in clusters such that the total galaxy count across the survey volume is lowered by a small fraction $y$, then the inferred amplitude of fluctuations across the rest of the volume are overestimated  by
\[
\hat{\delta}_g \simeq \delta_g^t (1+y) + y ,
\]
where $\delta_g^t$ is the true fractional perturbation.

\subsection{Redshift Space Distortions: Linear Model} \label{sec:kaiserRSD}

Whilst the true spatial distribution of galaxies is expected to be statistically isotropic, their redshift-inferred distances receive an additional displacement due to their peculiar velocities, generating a statistically anisotropic configuration.  This permits a measurement of $f \sigma_8$, where the logarithmic linear growth rate is given by $f \equiv \ud \ln \delta / \ud \ln a$, and $\sigma_8$ defines the amplitude of linear perturbations.  

An additional source of anisotropic clustering arises from inaccuracies in the assumed geometry of the Universe, which is required when converting the observed values of angles and redshifts into a Euclidean framework. This can potentially generate false measurements of the growth rate \cite{simpsonp09}. In this work we shall consider the background expansion to be fixed to a flat \lcdm model with $\om = 0.27$ unless specified otherwise.

Clipping in redshift space carries additional complications. The small scale velocity dispersion associated with the `Fingers of God' effect will tend to move galaxies out of the high density peaks, and potentially into a surrounding region that lies below the clipping threshold.  We should therefore expect the removal of non-linear effects to be less efficient in redshift space. The velocity dispersion also causes the power spectrum to steepen at larger wavenumbers along the line of sight. This strong spectral slope enhances the relative amplitude of the higher order terms in (\ref{eq:clipAnalyticAnisotropic}), so these should not be neglected. 
 
On all but the largest scales, the real space cosmological density field at low redshifts is not well described by a Gaussian field. It may instead be considered as a superposition of a Gaussian component $\delta^{r(1)}$ and an extra field $\delta^{X}$ representing the conglomeration of all nonlinear corrections.
\[
\delta_m^{r} = \delta_m^{r(1)} +  \delta_m^{X} \, .
\]
The $ \delta^{X}$ field is largest where the linear approximation is most strongly violated -- both from the truncation of higher order terms in perturbation theory and more fundamentally from the assumption of a single-valued and curl-free velocity field.  The matter density field is traced by the galaxy number density field, which again may be decomposed into a Gaussian component, with a linear bias factor $b$, and a residual term such that 
\[
\delta_g^{r} = b \delta_m^{r(1)} +  \delta_g^{X} \, .
\]

Now moving to redshift space, the Gaussian component is described by the Kaiser model  \cite{1987MNRAS.227....1K}, which relates the real space linear density perturbations with those in redshift space, which we couple with a Lorentzian model of the velocity dispersion:
\[ \label{eq:kaiserDelta}
\delta_g^{s(1)}(k, \mu) = \frac{b + f \mu^2}{\sqrt{1 + k^2 \sigma_v^2 \mu^2 / 2}} \delta_m^{r(1)}(k) \, ,
\]
where $\mu \equiv k_\parallel / k$ is defined as the relative fraction of the wavevector that extends along the line of sight. The pairwise velocity dispersion $\sigma_v$ effectively smooths the field along the line of sight. Theoretically this quantity is given by
\[ \label{eq:sigmav}
\sigma_v^2 = \frac{f^2 H^2_0}{3 H^2(z) \pi^2} \int P_{\theta \theta}(k) \ud k \, ,
\]
where $P_{\theta \theta}(k)$ is the velocity power spectrum \cite{Fisher1995}. In practice we shall treat $\sigma_v$ as a free parameter, due to the uncertain behaviour of nonlinear motions.

The clipped galaxy field $\delta_{cg}^{s}$ can be represented as the sum of a clipped Gaussian field and a residual term $\delta^{\mathrm{x}}$. Therefore the resulting power spectrum may be expressed as the sum of three terms, the two autocorrelations and the cross-spectrum
\[
P_c^{s}(k) = \Plinclipz +  2P_{c1\mathrm{x}}^{s}(k) + P_{\mathrm{xx}}^{s}(k) \, .
\]
Given that the nonlinear terms encapsulated by $\delta^{X}$ dominate the clustering statistics at larger values of $\delta$, it experiences a much greater loss of  power than the linear component.  Therefore we should expect that after clipping the first term remains the dominant contribution to the total power over a broader range of scales.  Our simplest model for the clipped galaxy power spectrum is therefore encapsulated by
\[\label{eq:clipKaiser}
\eqalign{
P_c^s(k, \mu) &\simeq \Plinclipz  \, , \cr
                       &=  \clip \left\{  P_g^s(k, \mu) ; u_0 \right\}  \, ,
}
\]
\noindent where $\clip$ denotes the transformation defined by (\ref{eq:clipAnalyticAnisotropic}), $u_0$ is the normalised clipping threshold experienced by the Gaussian field as given by (\ref{eq:alin}), and 
\[ \label{eq:KaiserLorentzian}
P_g^s(k, \mu)  =  \frac{b^2 (1 + \beta \mu^2)^2}{1 + (k \sigma_v \mu)^2/2}  \Plin  \, ,
\]
where $P_g^s(k, \mu)$ is the Gaussian contribution to the galaxy power spectrum in redshift space, $\Plin$ is the linear matter power spectrum in real space, and the anisotropy parameter $\beta \equiv f/b$ quantifies the level of anisotropy in the galaxy power spectrum induced by linear velocity perturbations.

For the case of a hybrid field such as the one defined in (\ref{eq:hybrid_field}), then as clipping is applied, and nonlinear contaminations are suppressed relative to the linear contributions, we should expect the recovered value of $\beta$ to be closer to the theoretical value for a given $k_{\rm max}$; alternatively, we should be able to achieve the same level of systematic error in $\beta$ at a higher $k_{\rm max}$. The actual level of error and/or smallest scale to probe must be determined empirically using simulations, as described in the following section.

\subsection{Redshift Space Distortions: Nonlinear Model} \label{sec:rsdNonlinear}

Our base model is defined by the set of four parameters  $\left\{ \beta, \amp, \sigma_v, \Alin \right\}$, and relies upon the recovery of the linear matter power spectrum. However the linear power spectrum decays rapidly towards higher wavenumbers, and by $k \sim 0.5 \hmpc$ is typically an order of magnitude lower than the contribution from the one-loop correction to the power spectrum. Therefore despite the suppression of higher order terms, the inclusion of a suppressed one-loop term substantially improves the model for the real space power spectra of matter and galaxies \cite{2013SimpsonClip}. Further motivated by the results of \S \ref{sec:hybrid}, we introduce an extended model with the additional parameter $\epsilon$, which accounts for a higher order contribution to the power spectrum
\[ \label{eq:oneloop}
P^s_c(k, \mu)  =  \Alin \frac{b^2(1 + \beta \mu^2)^2}{1 + (k \sigma_v \mu)^2/2} \left[  \Plin +  \epsilon \Poneloop  \right] \, .
\]

The parameter $\Alin$ is the coefficient of the linear power which contributes to the clipped power spectrum, as defined in  the model of  (\ref{eq:A11A22model}). In general the value of $\Alin$ cannot be determined a priori. It may be evaluated empirically by considering the fractional change in amplitude of the large scale clustering of the field,
\[
\Alin = \lim_{r \to\infty} \frac{\xi_c(r)}{\xi(r)} \, .
\]

\section{Simulations} \label{sec:sims}

In order to test our theoretical models, we construct mock density fields for both dark matter and galaxies. These are derived from numerical simulations, and transformed into redshift space using the distant observer approximation.   In this section we summarise our methods for generating and modelling the power spectra associated with clipped cosmological fields, and for the estimation of their covariance matrices.

\subsection{Number Density Fields in Redshift Space} \label{sec:dmz}

For our mock dark matter field we utilise the $z=0$ snapshot from the Horizon Run 2 simulation \cite{2011KimHorizon}, which consists of $6$,$000^3$ particles within a periodic box of size $7$,$200 \hinvmpc$.  The amplitude of linear perturbations is $\sigma_8 = 1/1.26$, with a matter density $\om = 0.26$. The redshift-space density field is defined by considering a $1920^3$ grid using a Nearest Grid Point (NGP) scheme, where the particles are displaced along one axis in accordance with their peculiar velocity. This leaves us with a grid size of $3.75 \hinvmpc$ corresponding to a Nyquist frequency of $k_{\mathrm{N}}=0.84 \hmpc$. 

The mock galaxy catalogues are taken from \citet{Guo2011}, which applies a semi-analytic model to the halo merger trees of the Millennium-I simulation  \cite{2005Natur.435..629S}. Following \citet{SimpsonClip} we use the two snapshots at $z=0$ and $z=0.687$ in order to explore different amplitudes of matter perturbations and growth rates. With $\sigma_8 = 0.9$, the $z=0$ snapshot possesses a slightly higher amplitude of fluctuations than the Horizon simulation. Applying a stellar mass cut of $\log_{10}(M_\ast/M_\odot h) \ge 9$ leaves us with a distribution resembling that found in our GAMA sample \cite{2011Taylor_GAMA_stellarmass}. Number density fields are formed on both $128^3$ and $256^3$ grids across the $500 \hinvmpc$ box, with Nyquist frequencies of $k_{\mathrm{N}}=0.8 \hmpc$ and $1.6 \hmpc$ respectively. The standard deviations are $3.1$ for the coarse grid and $6.0$ for the high resolution field. 

In order to evaluate the true value of $\beta$ we need the linear bias parameter. The ratio of the real space galaxy power spectrum with that of the Millennium simulation's dark matter field gives $b = 1.125 \pm 0.01$ at $z=0$, on the largest available scales. At $z=0.687$ this increases to $b = 1.29 \pm 0.02$, while the growth rate is well approximated by $f(z) \simeq \Omega_m^{0.55}(z)$. 

\subsection{Threshold Selection}

After constructing each number density field, we apply the clipping transformation defined by (\ref{eq:clip}). In order to select a suitable threshold value $\delta_0$ we require an appropriate metric for defining the strength of clipping. For the case of a Gaussian field, the normalised threshold $u_0 \equiv  \delta_0/\sqrt{2} \sigma$ provides a natural measure for this in relation to the standard deviation of the field. However when working with fields that are highly non-Gaussian, we do not know \emph{a priori} what impact a given threshold will have. Some degree of iteration is therefore required in order to reach the desired reduction in power, characterised by $\Alin$.  

When working with the simulations, rather than quoting absolute values of the threshold  $\dThresh$, we choose to work in terms of the fraction of mass (or galaxies) removed. This way fields with larger fluctuations naturally adopt higher thresholds $\dThresh$. 

The disadvantage of a stronger (lower) threshold is a larger drop in the amplitude of the power spectrum, which in turn reduces the maximum wavenumber available before the shot noise contributions dominate. Very strong thresholds also induce a large contribution from the cross-power $P_{12}(k)$, as given by (\ref{eq:P12}), further reducing the power. The optimal choice of threshold is therefore one that adequately suppresses contributions to the power from  nonlinear structure and nonlinear bias, while maintaining a relatively high amplitude of the linear power spectrum. As shown in Figures 6, 7, and 8 of \citet{SimpsonClip}, the decay of the higher order term rapidly outpaces the contribution from the linear power spectrum, such that a factor of two reduction in linear power is sufficient to eliminate approximately 80\% of the nonlinear power.

For each field we explore a range of threshold values, since it is important to verify that different thresholds generate consistent parameter constraints. Performing a likelihood analysis that combines the power spectra derived from different thresholds may further improve parameter constraints. However an estimation of the covariance between the different spectra is beyond the scope of this work, here we shall only consider the analysis of each threshold separately.

\subsection{Methods} \label{sec:sim_methods}

The theoretical model for the clipped power spectrum is evaluated with the following procedure

\begin{itemize}
\item{Apply the redshift space model defined in (\ref{eq:KaiserLorentzian}) to determine $\hat{P}^s(k, \mu)$ from the linear power spectrum evaluated by CAMB \cite{2000CAMB}} combined with the chosen parameter values of $b$, $f$, and $\sigma_v$.
\item{Interpolate the power spectrum onto a 3D grid matching the specifications of the NGP lattice generated from the simulations.}
\item{Transform to the 3D correlation function $\xi(\sigma, \pi)$}
\item{Evaluate the clipped 3D correlation function  $\xi_c(\sigma, \pi)$ using (\ref{eq:clipAnalytic}).  We find it sufficient to truncate the series expansion at $n_{\mathrm{max}} = 4$ in order to establish sub-percent accuracy. This step is discussed further in Appendix \ref{sec:nOrder}. 
}
\item{Apply the inverse Fourier transform to determine the expected 3D power spectrum as a function of  wavenumber magnitude $k$ and orientation $\mu$, $P_c^s(k, \mu)$}.  The power spectrum is assigned to linearly spaced bins, with widths of  $\Delta k = 0.025$, $\Delta \mu = 0.2$. We maintain the same binning scheme throughout this work. Note that the  mean wavenumber contributing to the bin is in general at higher values of $k$  than the bin centre, due to the abundance of modes within a spherical shell $\ud k$ scaling as $k^2$. 
\end{itemize}

A four-dimensional likelihood grid is constructed, $\mathcal{L}(P_c(k, \mu) | \Alin, \sigma_v, \beta, \amp)$, where the parameter set relates to the model given by (\ref{eq:clipKaiser}) and (\ref{eq:KaiserLorentzian}).  

With the clipped power spectrum alone as the only source of information, we would have little knowledge of how strong the applied clipping has been, and therefore $\Alin$ is poorly constrained. However the drop in power relative to the original (unclipped) power spectrum  provides us with some extra information that can assist in constraining the range of $\Alin$ values. We therefore make use of the fractional loss of power experienced at  $k < 0.1 \hmpc$ and $\mu < 0.6$, to provide some external information for the value of $\Alin$. At higher values of $k$ and $\mu$ the larger contributions from nonlinear structure mean that they experience a greater drop in power.

We assign flat priors of  $0 < \Alin < 1$, $0<\beta<2$, $0< \amp <2$, and for the velocity dispersion we use a broad Gaussian prior $\sigma_v/\sqrt{2} = 300 \pm 300$ $\mathrm{ km s}^{-1}$, as defined in (\ref{eq:KaiserLorentzian}). Our results are largely insensitive to the particular choice of priors. 

\subsection{Covariance Estimation} \label{sec:cov}

The large volume of the dark matter field in the Horizon Run 2 simulation, $(7$,$200 \hinvmpc)^3$, permits a direct estimation of the $70 \times 70$ 
covariance matrix associated with the $P(k, \mu)$ bins by considering the covariance of  power spectra evaluated from 350 subvolumes, each of size  $(960 \hinvmpc)^3$. To evaluate the covariance matrix we employ the Ledoit-Wolf shrinkage estimator \cite{2004LedoitWolf}, following the prescription
\[
\Sigma = \delta^{\star} {\mathrm F}  + (1 -  \delta^{\star} ) {\mathrm S},
\]
where $\Sigma$ is our estimated covariance matrix, and $\delta^{\star}$ is the shrinkage constant. S is the sample covariance matrix, defined as the ensemble average over $N = 350$ sub-volumes

\[
S_{ij} = \langle \Delta P_i \Delta P_j \rangle \, ,
\]
where $\Delta P_i$ represents the deviation in power in the $i$th $P(k, \mu)$ bin from the sample mean. The shrinkage target F is defined in terms of the sample covariance
\[
F_{ij} = 
\begin{cases}
    S_{ij},                                      &  (i =j) \\
     \bar{r} \sqrt{S_{ii} S_{jj}}  ,      &  (i \neq j) 
\end{cases}
\]

\[
\bar{r} = \frac{2}{(N-1)N} \sum_{i=1}^{N-1} \sum_{j=i+1}^N \frac{S_{ij}}{\sqrt{S_{ii} S_{jj}}} \, .
\]
where $N=350$. The shrinkage constant $\delta^{\star}$ is estimated using publicly available code \footnote{\protect\url{http://www.econ.uzh.ch/faculty/wolf/publications/covCor.m.zip}}. We emphasise that this approach yields a considerably improved estimate of the covariance matrix than simply using the sample covariance $S$ alone. 

For the galaxy sample associated with the Millennium simulation, the volume is considerably smaller and therefore a direct estimate of the covariance matrix would be prohibitively noisy. Instead we only estimate the diagonal terms of the covariance matrix, again using the variance of subvolumes. The estimation of off-diagonal terms is determined by using the dark matter covariance as a template, such that
\[
C_{ij} = \mathrm{corr}(i,j)   \sqrt{C_{ii} C_{jj}}  \, .
\]
where $\mathrm{corr}(i,j)$ is the correlation matrix from the dark matter power spectrum.  

\subsection{Shot Noise Estimation} \label{sec:shot}

For the conventional unclipped power spectrum, the discrete nature of sources leads to an additional contribution of power, which if we assume to be Poissonian in nature is given by $P_s = \bar{n}^{-1}$. As clipping smooths the field above the threshold value, the shot noise contribution is \emph{reduced}. Approximating the noise field as Gaussian allows us to utilise (\ref{eq:alin}) to estimate $P_s \simeq f_V^2 \bar{n}^{-1}$, where $f_V$ is the fraction of the volume of the field lying below the clipping threshold. In this work the typical volume fraction is of the order  $\sim 1\%$, and therefore the correction to the shot noise is negligible. 

The power spectra of clipped fields are highly robust to changes in the number density of sampled points \cite{2013SimpsonClip}. The only noticeable consequence appears to be that fields with higher shot noise possess noisier power spectra and can therefore not utilise as wide a range of wavenumbers.

\section{Results from Simulations} \label{sec:sim_results}

First we present the power spectra from the dark matter and galaxy fields, at different clipping thresholds, before reviewing the results of the likelihood analysis. 

\begin{figure*}
\includegraphics[width=170mm]{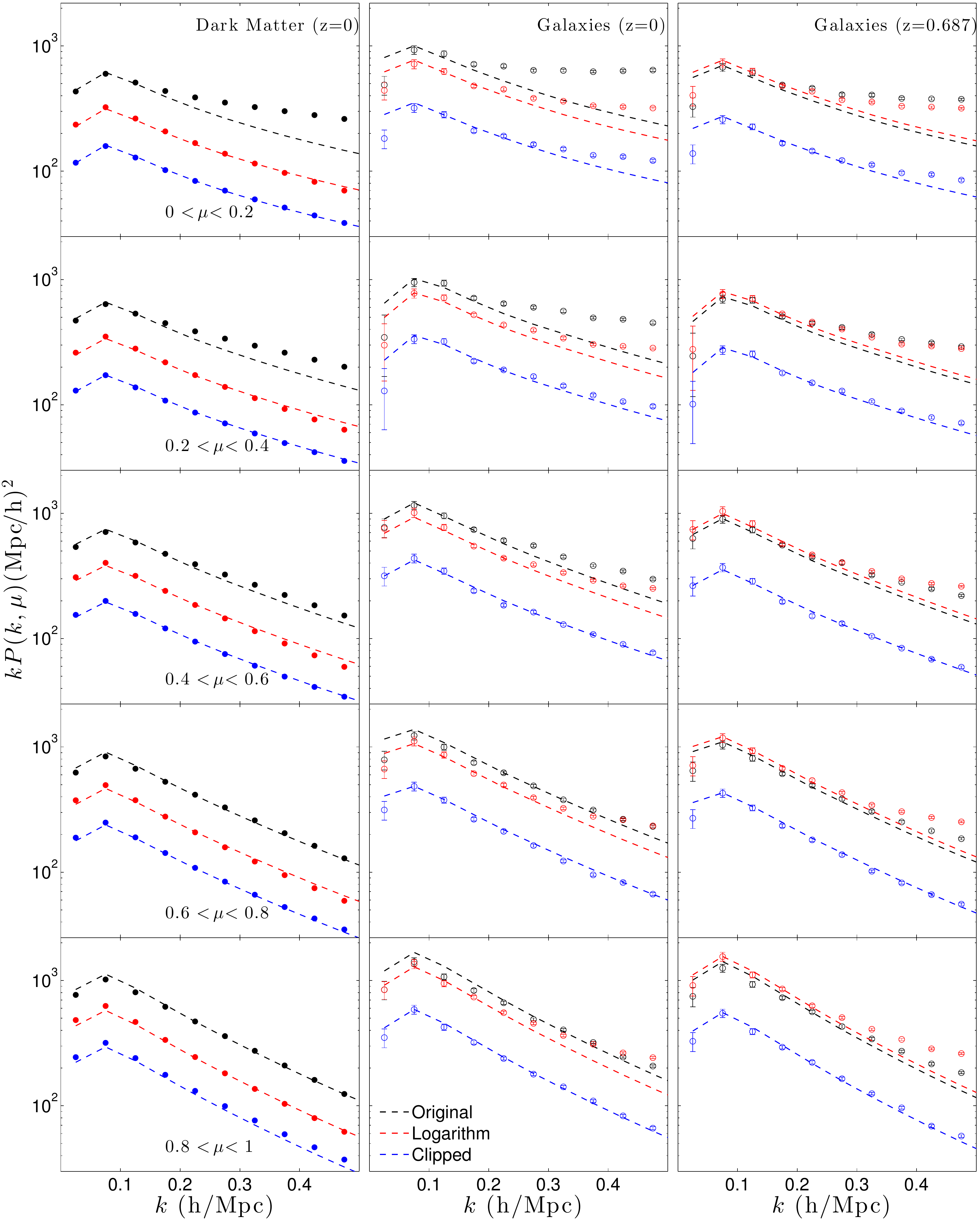} 
\caption{The redshift space power spectra  $P(k, \mu)$ from the simulations described in Section \ref{sec:sims}, divided into five angular bins. The three columns are dark matter from the Horizon simulation (left); galaxies at $z=0$ from the Millennium simulation (middle); and galaxies at $z=0.687$ from the Millennium simulation (right). Within each panel the original power spectrum is shown as black set of data points while the red and blue sets correspond to the power spectra generated after the number density field has been subject to logarithmic and clipping transformations respectively. The clipping threshold is chosen such that it lies below $20\%$ of the mass.
The dashed lines represent the linear model specified by (\ref{eq:KaiserLorentzian}). For the case of the transformed spectra, the amplitude of the model is rescaled to fit the data. \label{fig:linear_sim}}
\end{figure*}

\begin{figure*}
\includegraphics[width=170mm]{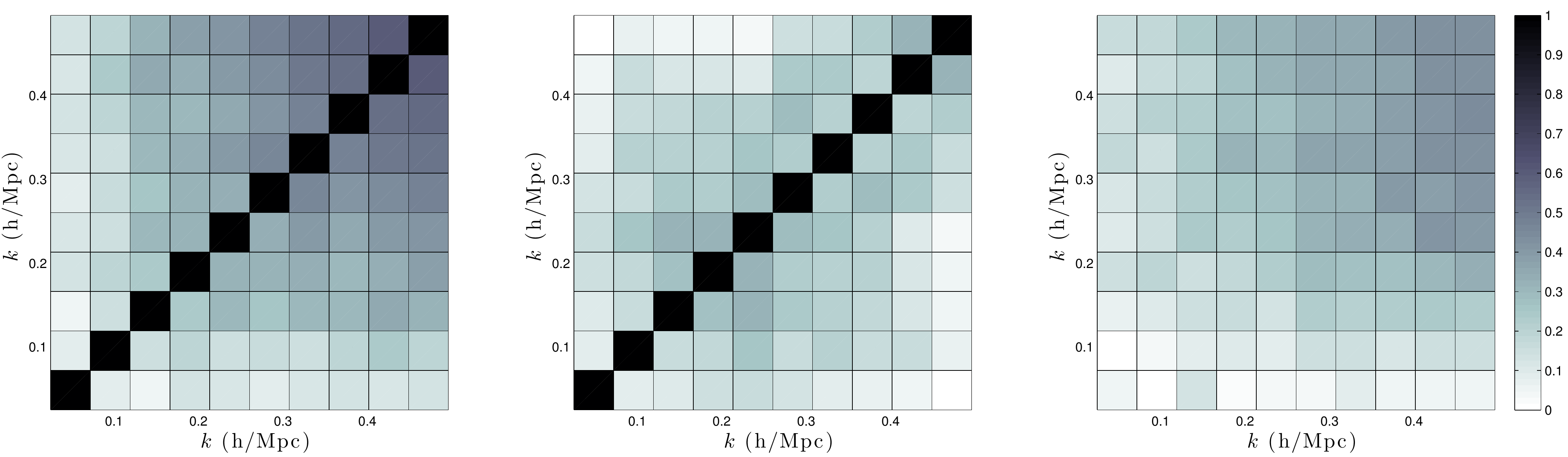}
\caption{Three subsections of the 70 parameter correlation matrix from the power spectrum $P(k, \mu)$ of the unclipped dark matter field in the Horizon simulation. The left hand panel corresponds to the lowest bin in wavevector orientation,  $0<\mu<0.2$, while the centre panels relates to the highest bin, $0.8<\mu<1$. Each section consists of ten bins in wavenumber $k$. The right hand panel is subsection of the correlation matrix associated with the cross-correlation between those two $\mu$ bins. The growing off-diagonal correlations towards smaller scales, particularly apparent in the lowest $\mu$ bin, is an indicator of the nonlinear structure that generates mode coupling. The correlation matrix was estimated by following the prescription described in \S \ref{sec:cov}.  \label{fig:corrMatrices}}
\end{figure*}

\subsection{Clipped Power Spectra} \label{sec:sim_pkmu}

The panels in the left hand column of Figure \ref{fig:linear_sim} compare the dark matter power spectrum from the Horizon Run 2 simulation with the linear theory prediction in each of five angular bins. Within each panel the uppermost set of points represent the power from the original unclipped field, while the central set is generated after applying a logarithmic transformation $\ln(1 + \delta)$. The lowest set of point corresponds to a field subject to a clipping transformations, (\ref{eq:clip}), with a threshold value $\delta_0$ chosen such that $20\%$ of the mass is removed.  Each dashed line corresponds to the model based on the linear power spectrum given by (\ref{eq:KaiserLorentzian}) where the amplitude is rescaled to fit the transformed spectra. The form of the real space linear power spectrum and the anisotropy parameter $\beta$ are assigned values according to the simulation parameters. The value of $\beta \equiv f/b$ is derived using the linear growth rate $f \simeq \Omega_m^{0.55}$ and the linear bias $b=1$ since we are working directly with the dark matter.  No error bars are displayed in these panels because the statistical error is considerably smaller than the marker size.   The bins in $k$ and $\mu$ were chosen to match the power spectra derived from the GAMA survey.

For such an evolved field the linear theory prescription given by (\ref{eq:KaiserLorentzian}) typically holds only on very large scales. Even at $k \sim 0.1 \hmpc$ the model overestimates the power in the highest $\mu$ bin by almost $10\%$, consistent with the findings of \citet{JenningsRSD2011}. However once either transformation is applied, the linear formalism of (\ref{eq:KaiserLorentzian}) provides a significantly improved description. Agreement with the clipped spectrum is better than $5\%$ within the range $(k < 0.5 \hmpc; \mu < 0.8)$.  This improvement in the modelling occurs due to the strong suppression of higher order terms in perturbation theory \cite{2013SimpsonClip}. The leading cause of tension with the model appears to be within the highest $\mu$ bin, which is perhaps unsurprising since these modes receive contributions from very small physical scales, due to the velocity dispersion of galaxies.  The central set of points illustrate the response of the power spectrum to another local transformation, the logarithm of the number density, $\ln(1 + \delta)$. \citet{Neyrinck2009} demonstrated that this can help linearise the power spectrum of the real space dark matter field. We find that considerable linearisation also occurs when applying the log transform to the dark matter field in redshift space. The shape of the linear theory power spectrum defined by (\ref{eq:KaiserLorentzian}) is recovered to better than $10\%$ for $k < 0.4 \hmpc$.

The central column of Figure \ref{fig:linear_sim} is in the same format as the left column, but illustrates the galaxy power spectrum from the Millennium Simulation at $z=0$ before and after applying transformations to the number density field. As before, the uppermost set of points in each panel represent the original unclipped field.  The lowest set of points are generated by clipping $20\%$ of the galaxies, which brings the data points closer to the shape of the linear model (\ref{eq:KaiserLorentzian}), as given by the dashed line. To obtain the value of $\beta$ for the model requires a combination of the linear bias, which is estimated from the amplitudes of the largest Fourier modes in the simulation to be $b=1.125$, and the growth rate $f(z\!=\!0) = 0.47$.  As quantified in \S \ref{sec:dmz},  the true value of the linear bias is only an estimate, however it remains a subdominant source of uncertainty.  The error bars in the central and right hand columns are significantly larger than those in the left hand column, reflecting the considerably smaller box of the Millennium-I Simulation (500 Mpc/h) compared to that of the Horizon Run (7,200 Mpc/h).

Deviations between the clipped spectrum and the linear model remain lower than $10\%$ for all data points at $k < 0.3 \hmpc$. Unlike the case of dark matter, it is the lowest $\mu$ bin that causes the greatest tension with the model. This may be due to the local motions of galaxies causing them to be displaced from their high density regions, which would make the clipping process less efficient, leaving behind a considerable proportion of the nonlinear contributions to the power spectrum.  The middle set of points in Figure \ref{fig:linear_sim} correspond to the logarithmic transformation, but now only a modest degree of linearisation is observed.  This reduced effectiveness can be attributed to the sampling noise from the galaxy fields with $\bar{n} = 0.01 \rm{ Mpc}^{-3}$. Due to this sensitivity to the level of shot noise, we shall focus on the clipping transform for the remainder of this work, which by contrast is largely insensitive to shot noise. 

The right hand column of Figure \ref{fig:linear_sim} explores the power spectra for a different sample of galaxies at a different redshift,  $z=0.687$. Here both the linear bias $b=1.29$ and growth rate $f(z\!=\!0.687)=0.77$ have changed from those of the central column, yet the outcome is similar, in that the application of clipping significantly improves the performance of the Kaiser model defined in (\ref{eq:KaiserLorentzian}).  As with the low redshift galaxy sample, the departure from linearity is less than $10\%$ for $k < 0.3 \hmpc$. 

Figure \ref{fig:corrMatrices} displays three subsections of the $70 \times 70$ correlation matrix (14 bins in $k$, 5 bins in $\mu$) associated with the unclipped dark matter field. The coupling of modes becomes particularly apparent towards higher wavenumbers, $k > 0.25 \hmpc$.   It is interesting to note that these off-diagonal terms fall by approximately $20 - 30\%$ after clipping has been applied. This decorrelation of neighbouring bins was previously observed in the real space power spectrum \cite{2013SimpsonClip}. 

The shrinkage constant also reduces considerably after the application of clipping. From the unclipped field we find a shrinkage constant of $\delta^{\star} = 0.25$. Thresholds selected to remove $5\%$ and $10\%$ of the dark matter yields shrinkage constants of $0.18$ and $0.14$ respectively, reflecting the increasingly Gaussian nature of these fields.  

\begin{figure*}
\includegraphics[width=170mm]{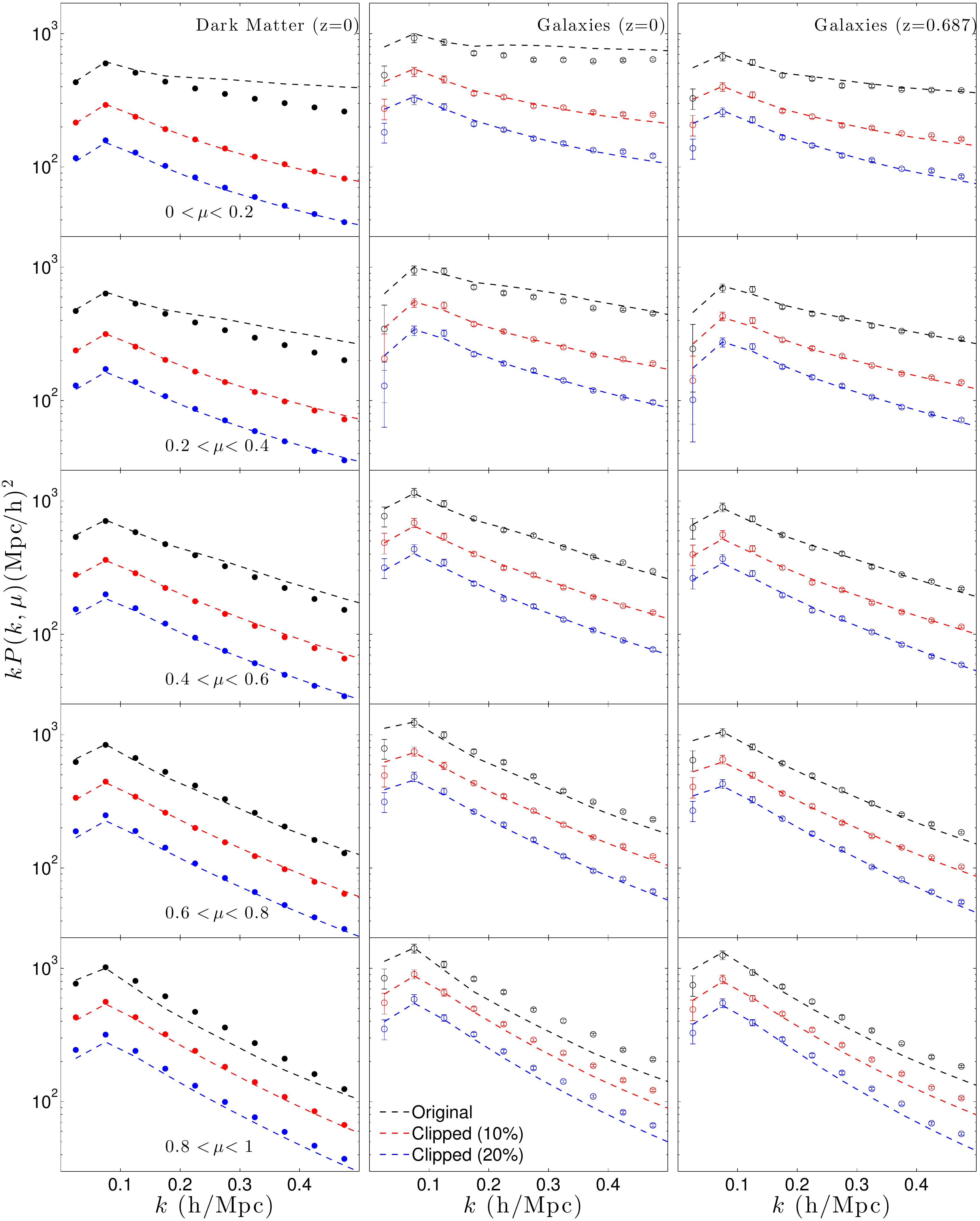} 
\caption{A similar format to Figure \ref{fig:linear_sim}, except here the dashed lines represent the best-fit non-linear model, as given by (\ref{eq:oneloop}). The central and lower sets of data points correspond to different clipping thresholds, removing $10\%$ and $20\%$ of the mass respectively. 
 \label{fig:nonlinear_sim}}
\end{figure*}

In Figure \ref{fig:nonlinear_sim} we explore the efficacy of fitting the power spectra of clipped fields with the non-linear model, specified by (\ref{eq:oneloop}), which is based on the model of the real space power spectrum presented in \citet{2013SimpsonClip}. The simulations in question are the same as those used in Figure \ref{fig:linear_sim}. The data points used to fit the model parameters are $k < 0.4 \hmpc$ and  $\mu < 0.8$. The upper set of data points corresponds to the original field, with the dashed line now making use of the one-loop power spectrum.  The middle and lower sets of data points relate to clipping $10\%$  and $20\%$ of the mass respectively. Now that an extra contribution from the one-loop power spectrum is included, the data points in the lower $\mu$ bins are much better accounted for, compared with the linear model in Figure \ref{fig:linear_sim}. However the highest $\mu$ bin appears significantly underestimated. 

We conducted further investigations by evaluating the power spectra of dark matter haloes in the Millennium simulation, before and after clipping, at the same two snapshots as the aforementioned galaxy catalogues. A very similar trend is found, whereby the clipped spectra are readily described by the model for $k < 0.4 \hmpc$. Since  these power spectra appear very similar to those displayed in Figure \ref{fig:nonlinear_sim}, they are not shown here. 

\begin{figure*}
\includegraphics[width=160mm]{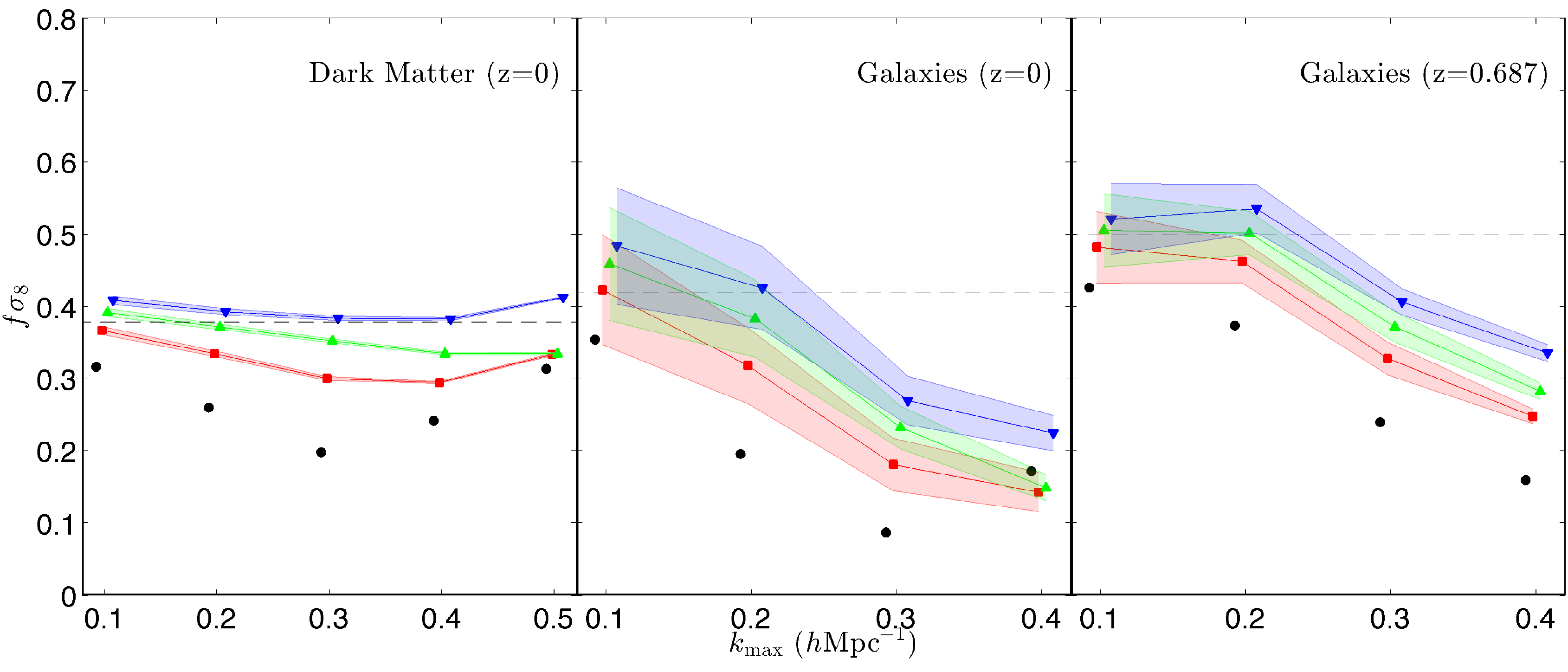}
\caption{Marginalised constraints on the normalised growth rate $f \sigma_8$ from the simulations,  when using only the linear power spectrum (\ref{eq:clipKaiser}) to model the data. The three shaded regions arise from the power spectra after clipping $5\%$ (squares), $10\%$ (triangles), and $15\%$ (inverted triangles) of the mass from each field. The horizontal dashed lines indicates the true values of  $f \sigma_8$ within each snapshot.  \label{fig:fsigma8Horizon}}
\end{figure*} 

\begin{figure*}
\includegraphics[width=160mm]{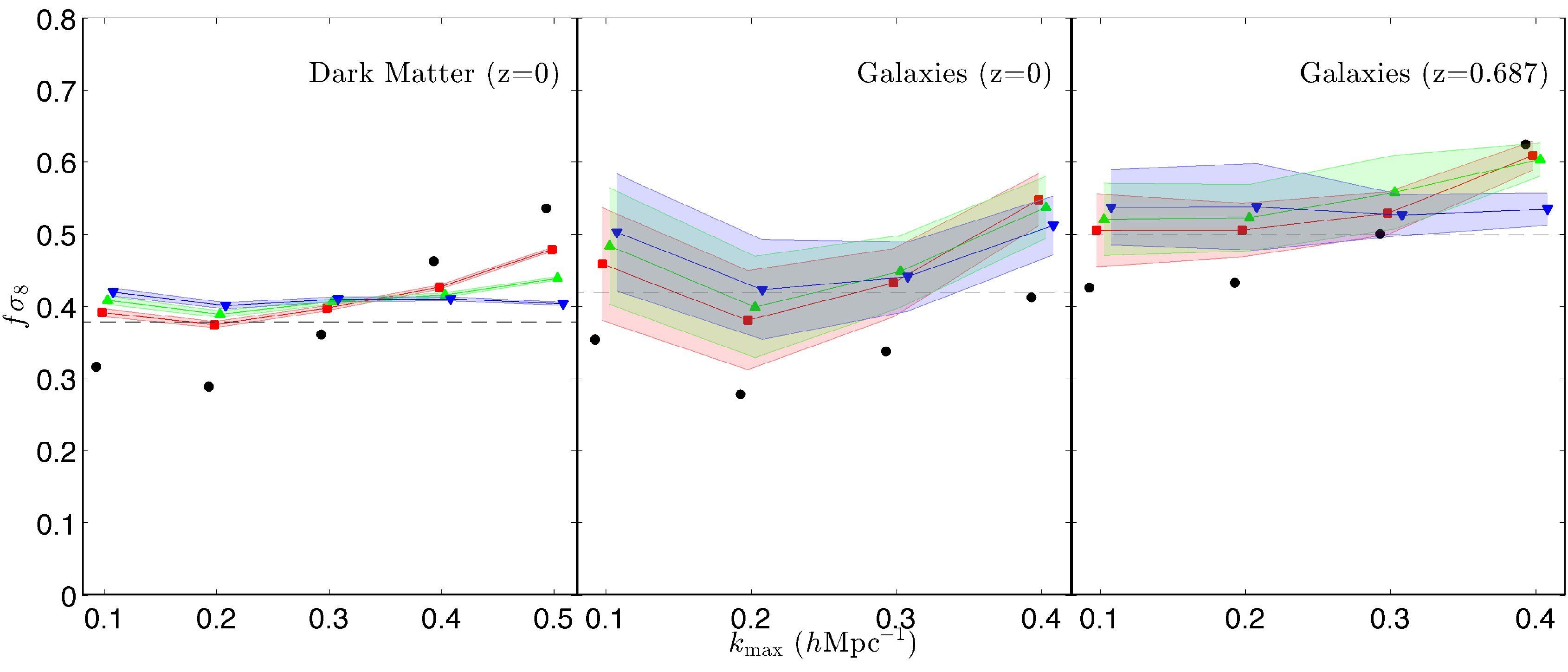}  
\caption{The same format as  Figure \ref{fig:fsigma8Horizon} except we now fit the data using the nonlinear model defined by (\ref{eq:oneloop}), with a uniform prior $-1<\epsilon<1$. }
\label{fig:fsigma8SimsOneLoop}
\end{figure*}

\begin{figure*}
\includegraphics[width=160mm]{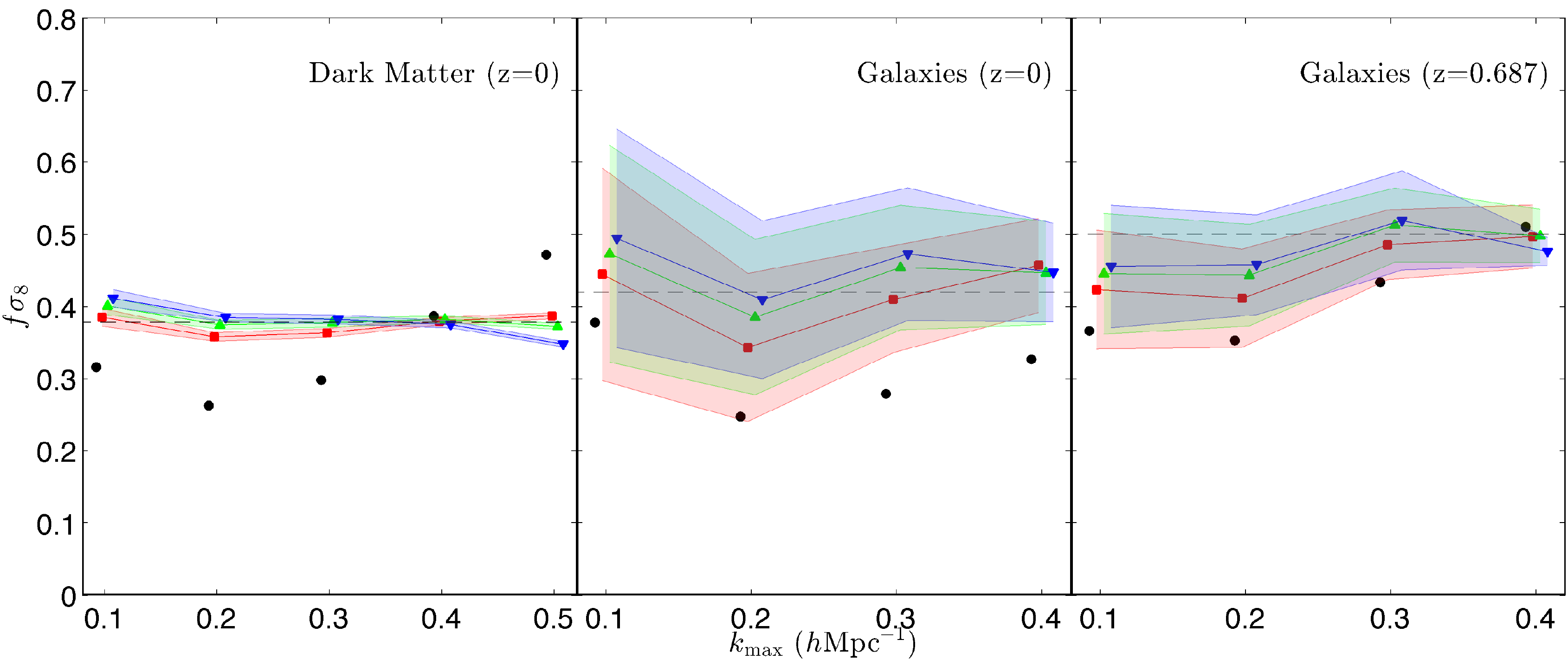}  
\caption{The same format as Figure \ref{fig:fsigma8SimsOneLoop} except here we only make use of intermediate wavevector angles $(0.2 < \mu < 0.8)$. }
\label{fig:fsigma8SimsOneLoopCut}
\end{figure*}

\subsection{Growth of Structure} \label{sec:sim_growth}

First we attempt to recover the correct cosmological parameters from the clipped power spectra, using only the linear model given by (\ref{eq:KaiserLorentzian}):
\[ \label{eq:clipKaiserGal}
P_c^s(k, \mu) =  \clip \left\{  \frac{b^2 (1 + \beta \mu^2)^2}{1 + (k \sigma_v \mu)^2/2} \Plin ; u_0  \right\} \, .
\]
Our basic set of parameters is  $\left\{ \beta, \amp, \sigma_v, \Alin \right\}$. The normalised growth rate $f \sigma_8$ may be derived from these via the following relation:
\[ \label{eq:fsigma8}
f \sigma_8 =    \beta \amp   \, .   
\]
Figure \ref{fig:fsigma8Horizon} shows the constraints on $f \sigma_8 $ when applying the linear model to the three different fields from the simulations, as a function of the maximum wavenumber $\kmax$. In each case the true value of $f \sigma_8$   is illustrated by a horizontal dashed line. The points have small horizontal offsets for clarity, and appear in order of increasing clipping strength.  The squares, triangles and inverted triangles correspond to clipping thresholds below $5$, $10$ and $15\%$ of the field respectively. The shaded regions represent their $68\%$ confidence limits.  For reference, the maximum likelihood points from the original unclipped field are shown as black circles.

Results from the dark matter field are shown in the left hand panel. Removing only $5\%$ of the mass is found to be sufficient to correct for much of the nonlinear behaviour on scales $k \lesssim 0.3 \hmpc$.   Similar behaviour is seen in the central panel of Figure \ref{fig:fsigma8Horizon}, which uses the $z=0$ galaxy sample, with a true value of $\sigma_8 = 0.9$.  In the right hand panel we find that the galaxy sample at  $z=0.687$ also significantly improves the recovery of the underlying cosmology when using the linear model on scales $k \lesssim 0.3 \hmpc$. The shaded regions are significantly broader in the central and right hand panels, reflecting the considerably smaller box of the Millennium-I Simulation ($500 \hinvmpc$) compared to that of the Horizon Run ($7,200 \hinvmpc$).

In the context of the more general model (\ref{eq:oneloop}), which invokes additional contribution controlled by the $\epsilon$ parameter, we know that for very weak clipping $\epsilon$ will be positive and when the clipping is very strong $\epsilon$ becomes negative as the contributions from cross spectra such as $P_{12}(k)$ dominate.  Therefore there is inevitably a threshold at which $\epsilon$ vanishes and the linear model offers a good fit to the data. This is represented by the inverted triangles in Figure \ref{fig:fsigma8Horizon}. 

\subsection{Nonlinear Model} \label{sec:sim_nonlinear}

We repeat our analysis using the extended model defined by (\ref{eq:oneloop}), which introduces an extra parameter $\epsilon$ to control the amplitude of the  clipped one-loop power spectrum, which is otherwise fully specified in terms of the linear power spectrum. Figure \ref{fig:fsigma8SimsOneLoop} shows the constraints derived on $f \sigma_8$ from the three simulated fields clipped with the same thresholds as Figure  \ref{fig:fsigma8Horizon}.  
The extra freedom in the real space matter power spectrum leads to significantly improved measurements at weaker clipping thresholds. In the case of dark matter we find that the maximum likelihood is within $10\%$ of the true value for each $\kmax \leq 0.4 \hmpc$ and for each threshold. Similarly in the galaxy field both clipping thresholds return more consistent constraints, and with only a modest loss of precision compared to the simpler linear model. However we find that the extreme values of $\mu$ are responsible for the bulk of the tension between the model and the data. It may be the case that more complex models such as those outlined by \citet{2013TaruyaRSD} may provide a better description of the anisotropies in the clipped power spectrum. An exploration of these models in the context of clipped fields is beyond the scope of this work. Restricting ourselves to $0.2 < \mu < 0.8$ improves agreement between the model and data to better than $2\%$ across all wavenumbers $k < 0.5 \hmpc$. This is reflected in Figure \ref{fig:fsigma8SimsOneLoopCut} where the tendency to overpredict  $f \sigma_8$ at the highest values of $\kmax$ is resolved.  

\section{Data} \label{sec:data}


The galaxy redshift surveys that have the greatest potential to benefit from clipping are those with a high number density of galaxies. A densely sampled field ensures that shot noise is low out to high wave numbers, even after the drop in the amplitude of the power spectrum due to clipping.  The GAMA survey provides an excellent basis for the first application of clipping to a real galaxy field. In this section we present details of the survey, and how the power spectra were generated. 

\subsection{The GAMA Survey} \label{sec:gama_survey}

The Galaxy and Mass Assembly (GAMA) project \cite{2010MNRAS.404...86B, 2010PASA...27...76R, 2011MNRAS.413..971D, Liske2015} is a multi-wavelength photometric and spectroscopic  survey.  The redshift survey, which has been carried out with the  Anglo-Australian Telescope (AAT), has provided a dense, highly-complete  sampling of large-scale structure up to redshift $z \sim 0.5$. The  primary target selection is $r < 19.8$ (where $r$ is an extinction-corrected SDSS Petrosian magnitude), using \texttt{TilingCatv41}.

Following \citet{2013MNRAS.436.3089B}, we analyzed a highly-complete subsample  of the survey dataset known as the GAMA II equatorial fields.  This subsample covers three $12 \times 5$ deg regions centred at 09h, 12h and  14h30m, which we refer to as G09, G12 and G15, respectively.  Galaxy  redshifts were obtained from the AAT spectra using a fully automatic  cross-correlation code that can robustly measure absorption and emission 
line redshifts \cite{2014MNRAS.441.2440B}.  We restricted the redshift  catalogue to galaxies with ``good'' redshifts (${\rm NQ} \ge 3$).  In 
order to obtain high-resolution measurements of the density field we restricted our analysis to the redshift range $0.002 \le z  \le 0.25$, where the galaxy number density exceeds $10^{-2} \, h^3$  Mpc$^{-3}$.  In the (G09, G12, G15) regions we utilized (32076, 37382,  36538) galaxies in our analysis. The comoving volume of each region is approximately $6.4 \times 10^6 (\! \hinvmpc$)$^3$ \cite{2013MNRAS.436.3089B}.

The survey selection function at each point, used in the calculation of the galaxy overdensity, was determined by combining the angular completeness map of the survey (which has a mean value of $97\%$ across  the three regions) with an empirical fit to the galaxy redshift distribution, performed after stacking together the data in the three regions to reduce fluctuations due to cosmic variance.  Full details of  the method are described in Section 3.2 of \citet{2013MNRAS.436.3089B}.

\subsection{Estimating the Clipped Power Spectrum} \label{sec:gama_clip}

The clipped power spectra for each GAMA region were determined 
for a given overdensity threshold $\delta_{\rm clip}$ as follows:

\begin{enumerate}

\item The galaxy distribution was binned on a common 3D grid to the selection function, with a resolution of $2 \, h^{-1}$ Mpc.  We denote the gridded distributions from the data and random samples as  $D(\vec{x})$ and $R(\vec{x})$, respectively. The random catalogues are sampled from the selection function constructed for the GAMA survey data, which combines the angular completeness in each survey region with an empirical smooth redshift distribution fit to a combination of the three regions.

\item The distributions were smoothed using a Gaussian kernel 
$G(\vec{x}) = e^{-(\vec{x}.\vec{x})/2\lambda^2}$.  We take $\lambda = 
2 \, h^{-1}$ Mpc for our analysis.  We denote the smoothed fields by
${\rm sm}(D)$ and ${\rm sm}(R)$.

\item The overdensity field for each region was estimated as 
$\delta(\vec{x}) = {\rm sm}(D)/{\rm sm}(R) - 1$, where the normalization 
of $R$ was fixed such that $\langle \delta \rangle = 0$.

\item The mean overdensity of each region in the redshift range $z < 0.25$, 
relative to the average of all three regions, was estimated using the 
measured number of galaxies as $\delta_{\rm reg} = (-0.0922, 0.0580, 
0.0341)$ for (G09, G12, G15).  The effective clipping threshold applied to the locally defined fluctuations within each 
region was then adjusted to $\delta_{\rm clip,eff}$ to accommodate these 
mean density fluctuations, where $1 + \delta_{\rm clip,eff} = (1 + 
\delta_{\rm clip})/(1 + \delta_{\rm reg})$.

\item For any grid cell with $\delta > \delta_{\rm clip,eff}$, the 
unsmoothed gridded data value was lowered to $D = R \, (1 + \delta_{\rm 
clip,eff})$.

\item The power spectrum of the clipped gridded data field $D$ was 
measured using Fast Fourier Transform (FFT) techniques following Section 
3.3 of \citet{2013MNRAS.436.3089B}.  The optimal-weighting estimation scheme of 
Feldman, Kaiser \& Peacock (1994) was applied, assuming a characteristic 
power spectrum amplitude $P(k) = 5000 \, h^{-3}$ Mpc$^3$.  We binned the 
power spectrum by $k$ and $\mu$, where $\mu$ is the cosine of the angle 
of the wavevector $\vec{k}$ with respect to the line-of-sight, using bin 
widths $\Delta k = 0.05 \, h$ Mpc$^{-1}$ and $\Delta \mu = 0.2$.  The 
integral constraint correction to the power spectrum was included in the 
estimation process (using the Fourier transform of the window function).

\item The amplitude of the measured power spectra was corrected for the 
mis-estimate of the mean density of the region, through multiplication 
by a factor $(1 + \delta_{\rm reg})^2$.

\item The convolution matrix, which is used to project a model power 
spectrum $P(k,\mu)$ to form a comparison with the data given the survey 
window function, was determined using the method outlined in Section 3.3 
of \citet{2013MNRAS.436.3089B}, in which the full FFT convolution is applied to 
a series of unit model vectors, and an equivalent matrix is constructed 
row-by-row.

\item The covariance matrix of the power spectrum measurement in 
$(k,\mu)$ bins was estimated by evaluating the sums described in Section 
3.4 of \citet{2013MNRAS.436.3089B}.  Initially the measured power spectrum in 
each bin was used to specify the cosmic variance component.  This produces an error estimate that is correlated with the data. To resolve this we modified the computation using an iterative procedure in which the 
best-fitting (convolved) theoretical model was determined and the covariance was 
re-estimated using that model.  Two iterations were used to ensure 
convergence.

\end{enumerate}

We repeated the above analysis for three different clipping thresholds $\delta_0 = \{8, 5, 4\}$. These values were selected on the basis of generating a suppression of linear power between $30$ and $60\%$. This provides an appropriate balance between the elimination of nonlinear structure and maintaining a high degree of signal to noise.  The three thresholds affect approximately $0.7\%, 2.1\%$, and $3.2\%$ of the field in terms of volume, and approximately $6.4\%, 15\%$, and $20\%$ in terms of galaxies. The latter quantity is defined as the fractional reduction in the value of $\sum D(\vec{x})$ due to clipping.

Unlike non-local transforms such as those used for reconstructing the baryon acoustic oscillations,  clipping commutes with the window function. This facilitates our interpretation of clipped power spectra, since the clipping transformation associated with the full (non-windowed) universe can be evaluated first, before compensating for the impact associated with the window function of the survey.

\section{Results from GAMA} \label{sec:results}

In this section we perform a likelihood analysis to estimate the normalised growth rate  $f \sigma_8$ at $z = 0.18$. The effective redshift of power spectrum measurements in the GAMA regions was determined by \citet{2013MNRAS.436.3089B}. The methodology from Section \ref{sec:sims}  is applied to each of the clipped galaxy power spectra from each of the three fields of the GAMA survey.   

\begin{figure}
\includegraphics[width=80mm]{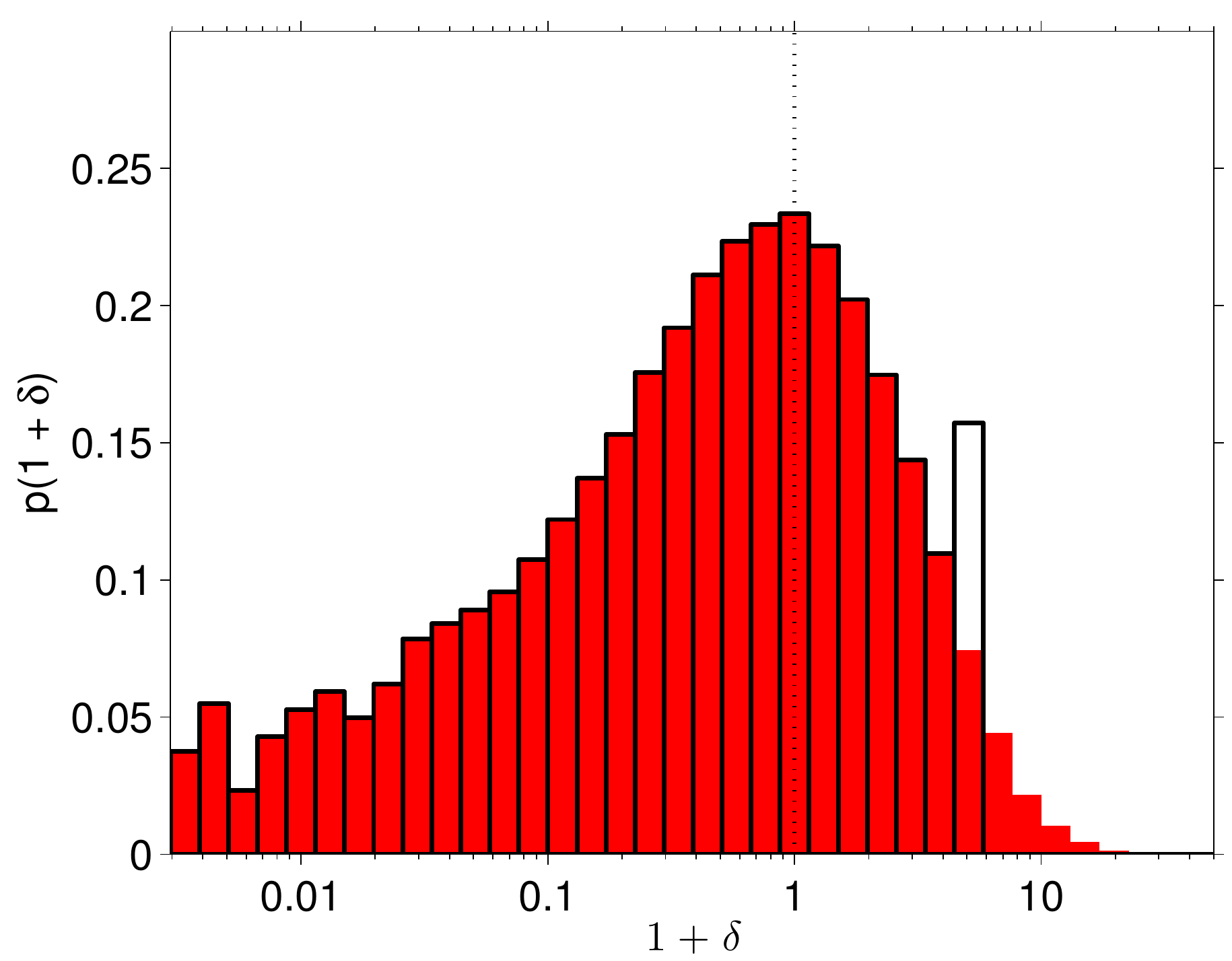}
\caption{The solid bars represent the probability density function for the galaxy density field within the G09 GAMA region. The hollow bars demonstrate the effect that applying a clipping threshold of $\delta_0 = 4$ would have on this field. A vertical dotted line demarcates the underdense and overdense regimes. \label{fig:GAMA_pdf}}
\end{figure}

Figure \ref{fig:GAMA_pdf} illustrates the effect of the clipping transformation. The solid bars represent the probability density function of the galaxy density field within the G09 region, defined in terms of 40 bins which are spaced equally in $\log (1 + \delta)$. The hollow bars show the resulting distribution function if we then apply a clipping transformation with a threshold of $\delta_0 = 4$.   The distribution function remains unaltered below the threshold value $\delta_0$, while all contributions from greater overdensities are compressed into the bin associated with the threshold value. 

\subsection{Clipped Power Spectra} \label{sec:gama_pkmu}

\begin{figure*}
\includegraphics[width=170mm]{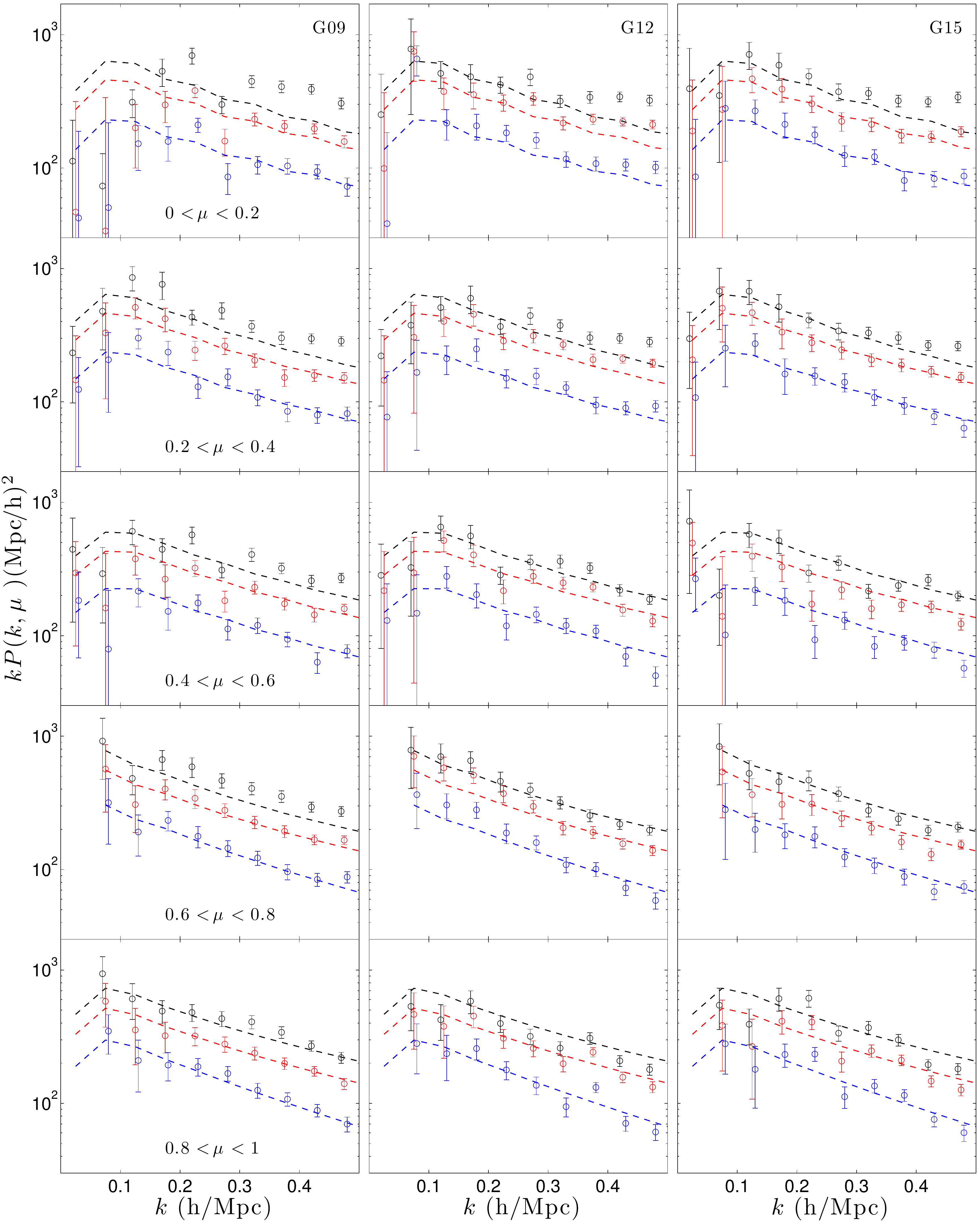}
\caption{The galaxy power spectra  $P_g(k, \mu)$ measured within the three GAMA regions, each divided into five angular bins. Within each panel the original power spectrum is shown as the upper set of data points while the middle and lower sets correspond to the power spectra generated after the number density field has been clipped with thresholds of $\delta_0 = 8$ and $\delta_0 = 4$ respectively. These thresholds generate a drop in large scale power of approximately $30\%$ and $60\%$. The upper dashed line is generated from the conventional model given by (\ref{eq:KaiserLorentzian}). The lower dashed line is the fiducial model for the clipped linear power (\ref{eq:clipKaiser}). At each clipping strength $\beta$ and $\sigma_v$ are set to their maximum likelihood values. The theoretical lines differ slightly between each region since each has been convolved with the window function of their respective region. 
For visual clarity a small horizontal offset is applied to the black and blue data points.  \label{fig:GAMAdata}}
\end{figure*}

\begin{figure}
\includegraphics[width=80mm]{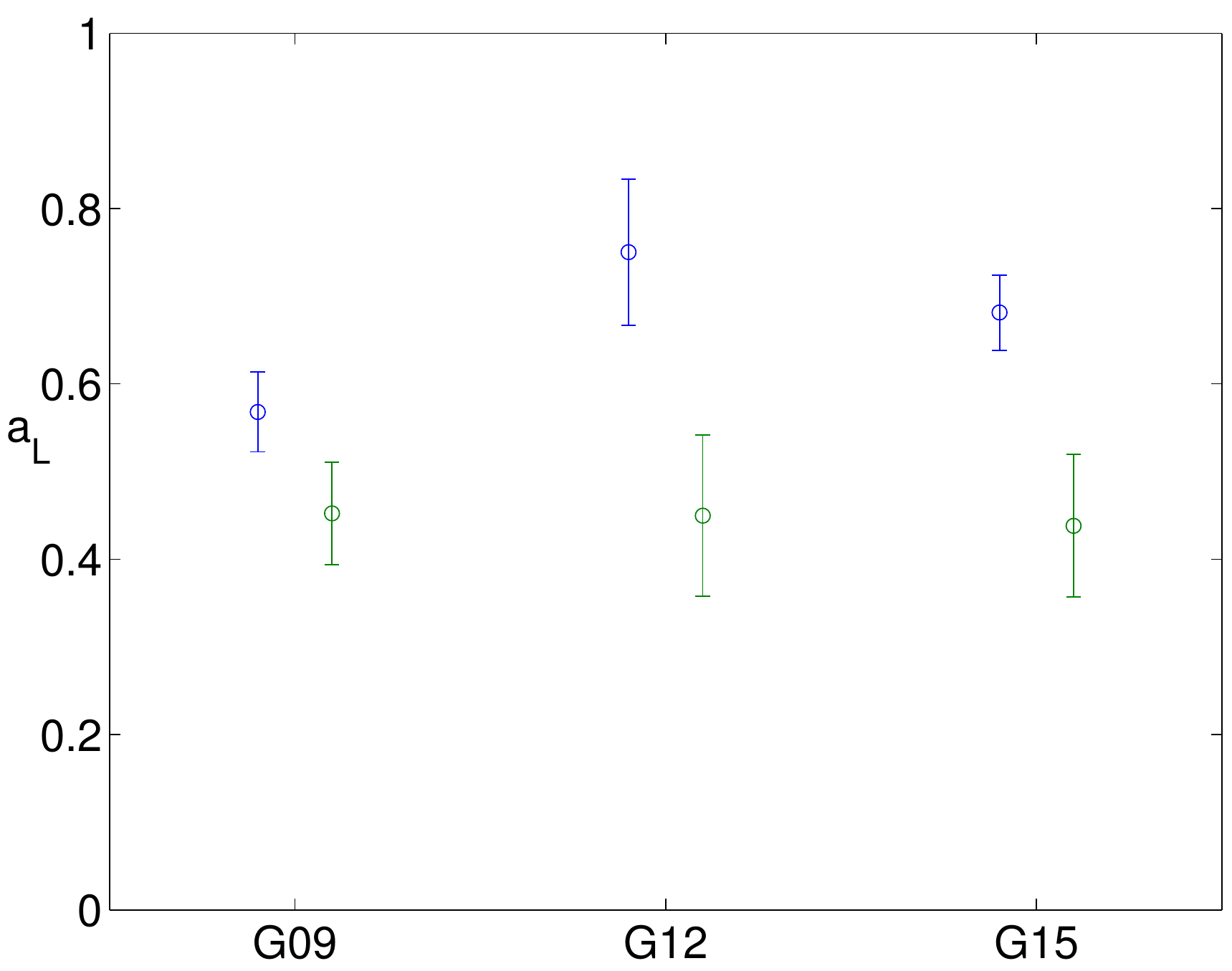}
\caption{Estimates of the fractional drop in linear power in each of the three GAMA regions as a result of applying clipping thresholds of $\delta_0 = 8$ (upper points) and  $\delta_0 = 4$ (lower points). \label{fig:GAMA_A11}}
\end{figure}

The panels in Figure \ref{fig:GAMAdata} illustrate the anisotropic power spectra derived from the three fields (G09, G12, and G15). As with the simulations, the power spectrum is divided into five equal bins in $\mu$, spanning $0 < \mu < 1$, while the wavenumber bin width is taken to be $\Delta k = 0.05$. Within each field the three sets of points correspond to the power in the field before (black) and after the application of clipping thresholds $\delta_0 = 8$ (red) and $\delta_0 = 4$ (blue).  At each clipping strength, the  dashed line reflects the linear model, with the maximum likelihood values of  $\beta$, $\sigma_v $, and $\amp$.   Estimates of the $\Alin$ parameter are shown in Figure \ref{fig:GAMA_A11}. These are determined by the fractional drop in $P(k, \mu)$ for $k < 0.1 \hmpc$ and $\mu < 0.6$, after clipping is applied. 

\begin{table}
\begin{tabular}{|c|c|c|c|c|}
\hline
$k_{\mathrm{max}}$& Unclipped&\textbf{$\delta_0 = 8$}&\textbf{$\delta_0 = 5$}&\textbf{$\delta_0 = 4$}\\\hline
0.1&${0.41 \pm 0.07}$&${0.34 \pm 0.19}$&${0.35 \pm 0.19}$&${0.32 \pm 0.19}$\\\hline
0.2&$0.43 \pm 0.08$&$\mathbf{0.23 \pm 0.10}$&$\mathbf{0.26 \pm 0.10}$&$\mathbf{0.24 \pm 0.11}$\\\hline
0.3&$0.25 \pm 0.05$&$0.19 \pm 0.07$&$\mathbf{0.22 \pm 0.08}$&$\mathbf{0.22 \pm 0.08}$\\\hline
0.4&$0.14 \pm 0.02$&$0.14 \pm 0.04$&$\mathbf{0.18 \pm 0.05}$&$\mathbf{0.19 \pm 0.06}$\\\hline
0.5&$0.02 \pm 0.01$&$0.06 \pm 0.04$&$0.16 \pm 0.05$&$0.22 \pm 0.06$\\\hline
\end{tabular}
 
\caption{ The maximum likelihood $f \sigma_8$ values and their associated $68\%$ confidence intervals, when fitting the linear model to the GAMA data. Each column represents a different clipping threshold while each row represents a different $\kmax$ condition. Table elements associated with a value of $\chi^2_{\mathrm{red}}<1.2$ are highlighted in bold. There is a strong covariance between nearby table elements,  so we do not attempt to combine constraints from different clipping strengths in this work. \label{tab:GAMA_fsigma8}}
\end{table}

\begin{table}
\begin{tabular}{|c|c|c|c|c|}
\hline
{$k_{\mathrm{max}}$}&{Unclipped}&{$\delta_0 = 8$}&{$\delta_0 = 5$}&{$\delta_0 = 4$}\\\hline
0.1&2.77&1.50&1.31&1.26\\\hline
0.2&3.68&1.00&0.88&0.73\\\hline
0.3&2.81&1.26&1.05&1.08\\\hline
0.4&2.66&1.35&1.01&1.00\\\hline
0.5&3.06&1.87&1.31&1.21\\\hline
\end{tabular}

\caption{ The reduced $\chi^2$ values for a variety of clipping thresholds and maximum wavenumbers, when fitting the linear model to the GAMA data. \label{tab:GAMA_chi2}}
\end{table}

\begin{figure*}
\includegraphics[width=160mm]{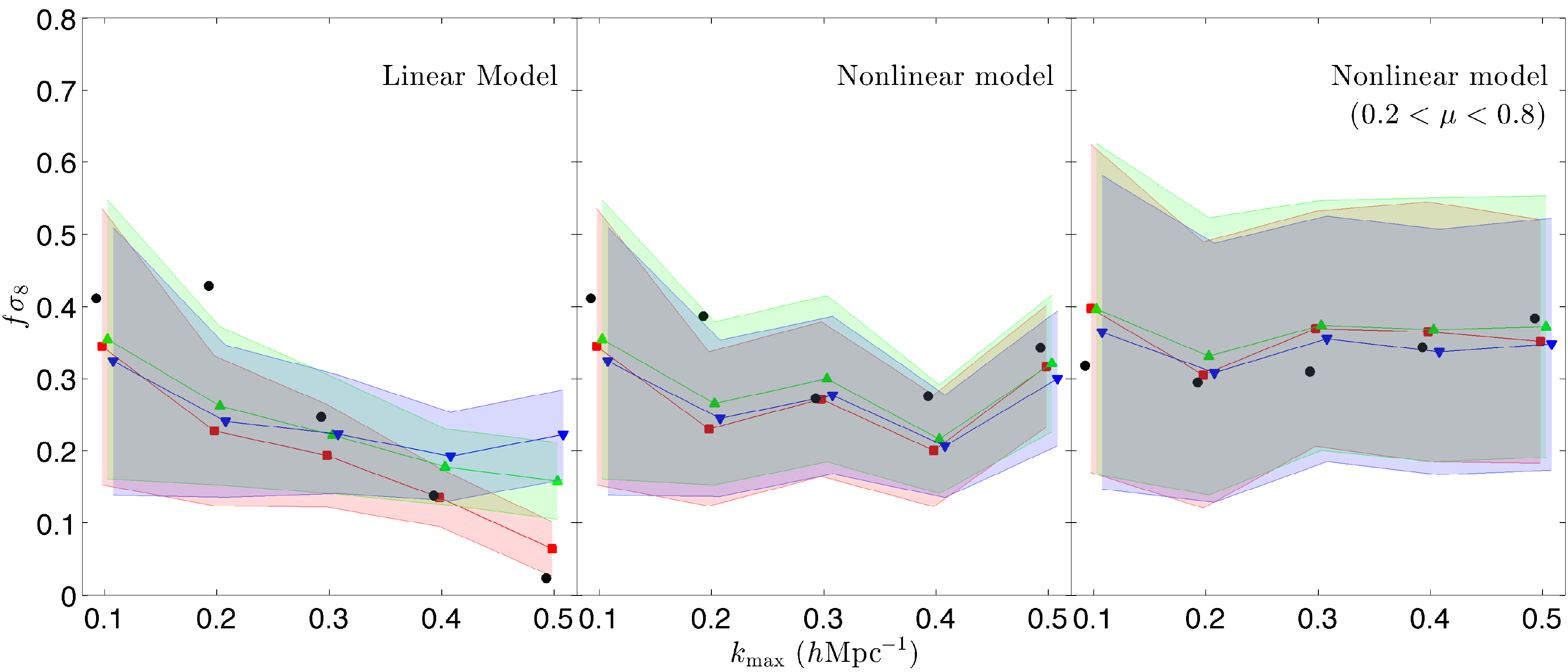}  
\caption{Constraints on $f \sigma_8$ from the power spectra of the galaxy field from GAMA. Results from the linear model are presented in the left hand panel, while those from the nonlinear model are illustrated in the central panel. The right hand panel uses the same model as the central panel but only includes intermediate wavevector angles $(0.2 < \mu < 0.8)$. 
Within each panel the squares, triangles and inverted triangles correspond to clipping thresholds $\dThresh$ of $8$, $5$, and $4$ respectively. The circles represent the maximum likelihood points when using the original field without clipping.  
} 
\label{fig:GAMAfsigma8Panels}
\end{figure*}

\begin{figure}
\includegraphics[width=80mm]{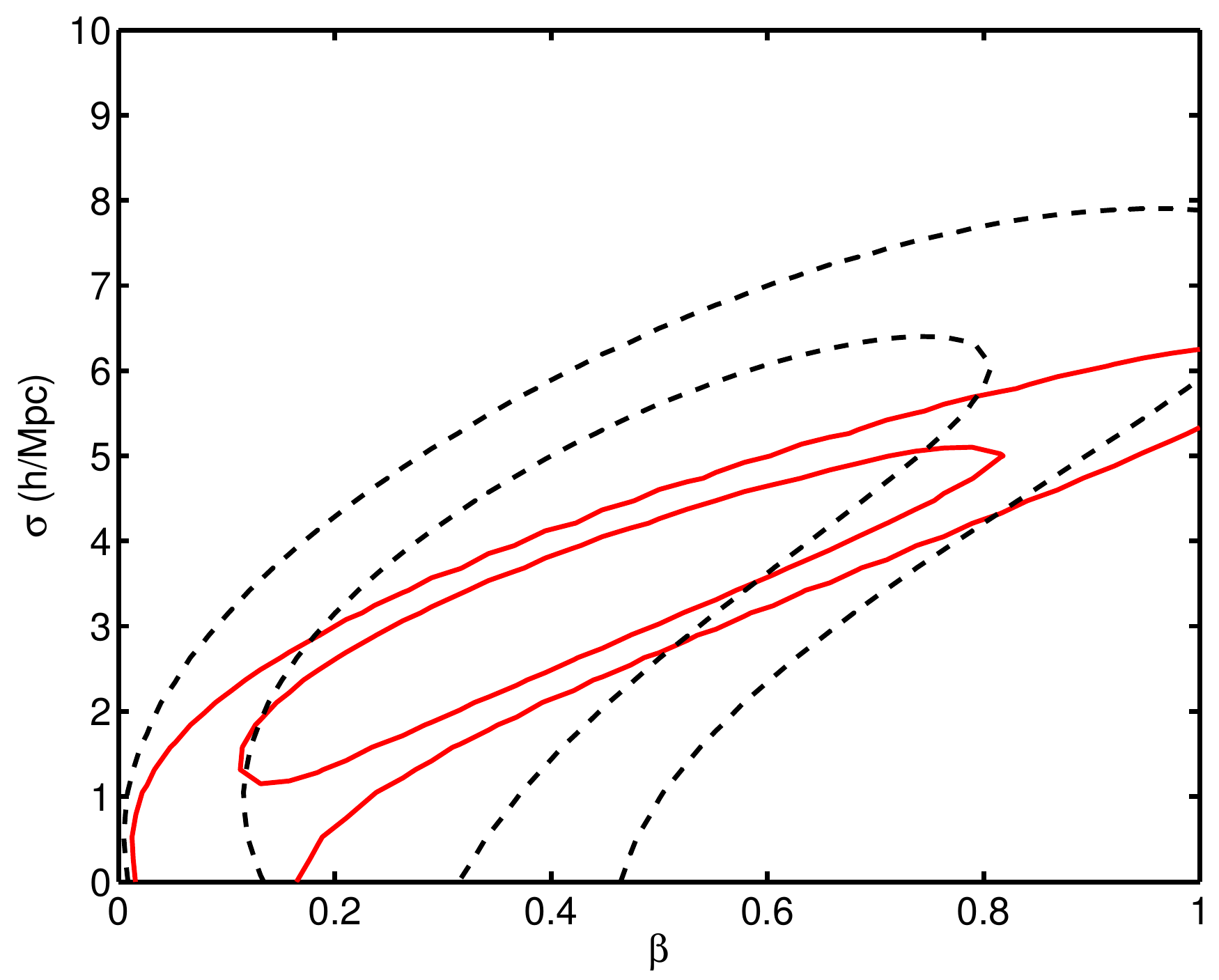}
\caption{The joint likelihood contours at 68\% and 95\% when evaluating the GAMA power spectra with $\dThresh = 4$, using the linear model. The degeneracy with the velocity dispersion $\sigma_v$ limits the gain in measuring the anisotropy parameter $\beta$  when utilising higher wave numbers.  The dashed contours show the degeneracy when fitting $\kmax =  0.3 \hmpc $, while the solid contours represent $\kmax =  0.5 \hmpc$.
\label{fig:sigma_contour}}
\end{figure}

\subsection{Linear Model}

Following the procedure outlined in Section \ref{sec:sim_growth}, we use the clipped power spectra from the three GAMA regions to measure the normalised growth rate $f \sigma_8 (z = \GAMAz)$.  Here we shall present results which combine the likelihoods of the three regions. Individual results from the three separate regions can be found in Appendix \ref{sec:GAMAfields}.  
 
First we employ the linear model defined by (\ref{eq:KaiserLorentzian}) and (\ref{eq:clipKaiser}), while fixing the shape of the linear power spectrum to the fiducial model. The left hand panel of Figure \ref{fig:GAMAfsigma8Panels} show the maximum likelihood values and $68\%$ error bars associated with $f \sigma_8$, under a range of different clipping thresholds and $\kmax$ values. As before, the squares, triangles and inverted triangles correspond to clipping thresholds of $\delta_0 = 8$, $5$ and $4$ respectively. The shaded regions represent their $68\%$ confidence limits.  For reference, the maximum likelihood points from the original unclipped field are shown as black circles. Their confidence limits are suppressed for clarity, as they do not provide an acceptable fit to the data. 

With the unclipped data the constraint on $f \sigma_8$ is highly sensitive to variations in $\kmax$, which is consistent with the 
behaviour found in the simulations. Since the model inevitably underestimates the amount of real space power towards larger $k$, this leads to an under-estimation of $\sigma_v$ which in turn biases the estimate of $\beta$ to be low.  Applying a high clipping threshold ($\delta_0 = 8$) show a modest improvement in terms of consistency and goodness of fit. Stronger thresholds of $\delta_0 = 5$ and $\delta_0 = 4$  provide a much improved agreement with the model, and more consistent results towards higher wavenumbers.  The maximum likelihood values are displayed in Table \ref{tab:GAMA_fsigma8}, and the reduced $\chi^2$ values are can be found in Table \ref{tab:GAMA_chi2}. The power spectra associated with the clipped fields are found to adhere to the linear theory prediction more closely than the original field.  From the simulations we expect a significant systematic error to arise at $\kmax \gtrsim 0.3 \hmpc$, so for the linear model we use $k < 0.2 \hmpc$ to find $f \sigma_8 (z = \GAMAz) = 0.26 \pm 0.10$.   Constraints from the clipped fields show a more consistent result across the range of wavenumbers than the original field, and also have a much improved goodness of fit.

Extracting robust constraints from higher wavenumbers requires a higher order model, since at these scales the amplitude of the linear power spectrum falls far below the non-linear contributions. 

\subsection{ Non-linear Model}

The central panel of Figure \ref{fig:GAMAfsigma8Panels} shows the constraints on $f \sigma_8$ when using the extended model defined by (\ref{eq:oneloop}). Again we find consistent behaviour between different clipping strengths, and across a variety different maximum wavenumbers. The extra degree of freedom does not appear to significantly weaken the constraints. Guided both by the performance of simulations, and the goodness of fit between the data and the model, we adopt our benchmark measurement to be $f\sigma_8(z= \GAMAz) = \GAMAfsig \pm \GAMAfsigError$, using $k < 0.3 \hmpc$ and $\delta_0 = 4$. This measurement is consistent with that derived from the linear model, and serves as the central result of this work. Our result is consistent with the findings of \citet{2013MNRAS.436.3089B}, who used the same (unclipped) galaxy field to determine $f\sigma_8(z= \GAMAz) = 0.36 \pm 0.09$. The full set of constraints on $f \sigma_8$ is presented in Table \ref{tab:GAMA_fsigma8_loop}, while the reduced $\chi^2$ values are displayed in Table \ref{tab:GAMA_chi2_loop}.

A more conservative approach is to restrict our analysis to intermediate wavevectors, $0.2 < \mu < 0.8$, and the results are shown in the right hand panel of Figure \ref{fig:GAMAfsigma8Panels}. While the susceptiblility to systematic errors has been reduced, there is also a substantial loss of precision. Therefore even when making use of the full range of wavenumbers, $k < 0.5 \hmpc$, the resulting confidence interval is found to be $f \sigma_8 = 0.35 \pm 0.17$. 

While a significant improvement in precision is achieved by increasing the $\kmax$ value up to $0.3 \hmpc$, thereafter the gain is not as great as one might expect from the increased abundance of Fourier modes.  One of the key limitations remains the degeneracy between the anisotropy parameter $\beta$ and the velocity dispersion $\sigma_v$. Their joint likelihood is illustrated in Figure \ref{fig:sigma_contour}, for two different values of $\kmax$. Clearly if additional information were available to measure or predict the value of $\sigma_v$,  substantial improvements in the measurement of $f \sigma_8$ could be made.   Another factor which limits the gains available from smaller scales is the shot noise. Its fractional importance is amplified by the reduction in the amplitude of the power, which becomes particularly apparent at the lowest threshold.

\begin{table}
\begin{tabular}{|c|c|c|c|c|}
\hline
{$k_{\mathrm{max}}$}&{Unclipped}&{$\delta_0 = 8$}&{$\delta_0 = 5$}&{$\delta_0 = 4$}\\\hline
0.1&${0.41 \pm 0.07}$&${0.34 \pm 0.19}$&${0.35 \pm 0.19}$&${0.32 \pm 0.19}$\\\hline
0.2&$0.39 \pm 0.08$&$\mathbf{0.23 \pm 0.11}$&$\mathbf{0.27 \pm 0.11}$&$\mathbf{0.25 \pm 0.11}$\\\hline
0.3&$0.27 \pm 0.07$&$0.27 \pm 0.11$ & $\mathbf{0.30 \pm 0.11}$&$\mathbf{0.29 \pm 0.10}$\\\hline
0.4&$0.28 \pm 0.05$&$0.20 \pm 0.08$ & $\mathbf{0.22 \pm 0.08}$&$\mathbf{0.21 \pm 0.07}$\\\hline
0.5&$0.34 \pm 0.04$&$0.32 \pm 0.08$ & $0.32 \pm 0.10$&$0.30 \pm 0.09$\\\hline
\end{tabular}

\caption{The same format as Table \ref{tab:GAMA_fsigma8} but now using the nonlinear model defined by (\ref{eq:oneloop}). \label{tab:GAMA_fsigma8_loop}}
\end{table}

\begin{table}
\begin{tabular}{|c|c|c|c|c|}
\hline
{$k_{\mathrm{max}}$}&{Unclipped}&{$\delta_0 = 8$}&{$\delta_0 = 5$}&{$\delta_0 = 4$}\\\hline
{0.1}&2.77&1.50&1.31&1.26\\\hline
0.2&3.12&0.96&0.88&0.76\\\hline
0.3&2.04&1.36&1.12&1.12\\\hline
0.4&2.34&1.39&1.02&1.02\\\hline
0.5&2.55&1.86&1.42&1.27\\\hline
\end{tabular}

\caption{ The reduced $\chi^2$ values associated with each of the maximum likelihood values presented in Table \ref{tab:GAMA_fsigma8_loop}. \label{tab:GAMA_chi2_loop}}
\end{table}

\section{Discussion} \label{sec:conclusions}

\begin{figure}
\includegraphics[width=80mm]{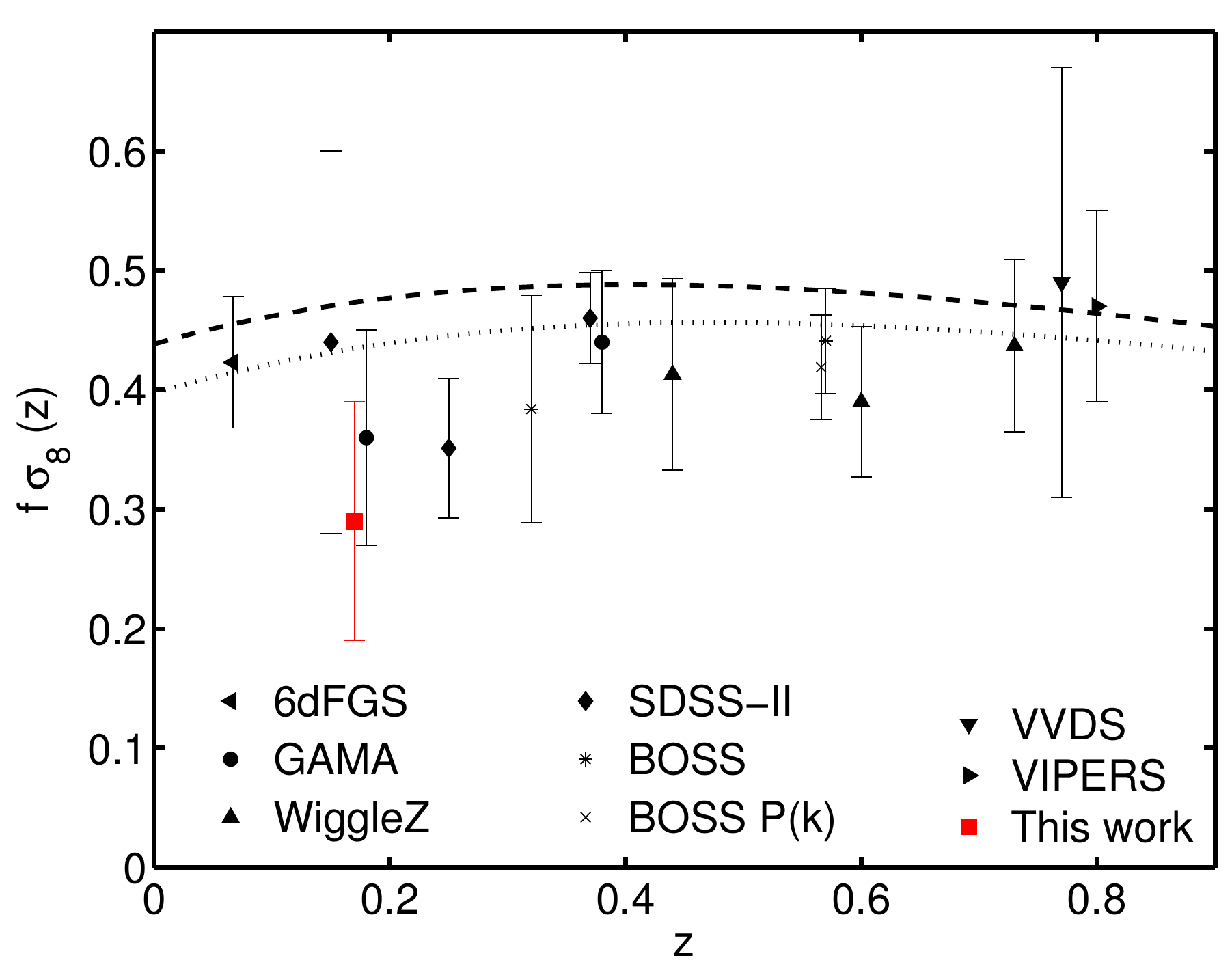}
\caption{Comparison of measurements of $f \sigma_8$ as measured from anisotropic clustering in galaxy redshift data.  These include 6dGFS \cite{Beutler2012MNRAS}; WiggleZ \cite{Blake2012MNRAS}; BOSS \cite{2013arXiv1312.4889C, 2014MNRAS.439.3504S}; BOSS P(k) \cite{2014MNRAS.443.1065B};  VVDS \cite{2008Natur.451..541G}; SDSS-II \cite{2014arXiv1409.3238H, 2012MNRAS.420.2102S};  VIPERS \cite{2013A&A...557A..54D}; GAMA \cite{2013MNRAS.436.3089B}. 
The dotted and dashed lines represent theoretical predictions for a flat LCDM cosmology, as given by [$\om$, $\sigma_8$] pairs from the maximum likelihood values from Planck \cite{2015PlanckParams} using only polarisation [0.286; 0.796], and the full dataset [0.3156; 0.831].
\label{fig:fsig8_surveys}}
\end{figure}

We have developed the clipping procedure proposed in \citet{SimpsonClip} to enable its application to anisotropic fields, and applied this new analysis technique to the $z < 0.25$ sample from the GAMA survey. A simple model based on the linear power spectrum is used to measure the normalised growth rate  $f \sigma_8 (z = \GAMAz) = 0.26 \pm 0.10$ at $\kmax = 0.2 \hmpc$. Employing a higher order model from perturbation theory allows the use of higher wave numbers, while still not requiring numerical simulations for calibration. For this case we find $\fsig(z= \GAMAz ) = \GAMAfsig \pm \GAMAfsigError$ when using  $\kmax = 0.3 \hmpc$ and density fluctuations $\delta_g < 4$. These results alone are not in significant tension with expectations from the Planck data \cite{2015PlanckParams} within the context of a standard \lcdm model. However, they do add to a growing body of evidence that appears to prefer a lower amplitude of density perturbations at low redshifts. Such evidence includes weak gravitational lensing \cite{syspaper}, galaxy clusters \cite{Xrays}, and a number of other measurements of redshift space distortions \cite{2013PhRvLMacaulay}.  This trend is also visibly apparent in Figure \ref{fig:fsig8_surveys}, but there are several possible explanations for this behaviour. One interpretation of this is the reduction in the quadrupole generated by nonlinear motions, relative to the Kaiser prediction, as illustrated in Figure 2 of \cite{JenningsRSD2011}. However our result would be largely insensitive to this effect. Another interpretation is the presence of nonlinear galaxy bias. An additional isotropic contribution to the power dilutes the strength of the anisotropic clustering signal. This effect can be seen in Figure \ref{fig:fsigma8SimsOneLoop} where, before clipping is applied, the inferred value of $\fsig$ is significantly lower than the correct value even when using  $k \lesssim 0.1 \hmpc$. The simplest forms of nonlinear bias are strongly suppressed by clipping, but others such as stochastic bias are likely to remain, and therefore merit further investigation. It is also important to note that most studies of redshift space distortions rely upon a prior on the range of possible background geometries based on results from WMAP. The portion of the error budget associated with the Alcock-Paczynski effect in each survey will therefore be highly correlated \cite{1979Natur.281..358A, 1996MNRAS.282..877B, simpsonp09}.

As was found to be the case in real space \cite{2013SimpsonClip}, the preferred range of clipping thresholds are typically those that reduce the linear power by around $25 - 50\%$. For our galaxy field this corresponded to thresholds in the range  $4 < \delta_0 < 8$. Higher thresholds lead to weaker clipping, which is less effective at removing the problematic contributions from nonlinear structure. Meanwhile stronger clipping from lower thresholds leads to reduced signal-to-noise, and may also induce a significant cross-correlation between the Gaussian and residual fields. 

At present the precision of our measurement of the growth of structure is limited by the degeneracy with the velocity dispersion $\sigma_v$. By applying a group finding algorithm to the galaxy catalogue it may be possible to reduce the influence of the Fingers of God. This would also improve the efficiency with which clipping removes peaks in the density field.  

Previous measurements of redshift space distortions either rely heavily on calibration from numerical simulations, or on more complex approaches to perturbation theory. Each of these rely on certain model-dependent assumptions, such as a linear bias model.  Clipping is a complementary approach as it can identify whether the galaxy bias is showing signs of scale-dependence.  The degree of covariance between parameter constraints obtained from the clipped analysis and a conventional non-linear analysis has yet to be quantified, however potentially these two approaches could be combined to yield considerable additional information within the same survey volume. It may also be beneficial to perform a combined analysis of power spectra from multiple clipping thresholds.

Clipping may also be applicable to a number of other cosmological fields, which we shall consider in turn.

\noindent \textit{Cosmic Microwave Background:}

\noindent In the early Universe, cosmological perturbations are know to be highly Gaussian. A recent analysis of the Cosmic Microwave Background (CMB) from Planck severely limits the amplitude of local departures from Gaussianity $f_\mathrm{NL} = 2.5 \pm 5.7$ \cite{2015Planck_fNL}. Applying a clipping transformation to the CMB anisotropies is unlikely to be beneficial, since the Gaussian component is already highly dominant. However it may be of interest to identify whether features such as the lack of power on large angular scales, and the power asymmetry on the sky, remain intact after clipping, or are exacerbated. 

\noindent \textit{Lyman-$\alpha$ forest:}

\noindent At later times, the cosmological perturbations are again detectable in the absorption lines of quasars. This technique has been used to detect the baryon acoustic oscillations in BOSS \cite{2014BOSSLyAlpha}. In this case the observed tracer already experiences a transformation similar to clipping, in that the highest density regions form damped Ly$\alpha$ systems. The results of Section \ref{sec:Theory} may be generalised to other local transformations. For example the transformation relating the local density to the observed flux $F$ is often approximated as

\[ \label{eq:transform}
g(\delta) =  \exp \left[-A(1 + \delta)^\alpha \right] \, ,
\]
which can be used in conjunction with (\ref{eq:price}) to directly compute the flux correlation function in terms of the linear power spectrum. 

%

\noindent \textit{Weak Gravitational Lensing:}

\noindent The shapes of high redshift galaxies are coherently distorted by the intervening matter perturbations. Cosmic shear offers the most direct insight into the dark matter distribution at lower redshifts, yet uncertainties in the small scale power spectrum limit the amount of cosmological information that may be extracted. 
Gaussianisation of the convergence field have been proposed by several authors, such as \citet{2011ApJ...729L..11S, 2011MNRAS.418..145J, 2011PhRvD..84b3523Y}.  However as highlighted in \cite{2011MNRAS.418..145J}, in the presence of shape noise the benefits of the transformation are minimal. This is due to the substantial reduction in the amplitude of the resulting power spectrum. Applying clipping here with a suitably high threshold may be advantageous as it can suppress the strongest sources of nonlinearity while still preserving a high level of signal to noise. But the observable field appears in a projected two-dimensional form, due to the broad lensing kernel, so the identification and suppression of peaks is a less efficient procedure compared with the full three-dimensional data that can be acquired from the distribution of galaxies. 

\noindent \textit{Galaxy Clustering:} 

\noindent Cosmological information from the galaxy power spectrum can be split into three categories: geometric information from the baryon acoustic oscillations; primordial information from the broader shape of the power spectrum; and gravitational information from the degree of anisotropic clustering.  

Local density transformations such as clipping ameliorate nonlinearities associated with high density regions. One form of nonlinearity for which this is not the case is that associated with the smoothing of the baryon acoustic peak in the galaxy correlation function.  Non-local transformations are more appropriate for this form of peak reconstruction, as demonstrated in \cite{EisSeoRecon, Padmanabhan2012, 2014_BOSS_BAO},  where the signal can be largely restored by reversing the inferred large scale displacements. 

The large-scale shape of the galaxy power spectrum is sensitive to a variety of cosmological parameters such as the matter density, the spectral index, and the neutrino mass. The precision  of these parameter measurements is limited by the uncertain nature of galaxy bias, which is expected to be linear on very large scales $k \lesssim 0.1 \hmpc$ but not on smaller scales where the vast majority of the information resides.  Clipping can greatly assist in linearising the galaxy bias, thereby ensuring the clipped galaxy power spectrum bears a close resemblance to the clipped dark matter power spectrum. Interpreting the shape is less straightforward since stronger clipping leads to a change in the shape of the power spectrum, but this is fully specified by the transformation defined in (\ref{eq:clipAnalytic}). Overall, then, we see considerable scope for further applications of the method presented here.
 
\noindent{\bf Acknowledgements}\\
FS acknowledges support by the European Research Council under the European Community's Seventh Framework Programme FP7-IDEAS-Phys.LSS 240117. 
CB acknowledges the support of the Australian Research Council through a Future Fellowship award.  This research was supported by the Australian Research Council Centre of Excellence for All-sky Astrophysics (CAASTRO), through project number CE110001020.
PN acknowledges the support of the Royal Society through the award of a University Research Fellowship and the European Research Council, through receipt of a Starting Grant (DEGAS-259586).
CH acknowledges support from the European Research Council under the EC FP7 grant number 240185.

GAMA is a joint European-Australasian project based around a spectroscopic campaign using the Anglo-Australian Telescope. The GAMA input catalogue is based on data taken from the Sloan Digital Sky Survey and the UKIRT Infrared Deep Sky Survey. Complementary imaging of the GAMA regions is being obtained by a number of independent survey programs including GALEX MIS, VST KiDS, VISTA VIKING, WISE, Herschel-ATLAS, GMRT and ASKAP providing UV to radio coverage. GAMA is funded by the STFC (UK), the ARC (Australia), the AAO, and the participating institutions. The GAMA website is http://www.gama-survey.org. 

\appendix

\begin{figure}
\includegraphics[width=80mm]{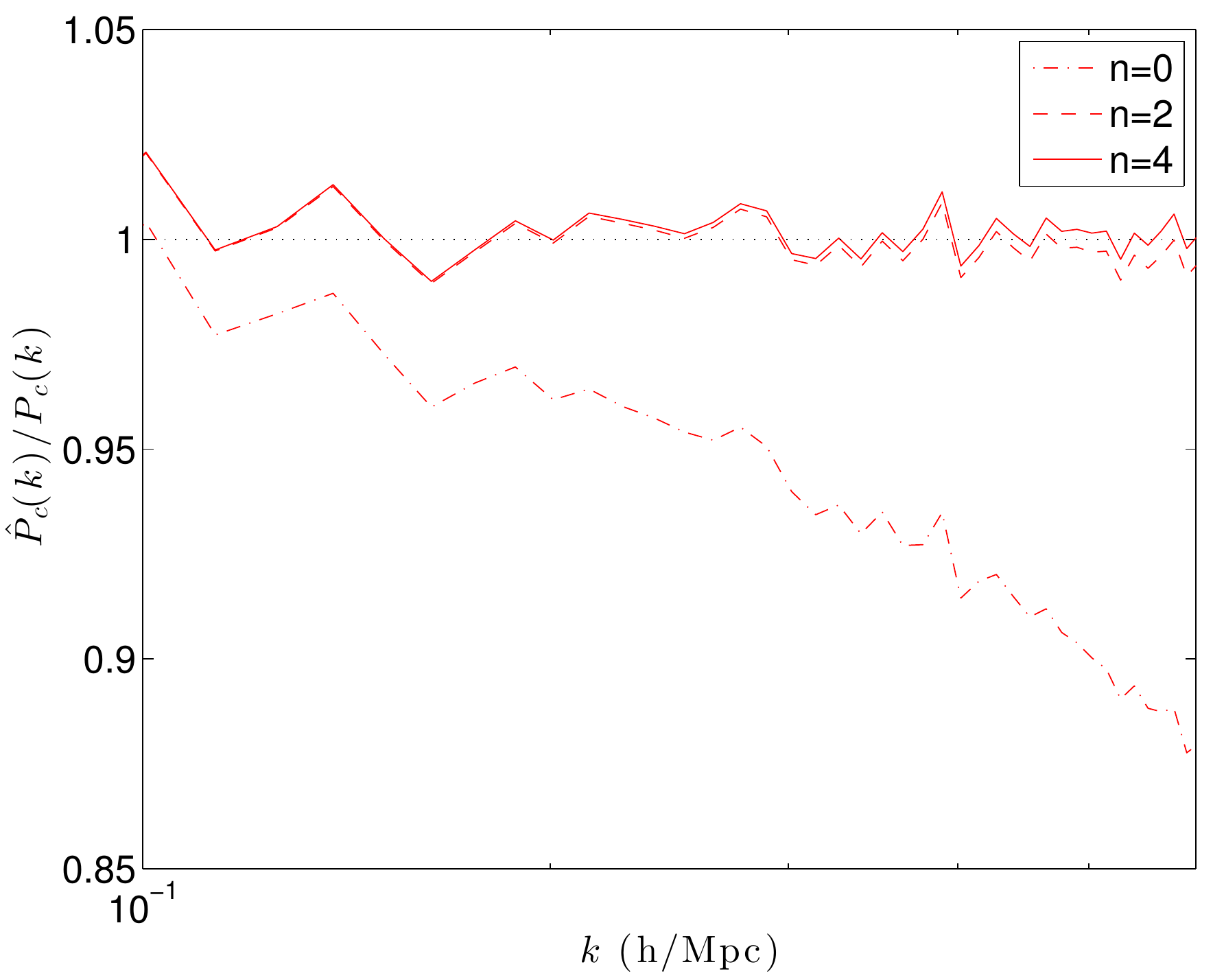}
\caption{The convergence of series given in (\ref{eq:clipAnalyticAnisotropic}) to the numerical solution, as the number of terms in the summation is increased from $n=0,2,4$. For this case the sample GRF was clipped at the $0.7 \sigma$ level, resulting in an approximately $40\%$  loss of power on large scales.  \label{fig:nOrder}}
\end{figure}

\sec{Spectral Distortion}  \label{sec:nOrder}
First we view how rapidly the  series expansion given by (\ref{eq:clipAnalyticAnisotropic}) converges on the numerical solution. Taking the $z=127$ snapshot as our fiducial Gaussian Random Field (GRF) we apply clipping and evaluate the true clipped power spectrum, $P_c(k)$. We then generate an estimate of the clipped power, $\hat{P}_c(k)$, by truncating the series of (\ref{eq:clipAnalyticAnisotropic}) at $n=0,2,4$. Figure \ref{fig:nOrder} illustrates the ratio of the true and estimated power spectra in each case. In order to establish sub-percent precision in the estimated power spectrum, it is sufficient to stop at $n=2$ provided the parameter $\Alin > 0.5$. Throughout this work we evaluate terms at $n \leq 4$. 

\sec{Clipped Anisotropic Fields}  \label{sec:testAnisotropicClipping}

\begin{figure}
\includegraphics[width=80mm]{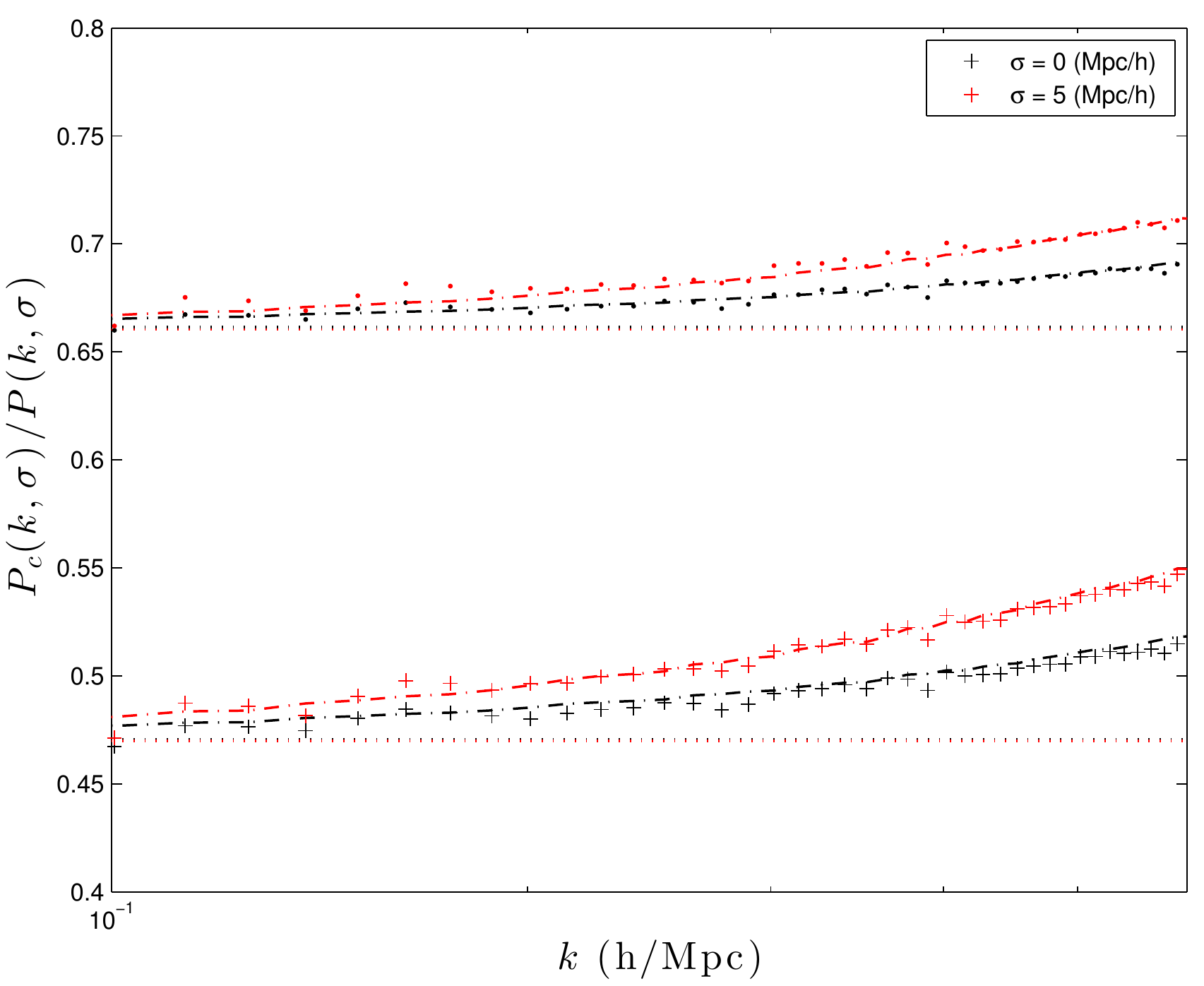}
\caption{The fractional change in the angle-averaged power spectrum of the field described by (\ref{eq:dispersion}), where we have set $\beta=0.5$ and we explore $\sigma=0$ and $\sigma=5$, each evaluated at two different clipping thresholds  (removing 5\% and 10\% of the mass). The real-space density field is taken from the $z=127$ snapshot of the Millennium-I simulation. Larger values of $\sigma$ can be seen to slightly reduce the effects of clipping on small scales. This arises from the contribution from terms in (\ref{eq:clipAnalyticAnisotropic}), which involve the self-convolution of the power spectrum. The horizontal dotted lines are the estimated clipped power when only using the first term from (\ref{eq:clipAnalyticAnisotropic}), while adding in the second leads to the dashed lines, which are in good agreement with the numerical results.  \label{fig:dispersion}}
\end{figure}

In this section we explore the consequences of (\ref{eq:clipAnalyticAnisotropic}) by performing numerical tests on anisotropic  fields. We take the $z=127$ Millennium-I density field in real space and impose a distortion along one axis consistent with the prescription of \cite{1987MNRAS.227....1K}:
\[ \label{eq:kaiserDeltatest}
\delta^s(k, \mu) = \delta^r(k) (1 + \beta \mu^2)   \frac{1}{\sqrt{1 + k^2 \mu^2 \sigma_v^2/2}} \, ,
\]
where $\mu$ is the cosine of the angle between the wave vector and the line of sight, and the longitudinal amplification factor $\beta \equiv f/b$.  This leaves us with an anisotropic GRF whose power spectrum recovers the standard form
\[ \label{eq:dispersion}
P(k, \mu) =  P^{r}(k)  (1 + \beta \mu^2)^2 \frac{1}{1 + k^2 \mu^2 \sigma_v^2/2}\; .
\]
In Figure  \ref{fig:dispersion} we can see that the fractional change in the angle-averaged power spectrum induced by clipping is slightly reduced when the velocity dispersion $\sigma_v$ is introduced. The larger gradient in $P(k)$ leads to a stronger contribution from higher order terms in (\ref{eq:clipAnalyticAnisotropic}).  However in all cases the dashed line of the model successfully reproduces the behaviour of the data points. 

\begin{figure*}
\includegraphics[width=110mm]{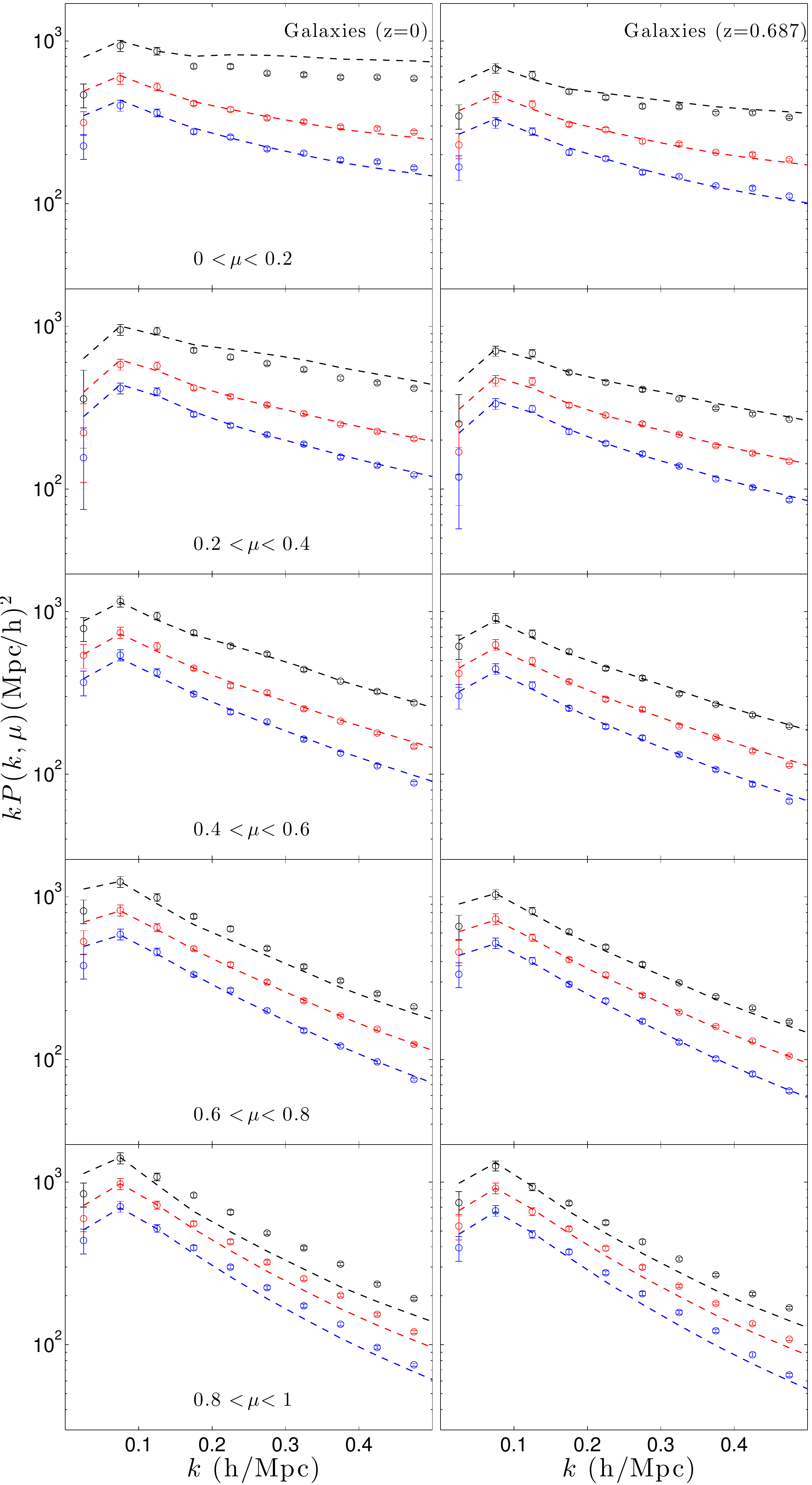}
\caption{The same format as Figure \ref{fig:nonlinear_sim} but we now use a finer grid to define the number density field, at $1.95 \, \mathrm{Mpc}/h$ instead of  $3.9 \, \mathrm{Mpc}/h$. Since the total variance of the field is greater, slightly less power is lost after removing the same fraction of galaxies as before.  \label{fig:hires_sims}}
\end{figure*}

Next we evaluate the power spectrum from the mock galaxy fields at lower redshifts. Figure \ref{fig:hires_sims} repeats the analysis of Figure \ref{fig:nonlinear_sim} but now with a smoothing length of $1.95 \, \mathrm{Mpc}/h$.  The sets of parameter constraints derived from these two different smoothing lengths are found to be highly consistent with each other. 

\begin{figure*}
\includegraphics[width=170mm]{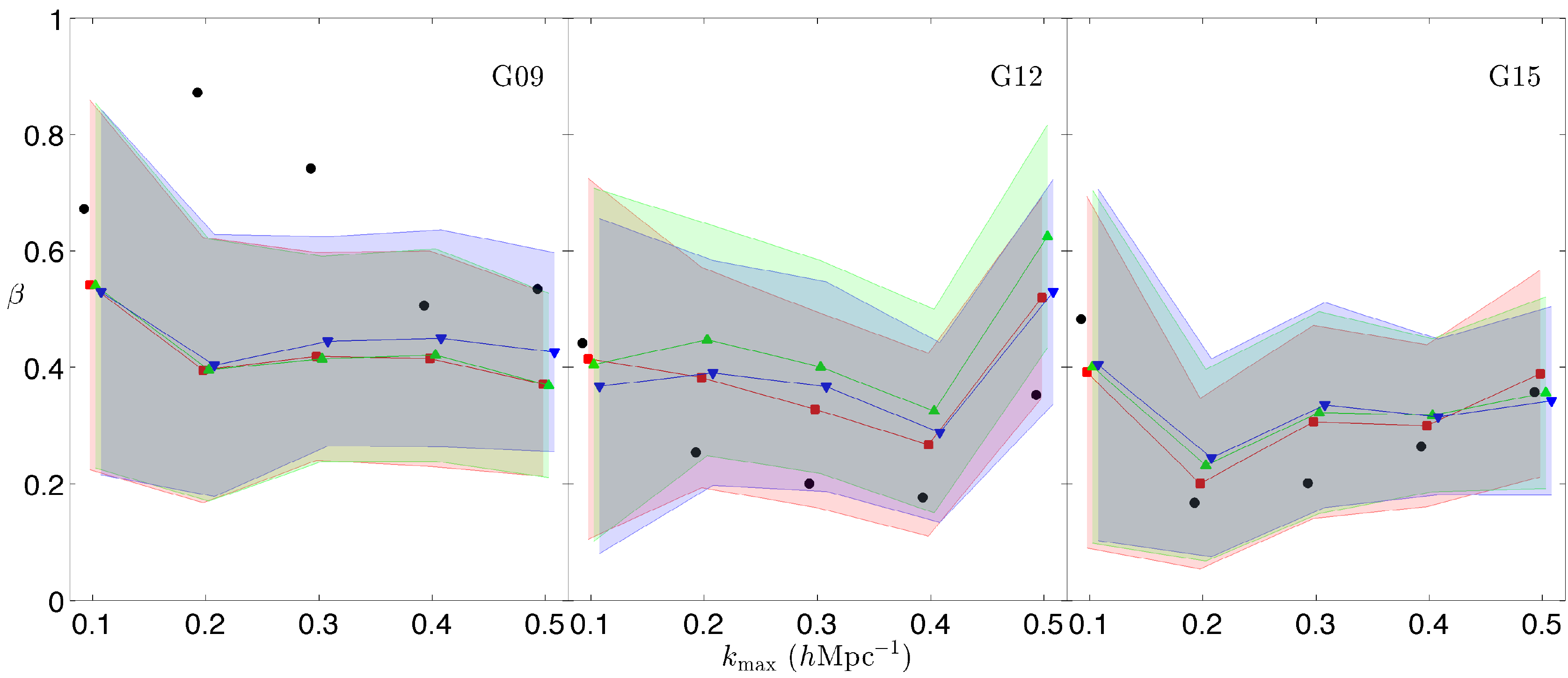}
\caption{Constraints on the anisotropy parameter $\beta$ from the three individual GAMA regions, from left to right: G09, G12, G15, for the case of the unclipped galaxy field and three clipping thresholds $\dThresh$ of 8, 5, and 4.\label{fig:betaGAMAfields}}
\end{figure*}

\begin{figure*}
\includegraphics[width=170mm]{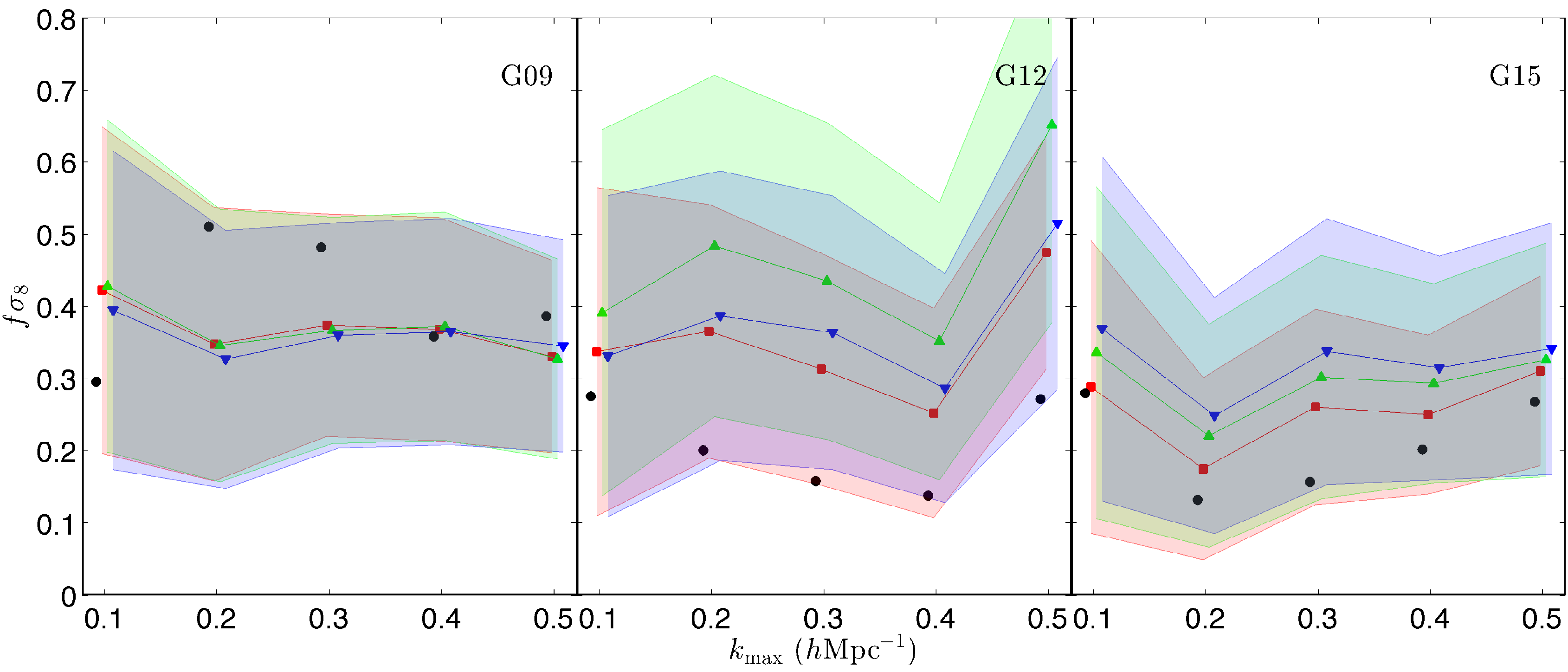}
\caption{Constraints on $\fsig$ from the three individual GAMA regions, from left to right: G09, G12, G15, for the case of the unclipped galaxy field and three clipping thresholds $\dThresh$ of 8, 5, and 4.\label{fig:fsigma8GAMAfields}}
\end{figure*}

\sec{The GAMA Regions}  \label{sec:GAMAfields}

Figures \ref{fig:betaGAMAfields} and \ref{fig:fsigma8GAMAfields} present constraints on $\beta$ and $f \sigma_8$ from the three separate GAMA regions. In each case the results appear consistent between the three regions, and across the three thresholds within each region. 

The clipping statistics for each field are presented in Table \ref{tab:GAMA_stats}.

\begin{table}
\begin{tabular}{|c|c|c|c|}
\hline
Region & $\delta_0$ & $f_V$ & $f_m$ \\\hline
9  & 4  &  0.025  & 0.169  \\\hline
12 & 4  &  0.037  & 0.210  \\\hline
15 & 4  &  0.035  & 0.217  \\\hline
9  & 5  &  0.016  & 0.119  \\\hline
12 & 5  &  0.024  & 0.157  \\\hline
15 & 5  &  0.024  & 0.165  \\\hline
9  & 8  &  0.005  & 0.047  \\\hline
12 & 8  &  0.008  & 0.065  \\\hline
15 & 8  &  0.008  & 0.081  \\\hline
\end{tabular}

\caption{ The fraction of the field volume $f_V$ and the effective fraction of the galaxies $f_m$  subject to clipping from each of the GAMA regions, for each of the three threshold values. 
\label{tab:GAMA_stats}}
\end{table}

\sec{Anisotropic Clustering}

Here we present constraints on the anisotropy parameter $\beta$ from the clipped galaxy power spectrum.  To do so we marginalise over the three model parameters ($\Alin$, $\sigma_v$, $\amp$) while fixing the shape of the linear power spectrum to the fiducial model.  

\begin{figure}
\includegraphics[width=80mm]{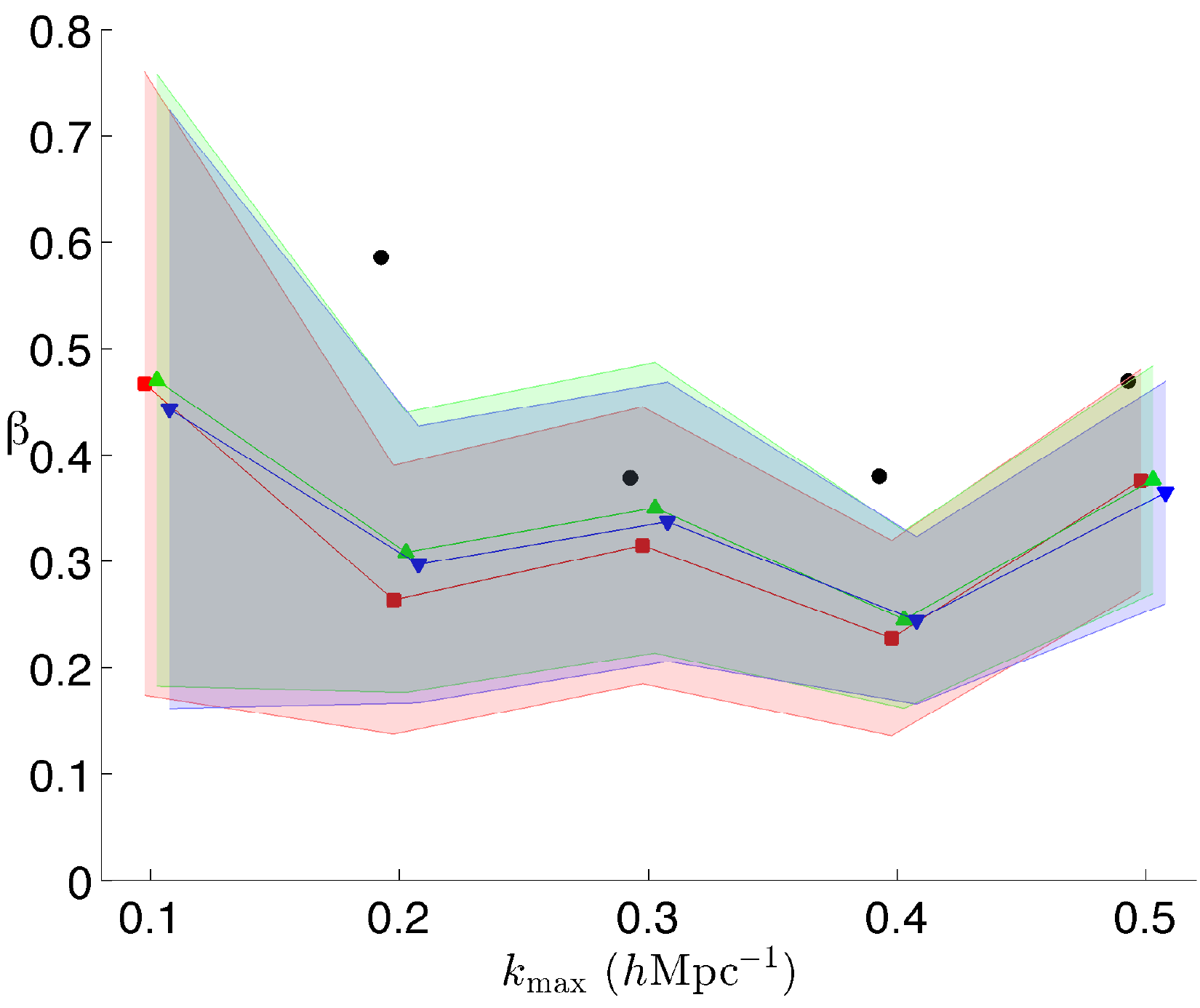}
\caption{Constraints on the anisotropy parameter $\beta$ from the combination of the three GAMA regions for a variety of $\kmax$ values, using the nonlinear model specified by (\ref{eq:oneloop}).  The squares, triangles and inverted triangles  correspond to clipping thresholds $\dThresh$ of $8$, $5$, and $4$ respectively. The circles represent the maximum likelihood points when using the original field without clipping. 
\label{fig:GAMA_beta}}
\end{figure}

Figure \ref{fig:GAMA_beta} illustrates the confidence intervals derived from the combination of the three fields.  The maximum likelihood values for $\beta$ are plotted as a function of the maximum wavenumber used to compare to the model.  The behaviour closely reflects that of $f \sigma_8$, as seen in the central panel of Figure \ref{fig:GAMAfsigma8Panels}. 

\bibliography{../../HomeSpace/Routines/dis}

\begin{thebibliography}{65}%
\makeatletter
\providecommand \@ifxundefined [1]{%
 \@ifx{#1\undefined}
}%
\providecommand \@ifnum [1]{%
 \ifnum #1\expandafter \@firstoftwo
 \else \expandafter \@secondoftwo
 \fi
}%
\providecommand \@ifx [1]{%
 \ifx #1\expandafter \@firstoftwo
 \else \expandafter \@secondoftwo
 \fi
}%
\providecommand \natexlab [1]{#1}%
\providecommand \enquote  [1]{``#1''}%
\providecommand \bibnamefont  [1]{#1}%
\providecommand \bibfnamefont [1]{#1}%
\providecommand \citenamefont [1]{#1}%
\providecommand \href@noop [0]{\@secondoftwo}%
\providecommand \href [0]{\begingroup \@sanitize@url \@href}%
\providecommand \@href[1]{\@@startlink{#1}\@@href}%
\providecommand \@@href[1]{\endgroup#1\@@endlink}%
\providecommand \@sanitize@url [0]{\catcode `\\12\catcode `\$12\catcode
  `\&12\catcode `\#12\catcode `\^12\catcode `\_12\catcode `\%12\relax}%
\providecommand \@@startlink[1]{}%
\providecommand \@@endlink[0]{}%
\providecommand \url  [0]{\begingroup\@sanitize@url \@url }%
\providecommand \@url [1]{\endgroup\@href {#1}{\urlprefix }}%
\providecommand \urlprefix  [0]{URL }%
\providecommand \Eprint [0]{\href }%
\providecommand \doibase [0]{http://dx.doi.org/}%
\providecommand \selectlanguage [0]{\@gobble}%
\providecommand \bibinfo  [0]{\@secondoftwo}%
\providecommand \bibfield  [0]{\@secondoftwo}%
\providecommand \translation [1]{[#1]}%
\providecommand \BibitemOpen [0]{}%
\providecommand \bibitemStop [0]{}%
\providecommand \bibitemNoStop [0]{.\EOS\space}%
\providecommand \EOS [0]{\spacefactor3000\relax}%
\providecommand \BibitemShut  [1]{\csname bibitem#1\endcsname}%
\let\auto@bib@innerbib\@empty
\bibitem [{\citenamefont {{Blake}}\ \emph {et~al.}(2011)\citenamefont
  {{Blake}}, \citenamefont {{Brough}}, \citenamefont {{Colless}}, \citenamefont
  {{Contreras}}, \citenamefont {{Couch}}, \citenamefont {{Croom}},
  \citenamefont {{Davis}}, \citenamefont {{Drinkwater}}, \citenamefont
  {{Forster}}, \citenamefont {{Gilbank}}, \citenamefont {{Gladders}},
  \citenamefont {{Glazebrook}}, \citenamefont {{Jelliffe}}, \citenamefont
  {{Jurek}}, \citenamefont {{Li}}, \citenamefont {{Madore}}, \citenamefont
  {{Martin}}, \citenamefont {{Pimbblet}}, \citenamefont {{Poole}},
  \citenamefont {{Pracy}}, \citenamefont {{Sharp}}, \citenamefont
  {{Wisnioski}}, \citenamefont {{Woods}}, \citenamefont {{Wyder}},\ and\
  \citenamefont {{Yee}}}]{BlakeWigglezRSD}%
  \BibitemOpen
  \bibfield  {author} {\bibinfo {author} {\bibfnamefont {C.}~\bibnamefont
  {{Blake}}}, \bibinfo {author} {\bibfnamefont {S.}~\bibnamefont {{Brough}}},
  \bibinfo {author} {\bibfnamefont {M.}~\bibnamefont {{Colless}}}, \bibinfo
  {author} {\bibfnamefont {C.}~\bibnamefont {{Contreras}}}, \bibinfo {author}
  {\bibfnamefont {W.}~\bibnamefont {{Couch}}}, \bibinfo {author} {\bibfnamefont
  {S.}~\bibnamefont {{Croom}}}, \bibinfo {author} {\bibfnamefont
  {T.}~\bibnamefont {{Davis}}}, \bibinfo {author} {\bibfnamefont {M.~J.}\
  \bibnamefont {{Drinkwater}}}, \bibinfo {author} {\bibfnamefont
  {K.}~\bibnamefont {{Forster}}}, \bibinfo {author} {\bibfnamefont
  {D.}~\bibnamefont {{Gilbank}}}, \bibinfo {author} {\bibfnamefont
  {M.}~\bibnamefont {{Gladders}}}, \bibinfo {author} {\bibfnamefont
  {K.}~\bibnamefont {{Glazebrook}}}, \bibinfo {author} {\bibfnamefont
  {B.}~\bibnamefont {{Jelliffe}}}, \bibinfo {author} {\bibfnamefont {R.~J.}\
  \bibnamefont {{Jurek}}}, \bibinfo {author} {\bibfnamefont {I.-H.}\
  \bibnamefont {{Li}}}, \bibinfo {author} {\bibfnamefont {B.}~\bibnamefont
  {{Madore}}}, \bibinfo {author} {\bibfnamefont {D.~C.}\ \bibnamefont
  {{Martin}}}, \bibinfo {author} {\bibfnamefont {K.}~\bibnamefont
  {{Pimbblet}}}, \bibinfo {author} {\bibfnamefont {G.~B.}\ \bibnamefont
  {{Poole}}}, \bibinfo {author} {\bibfnamefont {M.}~\bibnamefont {{Pracy}}},
  \bibinfo {author} {\bibfnamefont {R.}~\bibnamefont {{Sharp}}}, \bibinfo
  {author} {\bibfnamefont {E.}~\bibnamefont {{Wisnioski}}}, \bibinfo {author}
  {\bibfnamefont {D.}~\bibnamefont {{Woods}}}, \bibinfo {author} {\bibfnamefont
  {T.~K.}\ \bibnamefont {{Wyder}}}, \ and\ \bibinfo {author} {\bibfnamefont
  {H.~K.~C.}\ \bibnamefont {{Yee}}},\ }\href {\doibase
  10.1111/j.1365-2966.2011.18903.x} {\bibfield  {journal} {\bibinfo  {journal}
  {\mnras}\ }\textbf {\bibinfo {volume} {415}},\ \bibinfo {pages} {2876}
  (\bibinfo {year} {2011})},\ \Eprint {http://arxiv.org/abs/1104.2948}
  {arXiv:1104.2948 [astro-ph.CO]} \BibitemShut {NoStop}%
\bibitem [{\citenamefont {{Reid}}\ \emph {et~al.}(2012)\citenamefont {{Reid}},
  \citenamefont {{Samushia}}, \citenamefont {{White}}, \citenamefont
  {{Percival}}, \citenamefont {{Manera}}, \citenamefont {{Padmanabhan}},
  \citenamefont {{Ross}}, \citenamefont {{S{\'a}nchez}}, \citenamefont
  {{Bailey}}, \citenamefont {{Bizyaev}}, \citenamefont {{Bolton}},
  \citenamefont {{Brewington}}, \citenamefont {{Brinkmann}}, \citenamefont
  {{Brownstein}}, \citenamefont {{Cuesta}}, \citenamefont {{Eisenstein}},
  \citenamefont {{Gunn}}, \citenamefont {{Honscheid}}, \citenamefont
  {{Malanushenko}}, \citenamefont {{Malanushenko}}, \citenamefont {{Maraston}},
  \citenamefont {{McBride}}, \citenamefont {{Muna}}, \citenamefont {{Nichol}},
  \citenamefont {{Oravetz}}, \citenamefont {{Pan}}, \citenamefont {{de
  Putter}}, \citenamefont {{Roe}}, \citenamefont {{Ross}}, \citenamefont
  {{Schlegel}}, \citenamefont {{Schneider}}, \citenamefont {{Seo}},
  \citenamefont {{Shelden}}, \citenamefont {{Sheldon}}, \citenamefont
  {{Simmons}}, \citenamefont {{Skibba}}, \citenamefont {{Snedden}},
  \citenamefont {{Swanson}}, \citenamefont {{Thomas}}, \citenamefont
  {{Tinker}}, \citenamefont {{Tojeiro}}, \citenamefont {{Verde}}, \citenamefont
  {{Wake}}, \citenamefont {{Weaver}}, \citenamefont {{Weinberg}}, \citenamefont
  {{Zehavi}},\ and\ \citenamefont {{Zhao}}}]{BOSSRSD2012}%
  \BibitemOpen
  \bibfield  {author} {\bibinfo {author} {\bibfnamefont {B.~A.}\ \bibnamefont
  {{Reid}}}, \bibinfo {author} {\bibfnamefont {L.}~\bibnamefont {{Samushia}}},
  \bibinfo {author} {\bibfnamefont {M.}~\bibnamefont {{White}}}, \bibinfo
  {author} {\bibfnamefont {W.~J.}\ \bibnamefont {{Percival}}}, \bibinfo
  {author} {\bibfnamefont {M.}~\bibnamefont {{Manera}}}, \bibinfo {author}
  {\bibfnamefont {N.}~\bibnamefont {{Padmanabhan}}}, \bibinfo {author}
  {\bibfnamefont {A.~J.}\ \bibnamefont {{Ross}}}, \bibinfo {author}
  {\bibfnamefont {A.~G.}\ \bibnamefont {{S{\'a}nchez}}}, \bibinfo {author}
  {\bibfnamefont {S.}~\bibnamefont {{Bailey}}}, \bibinfo {author}
  {\bibfnamefont {D.}~\bibnamefont {{Bizyaev}}}, \bibinfo {author}
  {\bibfnamefont {A.~S.}\ \bibnamefont {{Bolton}}}, \bibinfo {author}
  {\bibfnamefont {H.}~\bibnamefont {{Brewington}}}, \bibinfo {author}
  {\bibfnamefont {J.}~\bibnamefont {{Brinkmann}}}, \bibinfo {author}
  {\bibfnamefont {J.~R.}\ \bibnamefont {{Brownstein}}}, \bibinfo {author}
  {\bibfnamefont {A.~J.}\ \bibnamefont {{Cuesta}}}, \bibinfo {author}
  {\bibfnamefont {D.~J.}\ \bibnamefont {{Eisenstein}}}, \bibinfo {author}
  {\bibfnamefont {J.~E.}\ \bibnamefont {{Gunn}}}, \bibinfo {author}
  {\bibfnamefont {K.}~\bibnamefont {{Honscheid}}}, \bibinfo {author}
  {\bibfnamefont {E.}~\bibnamefont {{Malanushenko}}}, \bibinfo {author}
  {\bibfnamefont {V.}~\bibnamefont {{Malanushenko}}}, \bibinfo {author}
  {\bibfnamefont {C.}~\bibnamefont {{Maraston}}}, \bibinfo {author}
  {\bibfnamefont {C.~K.}\ \bibnamefont {{McBride}}}, \bibinfo {author}
  {\bibfnamefont {D.}~\bibnamefont {{Muna}}}, \bibinfo {author} {\bibfnamefont
  {R.~C.}\ \bibnamefont {{Nichol}}}, \bibinfo {author} {\bibfnamefont
  {D.}~\bibnamefont {{Oravetz}}}, \bibinfo {author} {\bibfnamefont
  {K.}~\bibnamefont {{Pan}}}, \bibinfo {author} {\bibfnamefont
  {R.}~\bibnamefont {{de Putter}}}, \bibinfo {author} {\bibfnamefont {N.~A.}\
  \bibnamefont {{Roe}}}, \bibinfo {author} {\bibfnamefont {N.~P.}\ \bibnamefont
  {{Ross}}}, \bibinfo {author} {\bibfnamefont {D.~J.}\ \bibnamefont
  {{Schlegel}}}, \bibinfo {author} {\bibfnamefont {D.~P.}\ \bibnamefont
  {{Schneider}}}, \bibinfo {author} {\bibfnamefont {H.-J.}\ \bibnamefont
  {{Seo}}}, \bibinfo {author} {\bibfnamefont {A.}~\bibnamefont {{Shelden}}},
  \bibinfo {author} {\bibfnamefont {E.~S.}\ \bibnamefont {{Sheldon}}}, \bibinfo
  {author} {\bibfnamefont {A.}~\bibnamefont {{Simmons}}}, \bibinfo {author}
  {\bibfnamefont {R.~A.}\ \bibnamefont {{Skibba}}}, \bibinfo {author}
  {\bibfnamefont {S.}~\bibnamefont {{Snedden}}}, \bibinfo {author}
  {\bibfnamefont {M.~E.~C.}\ \bibnamefont {{Swanson}}}, \bibinfo {author}
  {\bibfnamefont {D.}~\bibnamefont {{Thomas}}}, \bibinfo {author}
  {\bibfnamefont {J.}~\bibnamefont {{Tinker}}}, \bibinfo {author}
  {\bibfnamefont {R.}~\bibnamefont {{Tojeiro}}}, \bibinfo {author}
  {\bibfnamefont {L.}~\bibnamefont {{Verde}}}, \bibinfo {author} {\bibfnamefont
  {D.~A.}\ \bibnamefont {{Wake}}}, \bibinfo {author} {\bibfnamefont {B.~A.}\
  \bibnamefont {{Weaver}}}, \bibinfo {author} {\bibfnamefont {D.~H.}\
  \bibnamefont {{Weinberg}}}, \bibinfo {author} {\bibfnamefont
  {I.}~\bibnamefont {{Zehavi}}}, \ and\ \bibinfo {author} {\bibfnamefont
  {G.-B.}\ \bibnamefont {{Zhao}}},\ }\href {\doibase
  10.1111/j.1365-2966.2012.21779.x} {\bibfield  {journal} {\bibinfo  {journal}
  {\mnras}\ }\textbf {\bibinfo {volume} {426}},\ \bibinfo {pages} {2719}
  (\bibinfo {year} {2012})},\ \Eprint {http://arxiv.org/abs/1203.6641}
  {arXiv:1203.6641 [astro-ph.CO]} \BibitemShut {NoStop}%
\bibitem [{\citenamefont {{Kwan}}\ \emph {et~al.}(2012)\citenamefont {{Kwan}},
  \citenamefont {{Lewis}},\ and\ \citenamefont {{Linder}}}]{2012KLL}%
  \BibitemOpen
  \bibfield  {author} {\bibinfo {author} {\bibfnamefont {J.}~\bibnamefont
  {{Kwan}}}, \bibinfo {author} {\bibfnamefont {G.~F.}\ \bibnamefont {{Lewis}}},
  \ and\ \bibinfo {author} {\bibfnamefont {E.~V.}\ \bibnamefont {{Linder}}},\
  }\href {\doibase 10.1088/0004-637X/748/2/78} {\bibfield  {journal} {\bibinfo
  {journal} {\apj}\ }\textbf {\bibinfo {volume} {748}},\ \bibinfo {eid} {78}
  (\bibinfo {year} {2012})},\ \Eprint {http://arxiv.org/abs/1105.1194}
  {arXiv:1105.1194 [astro-ph.CO]} \BibitemShut {NoStop}%
\bibitem [{\citenamefont {{Linder}}\ and\ \citenamefont
  {{Samsing}}(2013)}]{2013LinderRSD}%
  \BibitemOpen
  \bibfield  {author} {\bibinfo {author} {\bibfnamefont {E.~V.}\ \bibnamefont
  {{Linder}}}\ and\ \bibinfo {author} {\bibfnamefont {J.}~\bibnamefont
  {{Samsing}}},\ }\href {\doibase 10.1088/1475-7516/2013/02/025} {\bibfield
  {journal} {\bibinfo  {journal} {JCAP}\ }\textbf {\bibinfo {volume} {2}},\
  \bibinfo {eid} {025} (\bibinfo {year} {2013})},\ \Eprint
  {http://arxiv.org/abs/1211.2274} {arXiv:1211.2274 [astro-ph.CO]} \BibitemShut
  {NoStop}%
\bibitem [{\citenamefont {{Reid}}\ \emph {et~al.}(2014)\citenamefont {{Reid}},
  \citenamefont {{Seo}}, \citenamefont {{Leauthaud}}, \citenamefont
  {{Tinker}},\ and\ \citenamefont {{White}}}]{2014ReidBOSSRSD}%
  \BibitemOpen
  \bibfield  {author} {\bibinfo {author} {\bibfnamefont {B.~A.}\ \bibnamefont
  {{Reid}}}, \bibinfo {author} {\bibfnamefont {H.-J.}\ \bibnamefont {{Seo}}},
  \bibinfo {author} {\bibfnamefont {A.}~\bibnamefont {{Leauthaud}}}, \bibinfo
  {author} {\bibfnamefont {J.~L.}\ \bibnamefont {{Tinker}}}, \ and\ \bibinfo
  {author} {\bibfnamefont {M.}~\bibnamefont {{White}}},\ }\href {\doibase
  10.1093/mnras/stu1391} {\bibfield  {journal} {\bibinfo  {journal} {\mnras}\
  }\textbf {\bibinfo {volume} {444}},\ \bibinfo {pages} {476} (\bibinfo {year}
  {2014})},\ \Eprint {http://arxiv.org/abs/1404.3742} {arXiv:1404.3742}
  \BibitemShut {NoStop}%
\bibitem [{\citenamefont {{Weinberg}}(1992)}]{1992MNRAS.254..315W}%
  \BibitemOpen
  \bibfield  {author} {\bibinfo {author} {\bibfnamefont {D.~H.}\ \bibnamefont
  {{Weinberg}}},\ }\href@noop {} {\bibfield  {journal} {\bibinfo  {journal}
  {\mnras}\ }\textbf {\bibinfo {volume} {254}},\ \bibinfo {pages} {315}
  (\bibinfo {year} {1992})}\BibitemShut {NoStop}%
\bibitem [{\citenamefont {{McCullagh}}\ \emph {et~al.}(2015)\citenamefont
  {{McCullagh}}, \citenamefont {{Neyrinck}}, \citenamefont {{Norberg}},\ and\
  \citenamefont {{Cole}}}]{2015arXiv151102034M}%
  \BibitemOpen
  \bibfield  {author} {\bibinfo {author} {\bibfnamefont {N.}~\bibnamefont
  {{McCullagh}}}, \bibinfo {author} {\bibfnamefont {M.}~\bibnamefont
  {{Neyrinck}}}, \bibinfo {author} {\bibfnamefont {P.}~\bibnamefont
  {{Norberg}}}, \ and\ \bibinfo {author} {\bibfnamefont {S.}~\bibnamefont
  {{Cole}}},\ }\href@noop {} {\bibfield  {journal} {\bibinfo  {journal} {ArXiv
  e-prints}\ } (\bibinfo {year} {2015})},\ \Eprint
  {http://arxiv.org/abs/1511.02034} {arXiv:1511.02034} \BibitemShut {NoStop}%
\bibitem [{\citenamefont {{Neyrinck}}\ \emph {et~al.}(2009)\citenamefont
  {{Neyrinck}}, \citenamefont {{Szapudi}},\ and\ \citenamefont
  {{Szalay}}}]{Neyrinck2009}%
  \BibitemOpen
  \bibfield  {author} {\bibinfo {author} {\bibfnamefont {M.~C.}\ \bibnamefont
  {{Neyrinck}}}, \bibinfo {author} {\bibfnamefont {I.}~\bibnamefont
  {{Szapudi}}}, \ and\ \bibinfo {author} {\bibfnamefont {A.~S.}\ \bibnamefont
  {{Szalay}}},\ }\href {\doibase 10.1088/0004-637X/698/2/L90} {\bibfield
  {journal} {\bibinfo  {journal} {\apjl}\ }\textbf {\bibinfo {volume} {698}},\
  \bibinfo {pages} {L90} (\bibinfo {year} {2009})},\ \Eprint
  {http://arxiv.org/abs/0903.4693} {arXiv:0903.4693 [astro-ph.CO]} \BibitemShut
  {NoStop}%
\bibitem [{\citenamefont {{Simpson}}\ \emph {et~al.}(2013)\citenamefont
  {{Simpson}}, \citenamefont {{Heavens}},\ and\ \citenamefont
  {{Heymans}}}]{2013SimpsonClip}%
  \BibitemOpen
  \bibfield  {author} {\bibinfo {author} {\bibfnamefont {F.}~\bibnamefont
  {{Simpson}}}, \bibinfo {author} {\bibfnamefont {A.~F.}\ \bibnamefont
  {{Heavens}}}, \ and\ \bibinfo {author} {\bibfnamefont {C.}~\bibnamefont
  {{Heymans}}},\ }\href {\doibase 10.1103/PhysRevD.88.083510} {\bibfield
  {journal} {\bibinfo  {journal} {\prd}\ }\textbf {\bibinfo {volume} {88}},\
  \bibinfo {eid} {083510} (\bibinfo {year} {2013})},\ \Eprint
  {http://arxiv.org/abs/1306.6349} {arXiv:1306.6349 [astro-ph.CO]} \BibitemShut
  {NoStop}%
\bibitem [{\citenamefont {{Lombriser}}\ \emph {et~al.}(2015)\citenamefont
  {{Lombriser}}, \citenamefont {{Simpson}},\ and\ \citenamefont
  {{Mead}}}]{2015arXiv150104961L}%
  \BibitemOpen
  \bibfield  {author} {\bibinfo {author} {\bibfnamefont {L.}~\bibnamefont
  {{Lombriser}}}, \bibinfo {author} {\bibfnamefont {F.}~\bibnamefont
  {{Simpson}}}, \ and\ \bibinfo {author} {\bibfnamefont {A.}~\bibnamefont
  {{Mead}}},\ }\href {\doibase 10.1103/PhysRevLett.114.251101} {\bibfield
  {journal} {\bibinfo  {journal} {Physical Review Letters}\ }\textbf {\bibinfo
  {volume} {114}},\ \bibinfo {eid} {251101} (\bibinfo {year} {2015})},\ \Eprint
  {http://arxiv.org/abs/1501.04961} {arXiv:1501.04961} \BibitemShut {NoStop}%
\bibitem [{\citenamefont {{Rimes}}\ and\ \citenamefont
  {{Hamilton}}(2005)}]{Rimes_Hamilton_2005}%
  \BibitemOpen
  \bibfield  {author} {\bibinfo {author} {\bibfnamefont {C.~D.}\ \bibnamefont
  {{Rimes}}}\ and\ \bibinfo {author} {\bibfnamefont {A.~J.~S.}\ \bibnamefont
  {{Hamilton}}},\ }\href {\doibase 10.1111/j.1745-3933.2005.00051.x} {\bibfield
   {journal} {\bibinfo  {journal} {\mnras}\ }\textbf {\bibinfo {volume}
  {360}},\ \bibinfo {pages} {L82} (\bibinfo {year} {2005})},\ \Eprint
  {http://arxiv.org/abs/astro-ph/0502081} {astro-ph/0502081} \BibitemShut
  {NoStop}%
\bibitem [{\citenamefont {{Verde}}\ \emph {et~al.}(2002)\citenamefont
  {{Verde}}, \citenamefont {{Heavens}}, \citenamefont {{Percival}},
  \citenamefont {{Matarrese}}, \citenamefont {{Baugh}}, \citenamefont
  {{Bland-Hawthorn}}, \citenamefont {{Bridges}}, \citenamefont {{Cannon}},
  \citenamefont {{Cole}}, \citenamefont {{Colless}}, \citenamefont {{Collins}},
  \citenamefont {{Couch}}, \citenamefont {{Dalton}}, \citenamefont {{De
  Propris}}, \citenamefont {{Driver}}, \citenamefont {{Efstathiou}},
  \citenamefont {{Ellis}}, \citenamefont {{Frenk}}, \citenamefont
  {{Glazebrook}}, \citenamefont {{Jackson}}, \citenamefont {{Lahav}},
  \citenamefont {{Lewis}}, \citenamefont {{Lumsden}}, \citenamefont {{Maddox}},
  \citenamefont {{Madgwick}}, \citenamefont {{Norberg}}, \citenamefont
  {{Peacock}}, \citenamefont {{Peterson}}, \citenamefont {{Sutherland}},\ and\
  \citenamefont {{Taylor}}}]{2002MNRAS.335..432V}%
  \BibitemOpen
  \bibfield  {author} {\bibinfo {author} {\bibfnamefont {L.}~\bibnamefont
  {{Verde}}}, \bibinfo {author} {\bibfnamefont {A.~F.}\ \bibnamefont
  {{Heavens}}}, \bibinfo {author} {\bibfnamefont {W.~J.}\ \bibnamefont
  {{Percival}}}, \bibinfo {author} {\bibfnamefont {S.}~\bibnamefont
  {{Matarrese}}}, \bibinfo {author} {\bibfnamefont {C.~M.}\ \bibnamefont
  {{Baugh}}}, \bibinfo {author} {\bibfnamefont {J.}~\bibnamefont
  {{Bland-Hawthorn}}}, \bibinfo {author} {\bibfnamefont {T.}~\bibnamefont
  {{Bridges}}}, \bibinfo {author} {\bibfnamefont {R.}~\bibnamefont {{Cannon}}},
  \bibinfo {author} {\bibfnamefont {S.}~\bibnamefont {{Cole}}}, \bibinfo
  {author} {\bibfnamefont {M.}~\bibnamefont {{Colless}}}, \bibinfo {author}
  {\bibfnamefont {C.}~\bibnamefont {{Collins}}}, \bibinfo {author}
  {\bibfnamefont {W.}~\bibnamefont {{Couch}}}, \bibinfo {author} {\bibfnamefont
  {G.}~\bibnamefont {{Dalton}}}, \bibinfo {author} {\bibfnamefont
  {R.}~\bibnamefont {{De Propris}}}, \bibinfo {author} {\bibfnamefont {S.~P.}\
  \bibnamefont {{Driver}}}, \bibinfo {author} {\bibfnamefont {G.}~\bibnamefont
  {{Efstathiou}}}, \bibinfo {author} {\bibfnamefont {R.~S.}\ \bibnamefont
  {{Ellis}}}, \bibinfo {author} {\bibfnamefont {C.~S.}\ \bibnamefont
  {{Frenk}}}, \bibinfo {author} {\bibfnamefont {K.}~\bibnamefont
  {{Glazebrook}}}, \bibinfo {author} {\bibfnamefont {C.}~\bibnamefont
  {{Jackson}}}, \bibinfo {author} {\bibfnamefont {O.}~\bibnamefont {{Lahav}}},
  \bibinfo {author} {\bibfnamefont {I.}~\bibnamefont {{Lewis}}}, \bibinfo
  {author} {\bibfnamefont {S.}~\bibnamefont {{Lumsden}}}, \bibinfo {author}
  {\bibfnamefont {S.}~\bibnamefont {{Maddox}}}, \bibinfo {author}
  {\bibfnamefont {D.}~\bibnamefont {{Madgwick}}}, \bibinfo {author}
  {\bibfnamefont {P.}~\bibnamefont {{Norberg}}}, \bibinfo {author}
  {\bibfnamefont {J.~A.}\ \bibnamefont {{Peacock}}}, \bibinfo {author}
  {\bibfnamefont {B.~A.}\ \bibnamefont {{Peterson}}}, \bibinfo {author}
  {\bibfnamefont {W.}~\bibnamefont {{Sutherland}}}, \ and\ \bibinfo {author}
  {\bibfnamefont {K.}~\bibnamefont {{Taylor}}},\ }\href {\doibase
  10.1046/j.1365-8711.2002.05620.x} {\bibfield  {journal} {\bibinfo  {journal}
  {\mnras}\ }\textbf {\bibinfo {volume} {335}},\ \bibinfo {pages} {432}
  (\bibinfo {year} {2002})},\ \Eprint
  {http://arxiv.org/abs/arXiv:astro-ph/0112161} {arXiv:astro-ph/0112161}
  \BibitemShut {NoStop}%
\bibitem [{\citenamefont {{Mar{\'{\i}}n}}\ \emph {et~al.}(2013)\citenamefont
  {{Mar{\'{\i}}n}}, \citenamefont {{Blake}}, \citenamefont {{Poole}},
  \citenamefont {{McBride}}, \citenamefont {{Brough}}, \citenamefont
  {{Colless}}, \citenamefont {{Contreras}}, \citenamefont {{Couch}},
  \citenamefont {{Croton}}, \citenamefont {{Croom}}, \citenamefont {{Davis}},
  \citenamefont {{Drinkwater}}, \citenamefont {{Forster}}, \citenamefont
  {{Gilbank}}, \citenamefont {{Gladders}}, \citenamefont {{Glazebrook}},
  \citenamefont {{Jelliffe}}, \citenamefont {{Jurek}}, \citenamefont {{Li}},
  \citenamefont {{Madore}}, \citenamefont {{Martin}}, \citenamefont
  {{Pimbblet}}, \citenamefont {{Pracy}}, \citenamefont {{Sharp}}, \citenamefont
  {{Wisnioski}}, \citenamefont {{Woods}}, \citenamefont {{Wyder}},\ and\
  \citenamefont {{Yee}}}]{2013MarinBlake3pt}%
  \BibitemOpen
  \bibfield  {author} {\bibinfo {author} {\bibfnamefont {F.~A.}\ \bibnamefont
  {{Mar{\'{\i}}n}}}, \bibinfo {author} {\bibfnamefont {C.}~\bibnamefont
  {{Blake}}}, \bibinfo {author} {\bibfnamefont {G.~B.}\ \bibnamefont
  {{Poole}}}, \bibinfo {author} {\bibfnamefont {C.~K.}\ \bibnamefont
  {{McBride}}}, \bibinfo {author} {\bibfnamefont {S.}~\bibnamefont {{Brough}}},
  \bibinfo {author} {\bibfnamefont {M.}~\bibnamefont {{Colless}}}, \bibinfo
  {author} {\bibfnamefont {C.}~\bibnamefont {{Contreras}}}, \bibinfo {author}
  {\bibfnamefont {W.}~\bibnamefont {{Couch}}}, \bibinfo {author} {\bibfnamefont
  {D.~J.}\ \bibnamefont {{Croton}}}, \bibinfo {author} {\bibfnamefont
  {S.}~\bibnamefont {{Croom}}}, \bibinfo {author} {\bibfnamefont
  {T.}~\bibnamefont {{Davis}}}, \bibinfo {author} {\bibfnamefont {M.~J.}\
  \bibnamefont {{Drinkwater}}}, \bibinfo {author} {\bibfnamefont
  {K.}~\bibnamefont {{Forster}}}, \bibinfo {author} {\bibfnamefont
  {D.}~\bibnamefont {{Gilbank}}}, \bibinfo {author} {\bibfnamefont
  {M.}~\bibnamefont {{Gladders}}}, \bibinfo {author} {\bibfnamefont
  {K.}~\bibnamefont {{Glazebrook}}}, \bibinfo {author} {\bibfnamefont
  {B.}~\bibnamefont {{Jelliffe}}}, \bibinfo {author} {\bibfnamefont {R.~J.}\
  \bibnamefont {{Jurek}}}, \bibinfo {author} {\bibfnamefont {I.-h.}\
  \bibnamefont {{Li}}}, \bibinfo {author} {\bibfnamefont {B.}~\bibnamefont
  {{Madore}}}, \bibinfo {author} {\bibfnamefont {D.~C.}\ \bibnamefont
  {{Martin}}}, \bibinfo {author} {\bibfnamefont {K.}~\bibnamefont
  {{Pimbblet}}}, \bibinfo {author} {\bibfnamefont {M.}~\bibnamefont {{Pracy}}},
  \bibinfo {author} {\bibfnamefont {R.}~\bibnamefont {{Sharp}}}, \bibinfo
  {author} {\bibfnamefont {E.}~\bibnamefont {{Wisnioski}}}, \bibinfo {author}
  {\bibfnamefont {D.}~\bibnamefont {{Woods}}}, \bibinfo {author} {\bibfnamefont
  {T.~K.}\ \bibnamefont {{Wyder}}}, \ and\ \bibinfo {author} {\bibfnamefont
  {H.~K.~C.}\ \bibnamefont {{Yee}}},\ }\href {\doibase 10.1093/mnras/stt520}
  {\bibfield  {journal} {\bibinfo  {journal} {\mnras}\ }\textbf {\bibinfo
  {volume} {432}},\ \bibinfo {pages} {2654} (\bibinfo {year} {2013})},\ \Eprint
  {http://arxiv.org/abs/1303.6644} {arXiv:1303.6644 [astro-ph.CO]} \BibitemShut
  {NoStop}%
\bibitem [{\citenamefont {{Gil-Mar{\'{\i}}n}}\ \emph
  {et~al.}(2015)\citenamefont {{Gil-Mar{\'{\i}}n}}, \citenamefont
  {{Nore{\~n}a}}, \citenamefont {{Verde}}, \citenamefont {{Percival}},
  \citenamefont {{Wagner}}, \citenamefont {{Manera}},\ and\ \citenamefont
  {{Schneider}}}]{2014Hector}%
  \BibitemOpen
  \bibfield  {author} {\bibinfo {author} {\bibfnamefont {H.}~\bibnamefont
  {{Gil-Mar{\'{\i}}n}}}, \bibinfo {author} {\bibfnamefont {J.}~\bibnamefont
  {{Nore{\~n}a}}}, \bibinfo {author} {\bibfnamefont {L.}~\bibnamefont
  {{Verde}}}, \bibinfo {author} {\bibfnamefont {W.~J.}\ \bibnamefont
  {{Percival}}}, \bibinfo {author} {\bibfnamefont {C.}~\bibnamefont
  {{Wagner}}}, \bibinfo {author} {\bibfnamefont {M.}~\bibnamefont {{Manera}}},
  \ and\ \bibinfo {author} {\bibfnamefont {D.~P.}\ \bibnamefont
  {{Schneider}}},\ }\href {\doibase 10.1093/mnras/stv961} {\bibfield  {journal}
  {\bibinfo  {journal} {\mnras}\ }\textbf {\bibinfo {volume} {451}},\ \bibinfo
  {pages} {539} (\bibinfo {year} {2015})},\ \Eprint
  {http://arxiv.org/abs/1407.5668} {arXiv:1407.5668} \BibitemShut {NoStop}%
\bibitem [{\citenamefont {{Blake}}\ \emph {et~al.}(2014)\citenamefont
  {{Blake}}, \citenamefont {{James}},\ and\ \citenamefont
  {{Poole}}}]{2014BlakeBerian}%
  \BibitemOpen
  \bibfield  {author} {\bibinfo {author} {\bibfnamefont {C.}~\bibnamefont
  {{Blake}}}, \bibinfo {author} {\bibfnamefont {J.~B.}\ \bibnamefont
  {{James}}}, \ and\ \bibinfo {author} {\bibfnamefont {G.~B.}\ \bibnamefont
  {{Poole}}},\ }\href {\doibase 10.1093/mnras/stt2062} {\bibfield  {journal}
  {\bibinfo  {journal} {\mnras}\ }\textbf {\bibinfo {volume} {437}},\ \bibinfo
  {pages} {2488} (\bibinfo {year} {2014})},\ \Eprint
  {http://arxiv.org/abs/1310.6810} {arXiv:1310.6810 [astro-ph.CO]} \BibitemShut
  {NoStop}%
\bibitem [{\citenamefont {{Sutter}}\ \emph {et~al.}(2014)\citenamefont
  {{Sutter}}, \citenamefont {{Pisani}}, \citenamefont {{Wandelt}},\ and\
  \citenamefont {{Weinberg}}}]{2014SutterVoids}%
  \BibitemOpen
  \bibfield  {author} {\bibinfo {author} {\bibfnamefont {P.~M.}\ \bibnamefont
  {{Sutter}}}, \bibinfo {author} {\bibfnamefont {A.}~\bibnamefont {{Pisani}}},
  \bibinfo {author} {\bibfnamefont {B.~D.}\ \bibnamefont {{Wandelt}}}, \ and\
  \bibinfo {author} {\bibfnamefont {D.~H.}\ \bibnamefont {{Weinberg}}},\ }\href
  {\doibase 10.1093/mnras/stu1392} {\bibfield  {journal} {\bibinfo  {journal}
  {\mnras}\ }\textbf {\bibinfo {volume} {443}},\ \bibinfo {pages} {2983}
  (\bibinfo {year} {2014})},\ \Eprint {http://arxiv.org/abs/1404.5618}
  {arXiv:1404.5618} \BibitemShut {NoStop}%
\bibitem [{\citenamefont {{Wang}}\ \emph {et~al.}(2011)\citenamefont {{Wang}},
  \citenamefont {{Neyrinck}}, \citenamefont {{Szapudi}}, \citenamefont
  {{Szalay}}, \citenamefont {{Chen}}, \citenamefont {{Lesgourgues}},
  \citenamefont {{Riotto}},\ and\ \citenamefont {{Sloth}}}]{2011WangLog}%
  \BibitemOpen
  \bibfield  {author} {\bibinfo {author} {\bibfnamefont {X.}~\bibnamefont
  {{Wang}}}, \bibinfo {author} {\bibfnamefont {M.}~\bibnamefont {{Neyrinck}}},
  \bibinfo {author} {\bibfnamefont {I.}~\bibnamefont {{Szapudi}}}, \bibinfo
  {author} {\bibfnamefont {A.}~\bibnamefont {{Szalay}}}, \bibinfo {author}
  {\bibfnamefont {X.}~\bibnamefont {{Chen}}}, \bibinfo {author} {\bibfnamefont
  {J.}~\bibnamefont {{Lesgourgues}}}, \bibinfo {author} {\bibfnamefont
  {A.}~\bibnamefont {{Riotto}}}, \ and\ \bibinfo {author} {\bibfnamefont
  {M.}~\bibnamefont {{Sloth}}},\ }\href {\doibase 10.1088/0004-637X/735/1/32}
  {\bibfield  {journal} {\bibinfo  {journal} {\apj}\ }\textbf {\bibinfo
  {volume} {735}},\ \bibinfo {eid} {32} (\bibinfo {year} {2011})},\ \Eprint
  {http://arxiv.org/abs/1103.2166} {arXiv:1103.2166 [astro-ph.CO]} \BibitemShut
  {NoStop}%
\bibitem [{\citenamefont {{Carron}}\ and\ \citenamefont
  {{Szapudi}}(2013)}]{2013Carron}%
  \BibitemOpen
  \bibfield  {author} {\bibinfo {author} {\bibfnamefont {J.}~\bibnamefont
  {{Carron}}}\ and\ \bibinfo {author} {\bibfnamefont {I.}~\bibnamefont
  {{Szapudi}}},\ }\href {\doibase 10.1093/mnras/stt1215} {\bibfield  {journal}
  {\bibinfo  {journal} {\mnras}\ }\textbf {\bibinfo {volume} {434}},\ \bibinfo
  {pages} {2961} (\bibinfo {year} {2013})},\ \Eprint
  {http://arxiv.org/abs/1306.1230} {arXiv:1306.1230 [astro-ph.CO]} \BibitemShut
  {NoStop}%
\bibitem [{\citenamefont {{Carron}}\ and\ \citenamefont
  {{Szapudi}}(2014)}]{2014Carron}%
  \BibitemOpen
  \bibfield  {author} {\bibinfo {author} {\bibfnamefont {J.}~\bibnamefont
  {{Carron}}}\ and\ \bibinfo {author} {\bibfnamefont {I.}~\bibnamefont
  {{Szapudi}}},\ }\href {\doibase 10.1093/mnrasl/slt167} {\bibfield  {journal}
  {\bibinfo  {journal} {\mnras}\ }\textbf {\bibinfo {volume} {439}},\ \bibinfo
  {pages} {L11} (\bibinfo {year} {2014})},\ \Eprint
  {http://arxiv.org/abs/1310.6038} {arXiv:1310.6038 [astro-ph.CO]} \BibitemShut
  {NoStop}%
\bibitem [{\citenamefont {{Simpson}}\ \emph {et~al.}(2011)\citenamefont
  {{Simpson}}, \citenamefont {{James}}, \citenamefont {{Heavens}},\ and\
  \citenamefont {{Heymans}}}]{SimpsonClip}%
  \BibitemOpen
  \bibfield  {author} {\bibinfo {author} {\bibfnamefont {F.}~\bibnamefont
  {{Simpson}}}, \bibinfo {author} {\bibfnamefont {J.~B.}\ \bibnamefont
  {{James}}}, \bibinfo {author} {\bibfnamefont {A.~F.}\ \bibnamefont
  {{Heavens}}}, \ and\ \bibinfo {author} {\bibfnamefont {C.}~\bibnamefont
  {{Heymans}}},\ }\href {\doibase 10.1103/PhysRevLett.107.271301} {\bibfield
  {journal} {\bibinfo  {journal} {Physical Review Letters}\ }\textbf {\bibinfo
  {volume} {107}},\ \bibinfo {eid} {271301} (\bibinfo {year} {2011})},\ \Eprint
  {http://arxiv.org/abs/1107.5169} {arXiv:1107.5169 [astro-ph.CO]} \BibitemShut
  {NoStop}%
\bibitem [{\citenamefont {Price}(1958)}]{price1958useful}%
  \BibitemOpen
  \bibfield  {author} {\bibinfo {author} {\bibfnamefont {R.}~\bibnamefont
  {Price}},\ }\href@noop {} {\bibfield  {journal} {\bibinfo  {journal}
  {Information Theory, IRE Transactions on}\ }\textbf {\bibinfo {volume} {4}},\
  \bibinfo {pages} {69} (\bibinfo {year} {1958})}\BibitemShut {NoStop}%
\bibitem [{\citenamefont {Gross}\ and\ \citenamefont
  {Veeneman}(1994)}]{gross1994snr}%
  \BibitemOpen
  \bibfield  {author} {\bibinfo {author} {\bibfnamefont {R.}~\bibnamefont
  {Gross}}\ and\ \bibinfo {author} {\bibfnamefont {D.}~\bibnamefont
  {Veeneman}},\ }in\ \href@noop {} {\emph {\bibinfo {booktitle}
  {Communications, 1994. ICC'94, SUPERCOMM/ICC'94, Conference Record,'Serving
  Humanity Through Communications.'IEEE International Conference on}}}\
  (\bibinfo {organization} {IEEE},\ \bibinfo {year} {1994})\ pp.\ \bibinfo
  {pages} {843--847}\BibitemShut {NoStop}%
\bibitem [{\citenamefont {Fry}\ and\ \citenamefont
  {Gaztanaga}(1993)}]{fry1993biasing}%
  \BibitemOpen
  \bibfield  {author} {\bibinfo {author} {\bibfnamefont {J.~N.}\ \bibnamefont
  {Fry}}\ and\ \bibinfo {author} {\bibfnamefont {E.}~\bibnamefont
  {Gaztanaga}},\ }\href@noop {} {\bibfield  {journal} {\bibinfo  {journal} {The
  Astrophysical Journal}\ }\textbf {\bibinfo {volume} {413}},\ \bibinfo {pages}
  {447} (\bibinfo {year} {1993})}\BibitemShut {NoStop}%
\bibitem [{\citenamefont {{Baldauf}}\ \emph {et~al.}(2012)\citenamefont
  {{Baldauf}}, \citenamefont {{Seljak}}, \citenamefont {{Desjacques}},\ and\
  \citenamefont {{McDonald}}}]{2012Bald}%
  \BibitemOpen
  \bibfield  {author} {\bibinfo {author} {\bibfnamefont {T.}~\bibnamefont
  {{Baldauf}}}, \bibinfo {author} {\bibfnamefont {U.}~\bibnamefont {{Seljak}}},
  \bibinfo {author} {\bibfnamefont {V.}~\bibnamefont {{Desjacques}}}, \ and\
  \bibinfo {author} {\bibfnamefont {P.}~\bibnamefont {{McDonald}}},\ }\href
  {\doibase 10.1103/PhysRevD.86.083540} {\bibfield  {journal} {\bibinfo
  {journal} {\prd}\ }\textbf {\bibinfo {volume} {86}},\ \bibinfo {eid} {083540}
  (\bibinfo {year} {2012})},\ \Eprint {http://arxiv.org/abs/1201.4827}
  {arXiv:1201.4827 [astro-ph.CO]} \BibitemShut {NoStop}%
\bibitem [{\citenamefont {{Simpson}}\ and\ \citenamefont
  {{Peacock}}(2010)}]{simpsonp09}%
  \BibitemOpen
  \bibfield  {author} {\bibinfo {author} {\bibfnamefont {F.}~\bibnamefont
  {{Simpson}}}\ and\ \bibinfo {author} {\bibfnamefont {J.~A.}\ \bibnamefont
  {{Peacock}}},\ }\href {\doibase 10.1103/PhysRevD.81.043512} {\bibfield
  {journal} {\bibinfo  {journal} {\prd}\ }\textbf {\bibinfo {volume} {81}},\
  \bibinfo {pages} {043512} (\bibinfo {year} {2010})},\ \Eprint
  {http://arxiv.org/abs/0910.3834} {arXiv:0910.3834} \BibitemShut {NoStop}%
\bibitem [{\citenamefont {{Kaiser}}(1987)}]{1987MNRAS.227....1K}%
  \BibitemOpen
  \bibfield  {author} {\bibinfo {author} {\bibfnamefont {N.}~\bibnamefont
  {{Kaiser}}},\ }\href@noop {} {\bibfield  {journal} {\bibinfo  {journal}
  {\mnras}\ }\textbf {\bibinfo {volume} {227}},\ \bibinfo {pages} {1} (\bibinfo
  {year} {1987})}\BibitemShut {NoStop}%
\bibitem [{\citenamefont {{Fisher}}(1995)}]{Fisher1995}%
  \BibitemOpen
  \bibfield  {author} {\bibinfo {author} {\bibfnamefont {K.~B.}\ \bibnamefont
  {{Fisher}}},\ }\href {\doibase 10.1086/175980} {\bibfield  {journal}
  {\bibinfo  {journal} {\apj}\ }\textbf {\bibinfo {volume} {448}},\ \bibinfo
  {pages} {494} (\bibinfo {year} {1995})},\ \Eprint
  {http://arxiv.org/abs/astro-ph/9412081} {astro-ph/9412081} \BibitemShut
  {NoStop}%
\bibitem [{\citenamefont {{Kim}}\ \emph {et~al.}(2011)\citenamefont {{Kim}},
  \citenamefont {{Park}}, \citenamefont {{Rossi}}, \citenamefont {{Lee}},\ and\
  \citenamefont {{Gott}}}]{2011KimHorizon}%
  \BibitemOpen
  \bibfield  {author} {\bibinfo {author} {\bibfnamefont {J.}~\bibnamefont
  {{Kim}}}, \bibinfo {author} {\bibfnamefont {C.}~\bibnamefont {{Park}}},
  \bibinfo {author} {\bibfnamefont {G.}~\bibnamefont {{Rossi}}}, \bibinfo
  {author} {\bibfnamefont {S.~M.}\ \bibnamefont {{Lee}}}, \ and\ \bibinfo
  {author} {\bibfnamefont {J.~R.}\ \bibnamefont {{Gott}}, \bibfnamefont
  {III}},\ }\href {\doibase 10.5303/JKAS.2011.44.6.217} {\bibfield  {journal}
  {\bibinfo  {journal} {Journal of Korean Astronomical Society}\ }\textbf
  {\bibinfo {volume} {44}},\ \bibinfo {pages} {217} (\bibinfo {year} {2011})},\
  \Eprint {http://arxiv.org/abs/1112.1754} {arXiv:1112.1754 [astro-ph.CO]}
  \BibitemShut {NoStop}%
\bibitem [{\citenamefont {{Guo}}\ \emph {et~al.}(2011)\citenamefont {{Guo}},
  \citenamefont {{White}}, \citenamefont {{Boylan-Kolchin}}, \citenamefont {{De
  Lucia}}, \citenamefont {{Kauffmann}}, \citenamefont {{Lemson}}, \citenamefont
  {{Li}}, \citenamefont {{Springel}},\ and\ \citenamefont
  {{Weinmann}}}]{Guo2011}%
  \BibitemOpen
  \bibfield  {author} {\bibinfo {author} {\bibfnamefont {Q.}~\bibnamefont
  {{Guo}}}, \bibinfo {author} {\bibfnamefont {S.}~\bibnamefont {{White}}},
  \bibinfo {author} {\bibfnamefont {M.}~\bibnamefont {{Boylan-Kolchin}}},
  \bibinfo {author} {\bibfnamefont {G.}~\bibnamefont {{De Lucia}}}, \bibinfo
  {author} {\bibfnamefont {G.}~\bibnamefont {{Kauffmann}}}, \bibinfo {author}
  {\bibfnamefont {G.}~\bibnamefont {{Lemson}}}, \bibinfo {author}
  {\bibfnamefont {C.}~\bibnamefont {{Li}}}, \bibinfo {author} {\bibfnamefont
  {V.}~\bibnamefont {{Springel}}}, \ and\ \bibinfo {author} {\bibfnamefont
  {S.}~\bibnamefont {{Weinmann}}},\ }\href {\doibase
  10.1111/j.1365-2966.2010.18114.x} {\bibfield  {journal} {\bibinfo  {journal}
  {\mnras}\ }\textbf {\bibinfo {volume} {413}},\ \bibinfo {pages} {101}
  (\bibinfo {year} {2011})},\ \Eprint {http://arxiv.org/abs/1006.0106}
  {arXiv:1006.0106 [astro-ph.CO]} \BibitemShut {NoStop}%
\bibitem [{\citenamefont {{Springel}}\ \emph {et~al.}(2005)\citenamefont
  {{Springel}}, \citenamefont {{White}}, \citenamefont {{Jenkins}},
  \citenamefont {{Frenk}}, \citenamefont {{Yoshida}}, \citenamefont {{Gao}},
  \citenamefont {{Navarro}}, \citenamefont {{Thacker}}, \citenamefont
  {{Croton}}, \citenamefont {{Helly}}, \citenamefont {{Peacock}}, \citenamefont
  {{Cole}}, \citenamefont {{Thomas}}, \citenamefont {{Couchman}}, \citenamefont
  {{Evrard}}, \citenamefont {{Colberg}},\ and\ \citenamefont
  {{Pearce}}}]{2005Natur.435..629S}%
  \BibitemOpen
  \bibfield  {author} {\bibinfo {author} {\bibfnamefont {V.}~\bibnamefont
  {{Springel}}}, \bibinfo {author} {\bibfnamefont {S.~D.~M.}\ \bibnamefont
  {{White}}}, \bibinfo {author} {\bibfnamefont {A.}~\bibnamefont {{Jenkins}}},
  \bibinfo {author} {\bibfnamefont {C.~S.}\ \bibnamefont {{Frenk}}}, \bibinfo
  {author} {\bibfnamefont {N.}~\bibnamefont {{Yoshida}}}, \bibinfo {author}
  {\bibfnamefont {L.}~\bibnamefont {{Gao}}}, \bibinfo {author} {\bibfnamefont
  {J.}~\bibnamefont {{Navarro}}}, \bibinfo {author} {\bibfnamefont
  {R.}~\bibnamefont {{Thacker}}}, \bibinfo {author} {\bibfnamefont
  {D.}~\bibnamefont {{Croton}}}, \bibinfo {author} {\bibfnamefont
  {J.}~\bibnamefont {{Helly}}}, \bibinfo {author} {\bibfnamefont {J.~A.}\
  \bibnamefont {{Peacock}}}, \bibinfo {author} {\bibfnamefont {S.}~\bibnamefont
  {{Cole}}}, \bibinfo {author} {\bibfnamefont {P.}~\bibnamefont {{Thomas}}},
  \bibinfo {author} {\bibfnamefont {H.}~\bibnamefont {{Couchman}}}, \bibinfo
  {author} {\bibfnamefont {A.}~\bibnamefont {{Evrard}}}, \bibinfo {author}
  {\bibfnamefont {J.}~\bibnamefont {{Colberg}}}, \ and\ \bibinfo {author}
  {\bibfnamefont {F.}~\bibnamefont {{Pearce}}},\ }\href {\doibase
  10.1038/nature03597} {\bibfield  {journal} {\bibinfo  {journal} {\nat}\
  }\textbf {\bibinfo {volume} {435}},\ \bibinfo {pages} {629} (\bibinfo {year}
  {2005})},\ \Eprint {http://arxiv.org/abs/arXiv:astro-ph/0504097}
  {arXiv:astro-ph/0504097} \BibitemShut {NoStop}%
\bibitem [{\citenamefont {{Taylor}}\ \emph {et~al.}(2011)\citenamefont
  {{Taylor}}, \citenamefont {{Hopkins}}, \citenamefont {{Baldry}},
  \citenamefont {{Brown}}, \citenamefont {{Driver}}, \citenamefont {{Kelvin}},
  \citenamefont {{Hill}}, \citenamefont {{Robotham}}, \citenamefont
  {{Bland-Hawthorn}}, \citenamefont {{Jones}}, \citenamefont {{Sharp}},
  \citenamefont {{Thomas}}, \citenamefont {{Liske}}, \citenamefont {{Loveday}},
  \citenamefont {{Norberg}}, \citenamefont {{Peacock}}, \citenamefont
  {{Bamford}}, \citenamefont {{Brough}}, \citenamefont {{Colless}},
  \citenamefont {{Cameron}}, \citenamefont {{Conselice}}, \citenamefont
  {{Croom}}, \citenamefont {{Frenk}}, \citenamefont {{Gunawardhana}},
  \citenamefont {{Kuijken}}, \citenamefont {{Nichol}}, \citenamefont
  {{Parkinson}}, \citenamefont {{Phillipps}}, \citenamefont {{Pimbblet}},
  \citenamefont {{Popescu}}, \citenamefont {{Prescott}}, \citenamefont
  {{Sutherland}}, \citenamefont {{Tuffs}}, \citenamefont {{van Kampen}},\ and\
  \citenamefont {{Wijesinghe}}}]{2011Taylor_GAMA_stellarmass}%
  \BibitemOpen
  \bibfield  {author} {\bibinfo {author} {\bibfnamefont {E.~N.}\ \bibnamefont
  {{Taylor}}}, \bibinfo {author} {\bibfnamefont {A.~M.}\ \bibnamefont
  {{Hopkins}}}, \bibinfo {author} {\bibfnamefont {I.~K.}\ \bibnamefont
  {{Baldry}}}, \bibinfo {author} {\bibfnamefont {M.~J.~I.}\ \bibnamefont
  {{Brown}}}, \bibinfo {author} {\bibfnamefont {S.~P.}\ \bibnamefont
  {{Driver}}}, \bibinfo {author} {\bibfnamefont {L.~S.}\ \bibnamefont
  {{Kelvin}}}, \bibinfo {author} {\bibfnamefont {D.~T.}\ \bibnamefont
  {{Hill}}}, \bibinfo {author} {\bibfnamefont {A.~S.~G.}\ \bibnamefont
  {{Robotham}}}, \bibinfo {author} {\bibfnamefont {J.}~\bibnamefont
  {{Bland-Hawthorn}}}, \bibinfo {author} {\bibfnamefont {D.~H.}\ \bibnamefont
  {{Jones}}}, \bibinfo {author} {\bibfnamefont {R.~G.}\ \bibnamefont
  {{Sharp}}}, \bibinfo {author} {\bibfnamefont {D.}~\bibnamefont {{Thomas}}},
  \bibinfo {author} {\bibfnamefont {J.}~\bibnamefont {{Liske}}}, \bibinfo
  {author} {\bibfnamefont {J.}~\bibnamefont {{Loveday}}}, \bibinfo {author}
  {\bibfnamefont {P.}~\bibnamefont {{Norberg}}}, \bibinfo {author}
  {\bibfnamefont {J.~A.}\ \bibnamefont {{Peacock}}}, \bibinfo {author}
  {\bibfnamefont {S.~P.}\ \bibnamefont {{Bamford}}}, \bibinfo {author}
  {\bibfnamefont {S.}~\bibnamefont {{Brough}}}, \bibinfo {author}
  {\bibfnamefont {M.}~\bibnamefont {{Colless}}}, \bibinfo {author}
  {\bibfnamefont {E.}~\bibnamefont {{Cameron}}}, \bibinfo {author}
  {\bibfnamefont {C.~J.}\ \bibnamefont {{Conselice}}}, \bibinfo {author}
  {\bibfnamefont {S.~M.}\ \bibnamefont {{Croom}}}, \bibinfo {author}
  {\bibfnamefont {C.~S.}\ \bibnamefont {{Frenk}}}, \bibinfo {author}
  {\bibfnamefont {M.}~\bibnamefont {{Gunawardhana}}}, \bibinfo {author}
  {\bibfnamefont {K.}~\bibnamefont {{Kuijken}}}, \bibinfo {author}
  {\bibfnamefont {R.~C.}\ \bibnamefont {{Nichol}}}, \bibinfo {author}
  {\bibfnamefont {H.~R.}\ \bibnamefont {{Parkinson}}}, \bibinfo {author}
  {\bibfnamefont {S.}~\bibnamefont {{Phillipps}}}, \bibinfo {author}
  {\bibfnamefont {K.~A.}\ \bibnamefont {{Pimbblet}}}, \bibinfo {author}
  {\bibfnamefont {C.~C.}\ \bibnamefont {{Popescu}}}, \bibinfo {author}
  {\bibfnamefont {M.}~\bibnamefont {{Prescott}}}, \bibinfo {author}
  {\bibfnamefont {W.~J.}\ \bibnamefont {{Sutherland}}}, \bibinfo {author}
  {\bibfnamefont {R.~J.}\ \bibnamefont {{Tuffs}}}, \bibinfo {author}
  {\bibfnamefont {E.}~\bibnamefont {{van Kampen}}}, \ and\ \bibinfo {author}
  {\bibfnamefont {D.}~\bibnamefont {{Wijesinghe}}},\ }\href {\doibase
  10.1111/j.1365-2966.2011.19536.x} {\bibfield  {journal} {\bibinfo  {journal}
  {\mnras}\ }\textbf {\bibinfo {volume} {418}},\ \bibinfo {pages} {1587}
  (\bibinfo {year} {2011})},\ \Eprint {http://arxiv.org/abs/1108.0635}
  {arXiv:1108.0635 [astro-ph.CO]} \BibitemShut {NoStop}%
\bibitem [{\citenamefont {{Lewis}}\ \emph {et~al.}(2000)\citenamefont
  {{Lewis}}, \citenamefont {{Challinor}},\ and\ \citenamefont
  {{Lasenby}}}]{2000CAMB}%
  \BibitemOpen
  \bibfield  {author} {\bibinfo {author} {\bibfnamefont {A.}~\bibnamefont
  {{Lewis}}}, \bibinfo {author} {\bibfnamefont {A.}~\bibnamefont
  {{Challinor}}}, \ and\ \bibinfo {author} {\bibfnamefont {A.}~\bibnamefont
  {{Lasenby}}},\ }\href {\doibase 10.1086/309179} {\bibfield  {journal}
  {\bibinfo  {journal} {\apj}\ }\textbf {\bibinfo {volume} {538}},\ \bibinfo
  {pages} {473} (\bibinfo {year} {2000})},\ \Eprint
  {http://arxiv.org/abs/arXiv:astro-ph/9911177} {arXiv:astro-ph/9911177}
  \BibitemShut {NoStop}%
\bibitem [{\citenamefont {Ledoit}\ and\ \citenamefont
  {Wolf}(2004)}]{2004LedoitWolf}%
  \BibitemOpen
  \bibfield  {author} {\bibinfo {author} {\bibfnamefont {O.}~\bibnamefont
  {Ledoit}}\ and\ \bibinfo {author} {\bibfnamefont {M.}~\bibnamefont {Wolf}},\
  }\href@noop {} {\bibfield  {journal} {\bibinfo  {journal} {Journal of
  multivariate analysis}\ }\textbf {\bibinfo {volume} {88}},\ \bibinfo {pages}
  {365} (\bibinfo {year} {2004})}\BibitemShut {NoStop}%
\bibitem [{Note1()}]{Note1}%
  \BibitemOpen
  \bibinfo {note} {\protect \url
  {http://www.econ.uzh.ch/faculty/wolf/publications/covCor.m.zip}}\BibitemShut
  {NoStop}%
\bibitem [{\citenamefont {{Jennings}}\ \emph {et~al.}(2011)\citenamefont
  {{Jennings}}, \citenamefont {{Baugh}},\ and\ \citenamefont
  {{Pascoli}}}]{JenningsRSD2011}%
  \BibitemOpen
  \bibfield  {author} {\bibinfo {author} {\bibfnamefont {E.}~\bibnamefont
  {{Jennings}}}, \bibinfo {author} {\bibfnamefont {C.~M.}\ \bibnamefont
  {{Baugh}}}, \ and\ \bibinfo {author} {\bibfnamefont {S.}~\bibnamefont
  {{Pascoli}}},\ }\href {\doibase 10.1111/j.1365-2966.2010.17581.x} {\bibfield
  {journal} {\bibinfo  {journal} {\mnras}\ }\textbf {\bibinfo {volume} {410}},\
  \bibinfo {pages} {2081} (\bibinfo {year} {2011})},\ \Eprint
  {http://arxiv.org/abs/1003.4282} {arXiv:1003.4282 [astro-ph.CO]} \BibitemShut
  {NoStop}%
\bibitem [{\citenamefont {{Taruya}}\ \emph {et~al.}(2013)\citenamefont
  {{Taruya}}, \citenamefont {{Nishimichi}},\ and\ \citenamefont
  {{Bernardeau}}}]{2013TaruyaRSD}%
  \BibitemOpen
  \bibfield  {author} {\bibinfo {author} {\bibfnamefont {A.}~\bibnamefont
  {{Taruya}}}, \bibinfo {author} {\bibfnamefont {T.}~\bibnamefont
  {{Nishimichi}}}, \ and\ \bibinfo {author} {\bibfnamefont {F.}~\bibnamefont
  {{Bernardeau}}},\ }\href {\doibase 10.1103/PhysRevD.87.083509} {\bibfield
  {journal} {\bibinfo  {journal} {\prd}\ }\textbf {\bibinfo {volume} {87}},\
  \bibinfo {eid} {083509} (\bibinfo {year} {2013})},\ \Eprint
  {http://arxiv.org/abs/1301.3624} {arXiv:1301.3624 [astro-ph.CO]} \BibitemShut
  {NoStop}%
\bibitem [{\citenamefont {{Baldry}}\ \emph {et~al.}(2010)\citenamefont
  {{Baldry}}, \citenamefont {{Robotham}}, \citenamefont {{Hill}}, \citenamefont
  {{Driver}}, \citenamefont {{Liske}}, \citenamefont {{Norberg}}, \citenamefont
  {{Bamford}}, \citenamefont {{Hopkins}}, \citenamefont {{Loveday}},
  \citenamefont {{Peacock}}, \citenamefont {{Cameron}}, \citenamefont
  {{Croom}}, \citenamefont {{Cross}}, \citenamefont {{Doyle}}, \citenamefont
  {{Dye}}, \citenamefont {{Frenk}}, \citenamefont {{Jones}}, \citenamefont
  {{van Kampen}}, \citenamefont {{Kelvin}}, \citenamefont {{Nichol}},
  \citenamefont {{Parkinson}}, \citenamefont {{Popescu}}, \citenamefont
  {{Prescott}}, \citenamefont {{Sharp}}, \citenamefont {{Sutherland}},
  \citenamefont {{Thomas}},\ and\ \citenamefont
  {{Tuffs}}}]{2010MNRAS.404...86B}%
  \BibitemOpen
  \bibfield  {author} {\bibinfo {author} {\bibfnamefont {I.~K.}\ \bibnamefont
  {{Baldry}}}, \bibinfo {author} {\bibfnamefont {A.~S.~G.}\ \bibnamefont
  {{Robotham}}}, \bibinfo {author} {\bibfnamefont {D.~T.}\ \bibnamefont
  {{Hill}}}, \bibinfo {author} {\bibfnamefont {S.~P.}\ \bibnamefont
  {{Driver}}}, \bibinfo {author} {\bibfnamefont {J.}~\bibnamefont {{Liske}}},
  \bibinfo {author} {\bibfnamefont {P.}~\bibnamefont {{Norberg}}}, \bibinfo
  {author} {\bibfnamefont {S.~P.}\ \bibnamefont {{Bamford}}}, \bibinfo {author}
  {\bibfnamefont {A.~M.}\ \bibnamefont {{Hopkins}}}, \bibinfo {author}
  {\bibfnamefont {J.}~\bibnamefont {{Loveday}}}, \bibinfo {author}
  {\bibfnamefont {J.~A.}\ \bibnamefont {{Peacock}}}, \bibinfo {author}
  {\bibfnamefont {E.}~\bibnamefont {{Cameron}}}, \bibinfo {author}
  {\bibfnamefont {S.~M.}\ \bibnamefont {{Croom}}}, \bibinfo {author}
  {\bibfnamefont {N.~J.~G.}\ \bibnamefont {{Cross}}}, \bibinfo {author}
  {\bibfnamefont {I.~F.}\ \bibnamefont {{Doyle}}}, \bibinfo {author}
  {\bibfnamefont {S.}~\bibnamefont {{Dye}}}, \bibinfo {author} {\bibfnamefont
  {C.~S.}\ \bibnamefont {{Frenk}}}, \bibinfo {author} {\bibfnamefont {D.~H.}\
  \bibnamefont {{Jones}}}, \bibinfo {author} {\bibfnamefont {E.}~\bibnamefont
  {{van Kampen}}}, \bibinfo {author} {\bibfnamefont {L.~S.}\ \bibnamefont
  {{Kelvin}}}, \bibinfo {author} {\bibfnamefont {R.~C.}\ \bibnamefont
  {{Nichol}}}, \bibinfo {author} {\bibfnamefont {H.~R.}\ \bibnamefont
  {{Parkinson}}}, \bibinfo {author} {\bibfnamefont {C.~C.}\ \bibnamefont
  {{Popescu}}}, \bibinfo {author} {\bibfnamefont {M.}~\bibnamefont
  {{Prescott}}}, \bibinfo {author} {\bibfnamefont {R.~G.}\ \bibnamefont
  {{Sharp}}}, \bibinfo {author} {\bibfnamefont {W.~J.}\ \bibnamefont
  {{Sutherland}}}, \bibinfo {author} {\bibfnamefont {D.}~\bibnamefont
  {{Thomas}}}, \ and\ \bibinfo {author} {\bibfnamefont {R.~J.}\ \bibnamefont
  {{Tuffs}}},\ }\href {\doibase 10.1111/j.1365-2966.2010.16282.x} {\bibfield
  {journal} {\bibinfo  {journal} {\mnras}\ }\textbf {\bibinfo {volume} {404}},\
  \bibinfo {pages} {86} (\bibinfo {year} {2010})},\ \Eprint
  {http://arxiv.org/abs/0910.5120} {arXiv:0910.5120} \BibitemShut {NoStop}%
\bibitem [{\citenamefont {{Robotham}}\ \emph {et~al.}(2010)\citenamefont
  {{Robotham}}, \citenamefont {{Driver}}, \citenamefont {{Norberg}},
  \citenamefont {{Baldry}}, \citenamefont {{Bamford}}, \citenamefont
  {{Hopkins}}, \citenamefont {{Liske}}, \citenamefont {{Loveday}},
  \citenamefont {{Peacock}}, \citenamefont {{Cameron}}, \citenamefont
  {{Croom}}, \citenamefont {{Doyle}}, \citenamefont {{Frenk}}, \citenamefont
  {{Hill}}, \citenamefont {{Jones}}, \citenamefont {{van Kampen}},
  \citenamefont {{Kelvin}}, \citenamefont {{Kuijken}}, \citenamefont
  {{Nichol}}, \citenamefont {{Parkinson}}, \citenamefont {{Popescu}},
  \citenamefont {{Prescott}}, \citenamefont {{Sharp}}, \citenamefont
  {{Sutherland}}, \citenamefont {{Thomas}},\ and\ \citenamefont
  {{Tuffs}}}]{2010PASA...27...76R}%
  \BibitemOpen
  \bibfield  {author} {\bibinfo {author} {\bibfnamefont {A.}~\bibnamefont
  {{Robotham}}}, \bibinfo {author} {\bibfnamefont {S.~P.}\ \bibnamefont
  {{Driver}}}, \bibinfo {author} {\bibfnamefont {P.}~\bibnamefont {{Norberg}}},
  \bibinfo {author} {\bibfnamefont {I.~K.}\ \bibnamefont {{Baldry}}}, \bibinfo
  {author} {\bibfnamefont {S.~P.}\ \bibnamefont {{Bamford}}}, \bibinfo {author}
  {\bibfnamefont {A.~M.}\ \bibnamefont {{Hopkins}}}, \bibinfo {author}
  {\bibfnamefont {J.}~\bibnamefont {{Liske}}}, \bibinfo {author} {\bibfnamefont
  {J.}~\bibnamefont {{Loveday}}}, \bibinfo {author} {\bibfnamefont {J.~A.}\
  \bibnamefont {{Peacock}}}, \bibinfo {author} {\bibfnamefont {E.}~\bibnamefont
  {{Cameron}}}, \bibinfo {author} {\bibfnamefont {S.~M.}\ \bibnamefont
  {{Croom}}}, \bibinfo {author} {\bibfnamefont {I.~F.}\ \bibnamefont
  {{Doyle}}}, \bibinfo {author} {\bibfnamefont {C.~S.}\ \bibnamefont
  {{Frenk}}}, \bibinfo {author} {\bibfnamefont {D.~T.}\ \bibnamefont {{Hill}}},
  \bibinfo {author} {\bibfnamefont {D.~H.}\ \bibnamefont {{Jones}}}, \bibinfo
  {author} {\bibfnamefont {E.}~\bibnamefont {{van Kampen}}}, \bibinfo {author}
  {\bibfnamefont {L.~S.}\ \bibnamefont {{Kelvin}}}, \bibinfo {author}
  {\bibfnamefont {K.}~\bibnamefont {{Kuijken}}}, \bibinfo {author}
  {\bibfnamefont {R.~C.}\ \bibnamefont {{Nichol}}}, \bibinfo {author}
  {\bibfnamefont {H.~R.}\ \bibnamefont {{Parkinson}}}, \bibinfo {author}
  {\bibfnamefont {C.~C.}\ \bibnamefont {{Popescu}}}, \bibinfo {author}
  {\bibfnamefont {M.}~\bibnamefont {{Prescott}}}, \bibinfo {author}
  {\bibfnamefont {R.~G.}\ \bibnamefont {{Sharp}}}, \bibinfo {author}
  {\bibfnamefont {W.~J.}\ \bibnamefont {{Sutherland}}}, \bibinfo {author}
  {\bibfnamefont {D.}~\bibnamefont {{Thomas}}}, \ and\ \bibinfo {author}
  {\bibfnamefont {R.~J.}\ \bibnamefont {{Tuffs}}},\ }\href {\doibase
  10.1071/AS09053} {\bibfield  {journal} {\bibinfo  {journal} {PASA}\ }\textbf
  {\bibinfo {volume} {27}},\ \bibinfo {pages} {76} (\bibinfo {year} {2010})},\
  \Eprint {http://arxiv.org/abs/0910.5121} {arXiv:0910.5121} \BibitemShut
  {NoStop}%
\bibitem [{\citenamefont {{Driver}}\ \emph {et~al.}(2011)\citenamefont
  {{Driver}}, \citenamefont {{Hill}}, \citenamefont {{Kelvin}}, \citenamefont
  {{Robotham}}, \citenamefont {{Liske}}, \citenamefont {{Norberg}},
  \citenamefont {{Baldry}}, \citenamefont {{Bamford}}, \citenamefont
  {{Hopkins}}, \citenamefont {{Loveday}}, \citenamefont {{Peacock}},
  \citenamefont {{Andrae}}, \citenamefont {{Bland-Hawthorn}}, \citenamefont
  {{Brough}}, \citenamefont {{Brown}}, \citenamefont {{Cameron}}, \citenamefont
  {{Ching}}, \citenamefont {{Colless}}, \citenamefont {{Conselice}},
  \citenamefont {{Croom}}, \citenamefont {{Cross}}, \citenamefont {{de
  Propris}}, \citenamefont {{Dye}}, \citenamefont {{Drinkwater}}, \citenamefont
  {{Ellis}}, \citenamefont {{Graham}}, \citenamefont {{Grootes}}, \citenamefont
  {{Gunawardhana}}, \citenamefont {{Jones}}, \citenamefont {{van Kampen}},
  \citenamefont {{Maraston}}, \citenamefont {{Nichol}}, \citenamefont
  {{Parkinson}}, \citenamefont {{Phillipps}}, \citenamefont {{Pimbblet}},
  \citenamefont {{Popescu}}, \citenamefont {{Prescott}}, \citenamefont
  {{Roseboom}}, \citenamefont {{Sadler}}, \citenamefont {{Sansom}},
  \citenamefont {{Sharp}}, \citenamefont {{Smith}}, \citenamefont {{Taylor}},
  \citenamefont {{Thomas}}, \citenamefont {{Tuffs}}, \citenamefont
  {{Wijesinghe}}, \citenamefont {{Dunne}}, \citenamefont {{Frenk}},
  \citenamefont {{Jarvis}}, \citenamefont {{Madore}}, \citenamefont {{Meyer}},
  \citenamefont {{Seibert}}, \citenamefont {{Staveley-Smith}}, \citenamefont
  {{Sutherland}},\ and\ \citenamefont {{Warren}}}]{2011MNRAS.413..971D}%
  \BibitemOpen
  \bibfield  {author} {\bibinfo {author} {\bibfnamefont {S.~P.}\ \bibnamefont
  {{Driver}}}, \bibinfo {author} {\bibfnamefont {D.~T.}\ \bibnamefont
  {{Hill}}}, \bibinfo {author} {\bibfnamefont {L.~S.}\ \bibnamefont
  {{Kelvin}}}, \bibinfo {author} {\bibfnamefont {A.~S.~G.}\ \bibnamefont
  {{Robotham}}}, \bibinfo {author} {\bibfnamefont {J.}~\bibnamefont {{Liske}}},
  \bibinfo {author} {\bibfnamefont {P.}~\bibnamefont {{Norberg}}}, \bibinfo
  {author} {\bibfnamefont {I.~K.}\ \bibnamefont {{Baldry}}}, \bibinfo {author}
  {\bibfnamefont {S.~P.}\ \bibnamefont {{Bamford}}}, \bibinfo {author}
  {\bibfnamefont {A.~M.}\ \bibnamefont {{Hopkins}}}, \bibinfo {author}
  {\bibfnamefont {J.}~\bibnamefont {{Loveday}}}, \bibinfo {author}
  {\bibfnamefont {J.~A.}\ \bibnamefont {{Peacock}}}, \bibinfo {author}
  {\bibfnamefont {E.}~\bibnamefont {{Andrae}}}, \bibinfo {author}
  {\bibfnamefont {J.}~\bibnamefont {{Bland-Hawthorn}}}, \bibinfo {author}
  {\bibfnamefont {S.}~\bibnamefont {{Brough}}}, \bibinfo {author}
  {\bibfnamefont {M.~J.~I.}\ \bibnamefont {{Brown}}}, \bibinfo {author}
  {\bibfnamefont {E.}~\bibnamefont {{Cameron}}}, \bibinfo {author}
  {\bibfnamefont {J.~H.~Y.}\ \bibnamefont {{Ching}}}, \bibinfo {author}
  {\bibfnamefont {M.}~\bibnamefont {{Colless}}}, \bibinfo {author}
  {\bibfnamefont {C.~J.}\ \bibnamefont {{Conselice}}}, \bibinfo {author}
  {\bibfnamefont {S.~M.}\ \bibnamefont {{Croom}}}, \bibinfo {author}
  {\bibfnamefont {N.~J.~G.}\ \bibnamefont {{Cross}}}, \bibinfo {author}
  {\bibfnamefont {R.}~\bibnamefont {{de Propris}}}, \bibinfo {author}
  {\bibfnamefont {S.}~\bibnamefont {{Dye}}}, \bibinfo {author} {\bibfnamefont
  {M.~J.}\ \bibnamefont {{Drinkwater}}}, \bibinfo {author} {\bibfnamefont
  {S.}~\bibnamefont {{Ellis}}}, \bibinfo {author} {\bibfnamefont {A.~W.}\
  \bibnamefont {{Graham}}}, \bibinfo {author} {\bibfnamefont {M.~W.}\
  \bibnamefont {{Grootes}}}, \bibinfo {author} {\bibfnamefont {M.}~\bibnamefont
  {{Gunawardhana}}}, \bibinfo {author} {\bibfnamefont {D.~H.}\ \bibnamefont
  {{Jones}}}, \bibinfo {author} {\bibfnamefont {E.}~\bibnamefont {{van
  Kampen}}}, \bibinfo {author} {\bibfnamefont {C.}~\bibnamefont {{Maraston}}},
  \bibinfo {author} {\bibfnamefont {R.~C.}\ \bibnamefont {{Nichol}}}, \bibinfo
  {author} {\bibfnamefont {H.~R.}\ \bibnamefont {{Parkinson}}}, \bibinfo
  {author} {\bibfnamefont {S.}~\bibnamefont {{Phillipps}}}, \bibinfo {author}
  {\bibfnamefont {K.}~\bibnamefont {{Pimbblet}}}, \bibinfo {author}
  {\bibfnamefont {C.~C.}\ \bibnamefont {{Popescu}}}, \bibinfo {author}
  {\bibfnamefont {M.}~\bibnamefont {{Prescott}}}, \bibinfo {author}
  {\bibfnamefont {I.~G.}\ \bibnamefont {{Roseboom}}}, \bibinfo {author}
  {\bibfnamefont {E.~M.}\ \bibnamefont {{Sadler}}}, \bibinfo {author}
  {\bibfnamefont {A.~E.}\ \bibnamefont {{Sansom}}}, \bibinfo {author}
  {\bibfnamefont {R.~G.}\ \bibnamefont {{Sharp}}}, \bibinfo {author}
  {\bibfnamefont {D.~J.~B.}\ \bibnamefont {{Smith}}}, \bibinfo {author}
  {\bibfnamefont {E.}~\bibnamefont {{Taylor}}}, \bibinfo {author}
  {\bibfnamefont {D.}~\bibnamefont {{Thomas}}}, \bibinfo {author}
  {\bibfnamefont {R.~J.}\ \bibnamefont {{Tuffs}}}, \bibinfo {author}
  {\bibfnamefont {D.}~\bibnamefont {{Wijesinghe}}}, \bibinfo {author}
  {\bibfnamefont {L.}~\bibnamefont {{Dunne}}}, \bibinfo {author} {\bibfnamefont
  {C.~S.}\ \bibnamefont {{Frenk}}}, \bibinfo {author} {\bibfnamefont {M.~J.}\
  \bibnamefont {{Jarvis}}}, \bibinfo {author} {\bibfnamefont {B.~F.}\
  \bibnamefont {{Madore}}}, \bibinfo {author} {\bibfnamefont {M.~J.}\
  \bibnamefont {{Meyer}}}, \bibinfo {author} {\bibfnamefont {M.}~\bibnamefont
  {{Seibert}}}, \bibinfo {author} {\bibfnamefont {L.}~\bibnamefont
  {{Staveley-Smith}}}, \bibinfo {author} {\bibfnamefont {W.~J.}\ \bibnamefont
  {{Sutherland}}}, \ and\ \bibinfo {author} {\bibfnamefont {S.~J.}\
  \bibnamefont {{Warren}}},\ }\href {\doibase 10.1111/j.1365-2966.2010.18188.x}
  {\bibfield  {journal} {\bibinfo  {journal} {\mnras}\ }\textbf {\bibinfo
  {volume} {413}},\ \bibinfo {pages} {971} (\bibinfo {year} {2011})},\ \Eprint
  {http://arxiv.org/abs/1009.0614} {arXiv:1009.0614 [astro-ph.CO]} \BibitemShut
  {NoStop}%
\bibitem [{\citenamefont {{Liske}}\ \emph {et~al.}(2015)\citenamefont
  {{Liske}}, \citenamefont {{Baldry}}, \citenamefont {{Driver}}, \citenamefont
  {{Tuffs}}, \citenamefont {{Alpaslan}}, \citenamefont {{Andrae}},
  \citenamefont {{Brough}}, \citenamefont {{Cluver}}, \citenamefont
  {{Grootes}}, \citenamefont {{Gunawardhana}}, \citenamefont {{Kelvin}},
  \citenamefont {{Loveday}}, \citenamefont {{Robotham}}, \citenamefont
  {{Taylor}}, \citenamefont {{Bamford}}, \citenamefont {{Bland-Hawthorn}},
  \citenamefont {{Brown}}, \citenamefont {{Drinkwater}}, \citenamefont
  {{Hopkins}}, \citenamefont {{Meyer}}, \citenamefont {{Norberg}},
  \citenamefont {{Peacock}}, \citenamefont {{Agius}}, \citenamefont
  {{Andrews}}, \citenamefont {{Bauer}}, \citenamefont {{Ching}}, \citenamefont
  {{Colless}}, \citenamefont {{Conselice}}, \citenamefont {{Croom}},
  \citenamefont {{Davies}}, \citenamefont {{De Propris}}, \citenamefont
  {{Dunne}}, \citenamefont {{Eardley}}, \citenamefont {{Ellis}}, \citenamefont
  {{Foster}}, \citenamefont {{Frenk}}, \citenamefont {{H{\"a}u{\ss}ler}},
  \citenamefont {{Holwerda}}, \citenamefont {{Howlett}}, \citenamefont
  {{Ibarra}}, \citenamefont {{Jarvis}}, \citenamefont {{Jones}}, \citenamefont
  {{Kafle}}, \citenamefont {{Lacey}}, \citenamefont {{Lange}}, \citenamefont
  {{Lara-L{\'o}pez}}, \citenamefont {{L{\'o}pez-S{\'a}nchez}}, \citenamefont
  {{Maddox}}, \citenamefont {{Madore}}, \citenamefont {{McNaught-Roberts}},
  \citenamefont {{Moffett}}, \citenamefont {{Nichol}}, \citenamefont {{Owers}},
  \citenamefont {{Palamara}}, \citenamefont {{Penny}}, \citenamefont
  {{Phillipps}}, \citenamefont {{Pimbblet}}, \citenamefont {{Popescu}},
  \citenamefont {{Prescott}}, \citenamefont {{Proctor}}, \citenamefont
  {{Sadler}}, \citenamefont {{Sansom}}, \citenamefont {{Seibert}},
  \citenamefont {{Sharp}}, \citenamefont {{Sutherland}}, \citenamefont
  {{V{\'a}zquez-Mata}}, \citenamefont {{van Kampen}}, \citenamefont
  {{Wilkins}}, \citenamefont {{Williams}},\ and\ \citenamefont
  {{Wright}}}]{Liske2015}%
  \BibitemOpen
  \bibfield  {author} {\bibinfo {author} {\bibfnamefont {J.}~\bibnamefont
  {{Liske}}}, \bibinfo {author} {\bibfnamefont {I.~K.}\ \bibnamefont
  {{Baldry}}}, \bibinfo {author} {\bibfnamefont {S.~P.}\ \bibnamefont
  {{Driver}}}, \bibinfo {author} {\bibfnamefont {R.~J.}\ \bibnamefont
  {{Tuffs}}}, \bibinfo {author} {\bibfnamefont {M.}~\bibnamefont {{Alpaslan}}},
  \bibinfo {author} {\bibfnamefont {E.}~\bibnamefont {{Andrae}}}, \bibinfo
  {author} {\bibfnamefont {S.}~\bibnamefont {{Brough}}}, \bibinfo {author}
  {\bibfnamefont {M.~E.}\ \bibnamefont {{Cluver}}}, \bibinfo {author}
  {\bibfnamefont {M.~W.}\ \bibnamefont {{Grootes}}}, \bibinfo {author}
  {\bibfnamefont {M.~L.~P.}\ \bibnamefont {{Gunawardhana}}}, \bibinfo {author}
  {\bibfnamefont {L.~S.}\ \bibnamefont {{Kelvin}}}, \bibinfo {author}
  {\bibfnamefont {J.}~\bibnamefont {{Loveday}}}, \bibinfo {author}
  {\bibfnamefont {A.~S.~G.}\ \bibnamefont {{Robotham}}}, \bibinfo {author}
  {\bibfnamefont {E.~N.}\ \bibnamefont {{Taylor}}}, \bibinfo {author}
  {\bibfnamefont {S.~P.}\ \bibnamefont {{Bamford}}}, \bibinfo {author}
  {\bibfnamefont {J.}~\bibnamefont {{Bland-Hawthorn}}}, \bibinfo {author}
  {\bibfnamefont {M.~J.~I.}\ \bibnamefont {{Brown}}}, \bibinfo {author}
  {\bibfnamefont {M.~J.}\ \bibnamefont {{Drinkwater}}}, \bibinfo {author}
  {\bibfnamefont {A.~M.}\ \bibnamefont {{Hopkins}}}, \bibinfo {author}
  {\bibfnamefont {M.~J.}\ \bibnamefont {{Meyer}}}, \bibinfo {author}
  {\bibfnamefont {P.}~\bibnamefont {{Norberg}}}, \bibinfo {author}
  {\bibfnamefont {J.~A.}\ \bibnamefont {{Peacock}}}, \bibinfo {author}
  {\bibfnamefont {N.~K.}\ \bibnamefont {{Agius}}}, \bibinfo {author}
  {\bibfnamefont {S.~K.}\ \bibnamefont {{Andrews}}}, \bibinfo {author}
  {\bibfnamefont {A.~E.}\ \bibnamefont {{Bauer}}}, \bibinfo {author}
  {\bibfnamefont {J.~H.~Y.}\ \bibnamefont {{Ching}}}, \bibinfo {author}
  {\bibfnamefont {M.}~\bibnamefont {{Colless}}}, \bibinfo {author}
  {\bibfnamefont {C.~J.}\ \bibnamefont {{Conselice}}}, \bibinfo {author}
  {\bibfnamefont {S.~M.}\ \bibnamefont {{Croom}}}, \bibinfo {author}
  {\bibfnamefont {L.~J.~M.}\ \bibnamefont {{Davies}}}, \bibinfo {author}
  {\bibfnamefont {R.}~\bibnamefont {{De Propris}}}, \bibinfo {author}
  {\bibfnamefont {L.}~\bibnamefont {{Dunne}}}, \bibinfo {author} {\bibfnamefont
  {E.~M.}\ \bibnamefont {{Eardley}}}, \bibinfo {author} {\bibfnamefont
  {S.}~\bibnamefont {{Ellis}}}, \bibinfo {author} {\bibfnamefont
  {C.}~\bibnamefont {{Foster}}}, \bibinfo {author} {\bibfnamefont {C.~S.}\
  \bibnamefont {{Frenk}}}, \bibinfo {author} {\bibfnamefont {B.}~\bibnamefont
  {{H{\"a}u{\ss}ler}}}, \bibinfo {author} {\bibfnamefont {B.~W.}\ \bibnamefont
  {{Holwerda}}}, \bibinfo {author} {\bibfnamefont {C.}~\bibnamefont
  {{Howlett}}}, \bibinfo {author} {\bibfnamefont {H.}~\bibnamefont {{Ibarra}}},
  \bibinfo {author} {\bibfnamefont {M.~J.}\ \bibnamefont {{Jarvis}}}, \bibinfo
  {author} {\bibfnamefont {D.~H.}\ \bibnamefont {{Jones}}}, \bibinfo {author}
  {\bibfnamefont {P.~R.}\ \bibnamefont {{Kafle}}}, \bibinfo {author}
  {\bibfnamefont {C.~G.}\ \bibnamefont {{Lacey}}}, \bibinfo {author}
  {\bibfnamefont {R.}~\bibnamefont {{Lange}}}, \bibinfo {author} {\bibfnamefont
  {M.~A.}\ \bibnamefont {{Lara-L{\'o}pez}}}, \bibinfo {author} {\bibfnamefont
  {{\'A}.~R.}\ \bibnamefont {{L{\'o}pez-S{\'a}nchez}}}, \bibinfo {author}
  {\bibfnamefont {S.}~\bibnamefont {{Maddox}}}, \bibinfo {author}
  {\bibfnamefont {B.~F.}\ \bibnamefont {{Madore}}}, \bibinfo {author}
  {\bibfnamefont {T.}~\bibnamefont {{McNaught-Roberts}}}, \bibinfo {author}
  {\bibfnamefont {A.~J.}\ \bibnamefont {{Moffett}}}, \bibinfo {author}
  {\bibfnamefont {R.~C.}\ \bibnamefont {{Nichol}}}, \bibinfo {author}
  {\bibfnamefont {M.~S.}\ \bibnamefont {{Owers}}}, \bibinfo {author}
  {\bibfnamefont {D.}~\bibnamefont {{Palamara}}}, \bibinfo {author}
  {\bibfnamefont {S.~J.}\ \bibnamefont {{Penny}}}, \bibinfo {author}
  {\bibfnamefont {S.}~\bibnamefont {{Phillipps}}}, \bibinfo {author}
  {\bibfnamefont {K.~A.}\ \bibnamefont {{Pimbblet}}}, \bibinfo {author}
  {\bibfnamefont {C.~C.}\ \bibnamefont {{Popescu}}}, \bibinfo {author}
  {\bibfnamefont {M.}~\bibnamefont {{Prescott}}}, \bibinfo {author}
  {\bibfnamefont {R.}~\bibnamefont {{Proctor}}}, \bibinfo {author}
  {\bibfnamefont {E.~M.}\ \bibnamefont {{Sadler}}}, \bibinfo {author}
  {\bibfnamefont {A.~E.}\ \bibnamefont {{Sansom}}}, \bibinfo {author}
  {\bibfnamefont {M.}~\bibnamefont {{Seibert}}}, \bibinfo {author}
  {\bibfnamefont {R.}~\bibnamefont {{Sharp}}}, \bibinfo {author} {\bibfnamefont
  {W.}~\bibnamefont {{Sutherland}}}, \bibinfo {author} {\bibfnamefont {J.~A.}\
  \bibnamefont {{V{\'a}zquez-Mata}}}, \bibinfo {author} {\bibfnamefont
  {E.}~\bibnamefont {{van Kampen}}}, \bibinfo {author} {\bibfnamefont {S.~M.}\
  \bibnamefont {{Wilkins}}}, \bibinfo {author} {\bibfnamefont {R.}~\bibnamefont
  {{Williams}}}, \ and\ \bibinfo {author} {\bibfnamefont {A.~H.}\ \bibnamefont
  {{Wright}}},\ }\href {\doibase 10.1093/mnras/stv1436} {\bibfield  {journal}
  {\bibinfo  {journal} {\mnras}\ }\textbf {\bibinfo {volume} {452}},\ \bibinfo
  {pages} {2087} (\bibinfo {year} {2015})},\ \Eprint
  {http://arxiv.org/abs/1506.08222} {arXiv:1506.08222} \BibitemShut {NoStop}%
\bibitem [{\citenamefont {{Blake}}\ \emph {et~al.}(2013)\citenamefont
  {{Blake}}, \citenamefont {{Baldry}}, \citenamefont {{Bland-Hawthorn}},
  \citenamefont {{Christodoulou}}, \citenamefont {{Colless}}, \citenamefont
  {{Conselice}}, \citenamefont {{Driver}}, \citenamefont {{Hopkins}},
  \citenamefont {{Liske}}, \citenamefont {{Loveday}}, \citenamefont
  {{Norberg}}, \citenamefont {{Peacock}}, \citenamefont {{Poole}},\ and\
  \citenamefont {{Robotham}}}]{2013MNRAS.436.3089B}%
  \BibitemOpen
  \bibfield  {author} {\bibinfo {author} {\bibfnamefont {C.}~\bibnamefont
  {{Blake}}}, \bibinfo {author} {\bibfnamefont {I.~K.}\ \bibnamefont
  {{Baldry}}}, \bibinfo {author} {\bibfnamefont {J.}~\bibnamefont
  {{Bland-Hawthorn}}}, \bibinfo {author} {\bibfnamefont {L.}~\bibnamefont
  {{Christodoulou}}}, \bibinfo {author} {\bibfnamefont {M.}~\bibnamefont
  {{Colless}}}, \bibinfo {author} {\bibfnamefont {C.}~\bibnamefont
  {{Conselice}}}, \bibinfo {author} {\bibfnamefont {S.~P.}\ \bibnamefont
  {{Driver}}}, \bibinfo {author} {\bibfnamefont {A.~M.}\ \bibnamefont
  {{Hopkins}}}, \bibinfo {author} {\bibfnamefont {J.}~\bibnamefont {{Liske}}},
  \bibinfo {author} {\bibfnamefont {J.}~\bibnamefont {{Loveday}}}, \bibinfo
  {author} {\bibfnamefont {P.}~\bibnamefont {{Norberg}}}, \bibinfo {author}
  {\bibfnamefont {J.~A.}\ \bibnamefont {{Peacock}}}, \bibinfo {author}
  {\bibfnamefont {G.~B.}\ \bibnamefont {{Poole}}}, \ and\ \bibinfo {author}
  {\bibfnamefont {A.~S.~G.}\ \bibnamefont {{Robotham}}},\ }\href {\doibase
  10.1093/mnras/stt1791} {\bibfield  {journal} {\bibinfo  {journal} {\mnras}\
  }\textbf {\bibinfo {volume} {436}},\ \bibinfo {pages} {3089} (\bibinfo {year}
  {2013})},\ \Eprint {http://arxiv.org/abs/1309.5556} {arXiv:1309.5556
  [astro-ph.CO]} \BibitemShut {NoStop}%
\bibitem [{\citenamefont {{Baldry}}\ \emph {et~al.}(2014)\citenamefont
  {{Baldry}}, \citenamefont {{Alpaslan}}, \citenamefont {{Bauer}},
  \citenamefont {{Bland-Hawthorn}}, \citenamefont {{Brough}}, \citenamefont
  {{Cluver}}, \citenamefont {{Croom}}, \citenamefont {{Davies}}, \citenamefont
  {{Driver}}, \citenamefont {{Gunawardhana}}, \citenamefont {{Holwerda}},
  \citenamefont {{Hopkins}}, \citenamefont {{Kelvin}}, \citenamefont {{Liske}},
  \citenamefont {{L{\'o}pez-S{\'a}nchez}}, \citenamefont {{Loveday}},
  \citenamefont {{Norberg}}, \citenamefont {{Peacock}}, \citenamefont
  {{Robotham}},\ and\ \citenamefont {{Taylor}}}]{2014MNRAS.441.2440B}%
  \BibitemOpen
  \bibfield  {author} {\bibinfo {author} {\bibfnamefont {I.~K.}\ \bibnamefont
  {{Baldry}}}, \bibinfo {author} {\bibfnamefont {M.}~\bibnamefont
  {{Alpaslan}}}, \bibinfo {author} {\bibfnamefont {A.~E.}\ \bibnamefont
  {{Bauer}}}, \bibinfo {author} {\bibfnamefont {J.}~\bibnamefont
  {{Bland-Hawthorn}}}, \bibinfo {author} {\bibfnamefont {S.}~\bibnamefont
  {{Brough}}}, \bibinfo {author} {\bibfnamefont {M.~E.}\ \bibnamefont
  {{Cluver}}}, \bibinfo {author} {\bibfnamefont {S.~M.}\ \bibnamefont
  {{Croom}}}, \bibinfo {author} {\bibfnamefont {L.~J.~M.}\ \bibnamefont
  {{Davies}}}, \bibinfo {author} {\bibfnamefont {S.~P.}\ \bibnamefont
  {{Driver}}}, \bibinfo {author} {\bibfnamefont {M.~L.~P.}\ \bibnamefont
  {{Gunawardhana}}}, \bibinfo {author} {\bibfnamefont {B.~W.}\ \bibnamefont
  {{Holwerda}}}, \bibinfo {author} {\bibfnamefont {A.~M.}\ \bibnamefont
  {{Hopkins}}}, \bibinfo {author} {\bibfnamefont {L.~S.}\ \bibnamefont
  {{Kelvin}}}, \bibinfo {author} {\bibfnamefont {J.}~\bibnamefont {{Liske}}},
  \bibinfo {author} {\bibfnamefont {{\'A}.~R.}\ \bibnamefont
  {{L{\'o}pez-S{\'a}nchez}}}, \bibinfo {author} {\bibfnamefont
  {J.}~\bibnamefont {{Loveday}}}, \bibinfo {author} {\bibfnamefont
  {P.}~\bibnamefont {{Norberg}}}, \bibinfo {author} {\bibfnamefont
  {J.}~\bibnamefont {{Peacock}}}, \bibinfo {author} {\bibfnamefont {A.~S.~G.}\
  \bibnamefont {{Robotham}}}, \ and\ \bibinfo {author} {\bibfnamefont {E.~N.}\
  \bibnamefont {{Taylor}}},\ }\href {\doibase 10.1093/mnras/stu727} {\bibfield
  {journal} {\bibinfo  {journal} {\mnras}\ }\textbf {\bibinfo {volume} {441}},\
  \bibinfo {pages} {2440} (\bibinfo {year} {2014})},\ \Eprint
  {http://arxiv.org/abs/1404.2626} {arXiv:1404.2626 [astro-ph.IM]} \BibitemShut
  {NoStop}%
\bibitem [{\citenamefont {{Planck Collaboration}}(2015)}]{2015PlanckParams}%
  \BibitemOpen
  \bibfield  {author} {\bibinfo {author} {\bibnamefont {{Planck
  Collaboration}}},\ }\href@noop {} {\bibfield  {journal} {\bibinfo  {journal}
  {ArXiv e-prints}\ } (\bibinfo {year} {2015})},\ \Eprint
  {http://arxiv.org/abs/1502.01589} {arXiv:1502.01589} \BibitemShut {NoStop}%
\bibitem [{\citenamefont {{Heymans}}\ \emph {et~al.}(2013)\citenamefont
  {{Heymans}}, \citenamefont {{Grocutt}}, \citenamefont {{Heavens}},
  \citenamefont {{Kilbinger}}, \citenamefont {{Kitching}}, \citenamefont
  {{Simpson}}, \citenamefont {{Benjamin}}, \citenamefont {{Erben}},
  \citenamefont {{Hildebrandt}}, \citenamefont {{Hoekstra}}, \citenamefont
  {{Mellier}}, \citenamefont {{Miller}}, \citenamefont {{Van Waerbeke}},
  \citenamefont {{Brown}}, \citenamefont {{Coupon}}, \citenamefont {{Fu}},
  \citenamefont {{Harnois-D{\'e}raps}}, \citenamefont {{Hudson}}, \citenamefont
  {{Kuijken}}, \citenamefont {{Rowe}}, \citenamefont {{Schrabback}},
  \citenamefont {{Semboloni}}, \citenamefont {{Vafaei}},\ and\ \citenamefont
  {{Velander}}}]{syspaper}%
  \BibitemOpen
  \bibfield  {author} {\bibinfo {author} {\bibfnamefont {C.}~\bibnamefont
  {{Heymans}}}, \bibinfo {author} {\bibfnamefont {E.}~\bibnamefont
  {{Grocutt}}}, \bibinfo {author} {\bibfnamefont {A.}~\bibnamefont
  {{Heavens}}}, \bibinfo {author} {\bibfnamefont {M.}~\bibnamefont
  {{Kilbinger}}}, \bibinfo {author} {\bibfnamefont {T.~D.}\ \bibnamefont
  {{Kitching}}}, \bibinfo {author} {\bibfnamefont {F.}~\bibnamefont
  {{Simpson}}}, \bibinfo {author} {\bibfnamefont {J.}~\bibnamefont
  {{Benjamin}}}, \bibinfo {author} {\bibfnamefont {T.}~\bibnamefont {{Erben}}},
  \bibinfo {author} {\bibfnamefont {H.}~\bibnamefont {{Hildebrandt}}}, \bibinfo
  {author} {\bibfnamefont {H.}~\bibnamefont {{Hoekstra}}}, \bibinfo {author}
  {\bibfnamefont {Y.}~\bibnamefont {{Mellier}}}, \bibinfo {author}
  {\bibfnamefont {L.}~\bibnamefont {{Miller}}}, \bibinfo {author}
  {\bibfnamefont {L.}~\bibnamefont {{Van Waerbeke}}}, \bibinfo {author}
  {\bibfnamefont {M.~L.}\ \bibnamefont {{Brown}}}, \bibinfo {author}
  {\bibfnamefont {J.}~\bibnamefont {{Coupon}}}, \bibinfo {author}
  {\bibfnamefont {L.}~\bibnamefont {{Fu}}}, \bibinfo {author} {\bibfnamefont
  {J.}~\bibnamefont {{Harnois-D{\'e}raps}}}, \bibinfo {author} {\bibfnamefont
  {M.~J.}\ \bibnamefont {{Hudson}}}, \bibinfo {author} {\bibfnamefont
  {K.}~\bibnamefont {{Kuijken}}}, \bibinfo {author} {\bibfnamefont
  {B.}~\bibnamefont {{Rowe}}}, \bibinfo {author} {\bibfnamefont
  {T.}~\bibnamefont {{Schrabback}}}, \bibinfo {author} {\bibfnamefont
  {E.}~\bibnamefont {{Semboloni}}}, \bibinfo {author} {\bibfnamefont
  {S.}~\bibnamefont {{Vafaei}}}, \ and\ \bibinfo {author} {\bibfnamefont
  {M.}~\bibnamefont {{Velander}}},\ }\href {\doibase 10.1093/mnras/stt601}
  {\bibfield  {journal} {\bibinfo  {journal} {\mnras}\ }\textbf {\bibinfo
  {volume} {432}},\ \bibinfo {pages} {2433} (\bibinfo {year} {2013})},\ \Eprint
  {http://arxiv.org/abs/1303.1808} {arXiv:1303.1808 [astro-ph.CO]} \BibitemShut
  {NoStop}%
\bibitem [{\citenamefont {{Vikhlinin}}\ \emph {et~al.}(2009)\citenamefont
  {{Vikhlinin}}, \citenamefont {{Kravtsov}}, \citenamefont {{Burenin}},
  \citenamefont {{Ebeling}}, \citenamefont {{Forman}}, \citenamefont
  {{Hornstrup}}, \citenamefont {{Jones}}, \citenamefont {{Murray}},
  \citenamefont {{Nagai}}, \citenamefont {{Quintana}},\ and\ \citenamefont
  {{Voevodkin}}}]{Xrays}%
  \BibitemOpen
  \bibfield  {author} {\bibinfo {author} {\bibfnamefont {A.}~\bibnamefont
  {{Vikhlinin}}}, \bibinfo {author} {\bibfnamefont {A.~V.}\ \bibnamefont
  {{Kravtsov}}}, \bibinfo {author} {\bibfnamefont {R.~A.}\ \bibnamefont
  {{Burenin}}}, \bibinfo {author} {\bibfnamefont {H.}~\bibnamefont
  {{Ebeling}}}, \bibinfo {author} {\bibfnamefont {W.~R.}\ \bibnamefont
  {{Forman}}}, \bibinfo {author} {\bibfnamefont {A.}~\bibnamefont
  {{Hornstrup}}}, \bibinfo {author} {\bibfnamefont {C.}~\bibnamefont
  {{Jones}}}, \bibinfo {author} {\bibfnamefont {S.~S.}\ \bibnamefont
  {{Murray}}}, \bibinfo {author} {\bibfnamefont {D.}~\bibnamefont {{Nagai}}},
  \bibinfo {author} {\bibfnamefont {H.}~\bibnamefont {{Quintana}}}, \ and\
  \bibinfo {author} {\bibfnamefont {A.}~\bibnamefont {{Voevodkin}}},\ }\href
  {\doibase 10.1088/0004-637X/692/2/1060} {\bibfield  {journal} {\bibinfo
  {journal} {\apj}\ }\textbf {\bibinfo {volume} {692}},\ \bibinfo {pages}
  {1060} (\bibinfo {year} {2009})},\ \Eprint {http://arxiv.org/abs/0812.2720}
  {arXiv:0812.2720} \BibitemShut {NoStop}%
\bibitem [{\citenamefont {{Macaulay}}\ \emph {et~al.}(2013)\citenamefont
  {{Macaulay}}, \citenamefont {{Wehus}},\ and\ \citenamefont
  {{Eriksen}}}]{2013PhRvLMacaulay}%
  \BibitemOpen
  \bibfield  {author} {\bibinfo {author} {\bibfnamefont {E.}~\bibnamefont
  {{Macaulay}}}, \bibinfo {author} {\bibfnamefont {I.~K.}\ \bibnamefont
  {{Wehus}}}, \ and\ \bibinfo {author} {\bibfnamefont {H.~K.}\ \bibnamefont
  {{Eriksen}}},\ }\href {\doibase 10.1103/PhysRevLett.111.161301} {\bibfield
  {journal} {\bibinfo  {journal} {Physical Review Letters}\ }\textbf {\bibinfo
  {volume} {111}},\ \bibinfo {eid} {161301} (\bibinfo {year} {2013})},\ \Eprint
  {http://arxiv.org/abs/1303.6583} {arXiv:1303.6583 [astro-ph.CO]} \BibitemShut
  {NoStop}%
\bibitem [{\citenamefont {{Alcock}}\ and\ \citenamefont
  {{Paczynski}}(1979)}]{1979Natur.281..358A}%
  \BibitemOpen
  \bibfield  {author} {\bibinfo {author} {\bibfnamefont {C.}~\bibnamefont
  {{Alcock}}}\ and\ \bibinfo {author} {\bibfnamefont {B.}~\bibnamefont
  {{Paczynski}}},\ }\href {\doibase 10.1038/281358a0} {\bibfield  {journal}
  {\bibinfo  {journal} {\nat}\ }\textbf {\bibinfo {volume} {281}},\ \bibinfo
  {pages} {358} (\bibinfo {year} {1979})}\BibitemShut {NoStop}%
\bibitem [{\citenamefont {{Ballinger}}\ \emph {et~al.}(1996)\citenamefont
  {{Ballinger}}, \citenamefont {{Peacock}},\ and\ \citenamefont
  {{Heavens}}}]{1996MNRAS.282..877B}%
  \BibitemOpen
  \bibfield  {author} {\bibinfo {author} {\bibfnamefont {W.~E.}\ \bibnamefont
  {{Ballinger}}}, \bibinfo {author} {\bibfnamefont {J.~A.}\ \bibnamefont
  {{Peacock}}}, \ and\ \bibinfo {author} {\bibfnamefont {A.~F.}\ \bibnamefont
  {{Heavens}}},\ }\href@noop {} {\bibfield  {journal} {\bibinfo  {journal}
  {\mnras}\ }\textbf {\bibinfo {volume} {282}},\ \bibinfo {pages} {877}
  (\bibinfo {year} {1996})},\ \Eprint
  {http://arxiv.org/abs/arXiv:astro-ph/9605017} {arXiv:astro-ph/9605017}
  \BibitemShut {NoStop}%
\bibitem [{\citenamefont {{Planck Collaboration}}\ \emph
  {et~al.}(2015)\citenamefont {{Planck Collaboration}}, \citenamefont {{Ade}},
  \citenamefont {{Aghanim}}, \citenamefont {{Arnaud}}, \citenamefont
  {{Arroja}}, \citenamefont {{Ashdown}}, \citenamefont {{Aumont}},
  \citenamefont {{Baccigalupi}}, \citenamefont {{Ballardini}}, \citenamefont
  {{Banday}},\ and\ \citenamefont {et~al.}}]{2015Planck_fNL}%
  \BibitemOpen
  \bibfield  {author} {\bibinfo {author} {\bibnamefont {{Planck
  Collaboration}}}, \bibinfo {author} {\bibfnamefont {P.~A.~R.}\ \bibnamefont
  {{Ade}}}, \bibinfo {author} {\bibfnamefont {N.}~\bibnamefont {{Aghanim}}},
  \bibinfo {author} {\bibfnamefont {M.}~\bibnamefont {{Arnaud}}}, \bibinfo
  {author} {\bibfnamefont {F.}~\bibnamefont {{Arroja}}}, \bibinfo {author}
  {\bibfnamefont {M.}~\bibnamefont {{Ashdown}}}, \bibinfo {author}
  {\bibfnamefont {J.}~\bibnamefont {{Aumont}}}, \bibinfo {author}
  {\bibfnamefont {C.}~\bibnamefont {{Baccigalupi}}}, \bibinfo {author}
  {\bibfnamefont {M.}~\bibnamefont {{Ballardini}}}, \bibinfo {author}
  {\bibfnamefont {A.~J.}\ \bibnamefont {{Banday}}}, \ and\ \bibinfo {author}
  {\bibnamefont {et~al.}},\ }\href@noop {} {\bibfield  {journal} {\bibinfo
  {journal} {ArXiv e-prints}\ } (\bibinfo {year} {2015})},\ \Eprint
  {http://arxiv.org/abs/1502.01592} {arXiv:1502.01592} \BibitemShut {NoStop}%
\bibitem [{\citenamefont {{Delubac}}\ \emph {et~al.}(2015)\citenamefont
  {{Delubac}}, \citenamefont {{Bautista}}, \citenamefont {{Busca}},
  \citenamefont {{Rich}}, \citenamefont {{Kirkby}}, \citenamefont {{Bailey}},
  \citenamefont {{Font-Ribera}}, \citenamefont {{Slosar}}, \citenamefont
  {{Lee}}, \citenamefont {{Pieri}}, \citenamefont {{Hamilton}}, \citenamefont
  {{Aubourg}}, \citenamefont {{Blomqvist}}, \citenamefont {{Bovy}},
  \citenamefont {{Brinkmann}}, \citenamefont {{Carithers}}, \citenamefont
  {{Dawson}}, \citenamefont {{Eisenstein}}, \citenamefont {{Gontcho}},
  \citenamefont {{Kneib}}, \citenamefont {{Le Goff}}, \citenamefont
  {{Margala}}, \citenamefont {{Miralda-Escud{\'e}}}, \citenamefont {{Myers}},
  \citenamefont {{Nichol}}, \citenamefont {{Noterdaeme}}, \citenamefont
  {{O'Connell}}, \citenamefont {{Olmstead}}, \citenamefont
  {{Palanque-Delabrouille}}, \citenamefont {{P{\^a}ris}}, \citenamefont
  {{Petitjean}}, \citenamefont {{Ross}}, \citenamefont {{Rossi}}, \citenamefont
  {{Schlegel}}, \citenamefont {{Schneider}}, \citenamefont {{Weinberg}},
  \citenamefont {{Y{\`e}che}},\ and\ \citenamefont {{York}}}]{2014BOSSLyAlpha}%
  \BibitemOpen
  \bibfield  {author} {\bibinfo {author} {\bibfnamefont {T.}~\bibnamefont
  {{Delubac}}}, \bibinfo {author} {\bibfnamefont {J.~E.}\ \bibnamefont
  {{Bautista}}}, \bibinfo {author} {\bibfnamefont {N.~G.}\ \bibnamefont
  {{Busca}}}, \bibinfo {author} {\bibfnamefont {J.}~\bibnamefont {{Rich}}},
  \bibinfo {author} {\bibfnamefont {D.}~\bibnamefont {{Kirkby}}}, \bibinfo
  {author} {\bibfnamefont {S.}~\bibnamefont {{Bailey}}}, \bibinfo {author}
  {\bibfnamefont {A.}~\bibnamefont {{Font-Ribera}}}, \bibinfo {author}
  {\bibfnamefont {A.}~\bibnamefont {{Slosar}}}, \bibinfo {author}
  {\bibfnamefont {K.-G.}\ \bibnamefont {{Lee}}}, \bibinfo {author}
  {\bibfnamefont {M.~M.}\ \bibnamefont {{Pieri}}}, \bibinfo {author}
  {\bibfnamefont {J.-C.}\ \bibnamefont {{Hamilton}}}, \bibinfo {author}
  {\bibfnamefont {{\'E}.}~\bibnamefont {{Aubourg}}}, \bibinfo {author}
  {\bibfnamefont {M.}~\bibnamefont {{Blomqvist}}}, \bibinfo {author}
  {\bibfnamefont {J.}~\bibnamefont {{Bovy}}}, \bibinfo {author} {\bibfnamefont
  {J.}~\bibnamefont {{Brinkmann}}}, \bibinfo {author} {\bibfnamefont
  {W.}~\bibnamefont {{Carithers}}}, \bibinfo {author} {\bibfnamefont {K.~S.}\
  \bibnamefont {{Dawson}}}, \bibinfo {author} {\bibfnamefont {D.~J.}\
  \bibnamefont {{Eisenstein}}}, \bibinfo {author} {\bibfnamefont {S.~G.~A.}\
  \bibnamefont {{Gontcho}}}, \bibinfo {author} {\bibfnamefont {J.-P.}\
  \bibnamefont {{Kneib}}}, \bibinfo {author} {\bibfnamefont {J.-M.}\
  \bibnamefont {{Le Goff}}}, \bibinfo {author} {\bibfnamefont {D.}~\bibnamefont
  {{Margala}}}, \bibinfo {author} {\bibfnamefont {J.}~\bibnamefont
  {{Miralda-Escud{\'e}}}}, \bibinfo {author} {\bibfnamefont {A.~D.}\
  \bibnamefont {{Myers}}}, \bibinfo {author} {\bibfnamefont {R.~C.}\
  \bibnamefont {{Nichol}}}, \bibinfo {author} {\bibfnamefont {P.}~\bibnamefont
  {{Noterdaeme}}}, \bibinfo {author} {\bibfnamefont {R.}~\bibnamefont
  {{O'Connell}}}, \bibinfo {author} {\bibfnamefont {M.~D.}\ \bibnamefont
  {{Olmstead}}}, \bibinfo {author} {\bibfnamefont {N.}~\bibnamefont
  {{Palanque-Delabrouille}}}, \bibinfo {author} {\bibfnamefont
  {I.}~\bibnamefont {{P{\^a}ris}}}, \bibinfo {author} {\bibfnamefont
  {P.}~\bibnamefont {{Petitjean}}}, \bibinfo {author} {\bibfnamefont {N.~P.}\
  \bibnamefont {{Ross}}}, \bibinfo {author} {\bibfnamefont {G.}~\bibnamefont
  {{Rossi}}}, \bibinfo {author} {\bibfnamefont {D.~J.}\ \bibnamefont
  {{Schlegel}}}, \bibinfo {author} {\bibfnamefont {D.~P.}\ \bibnamefont
  {{Schneider}}}, \bibinfo {author} {\bibfnamefont {D.~H.}\ \bibnamefont
  {{Weinberg}}}, \bibinfo {author} {\bibfnamefont {C.}~\bibnamefont
  {{Y{\`e}che}}}, \ and\ \bibinfo {author} {\bibfnamefont {D.~G.}\ \bibnamefont
  {{York}}},\ }\href {\doibase 10.1051/0004-6361/201423969} {\bibfield
  {journal} {\bibinfo  {journal} {\aap}\ }\textbf {\bibinfo {volume} {574}},\
  \bibinfo {eid} {A59} (\bibinfo {year} {2015})},\ \Eprint
  {http://arxiv.org/abs/1404.1801} {arXiv:1404.1801} \BibitemShut {NoStop}%
\bibitem [{\citenamefont {{Beutler}}\ \emph {et~al.}(2012)\citenamefont
  {{Beutler}}, \citenamefont {{Blake}}, \citenamefont {{Colless}},
  \citenamefont {{Jones}}, \citenamefont {{Staveley-Smith}}, \citenamefont
  {{Poole}}, \citenamefont {{Campbell}}, \citenamefont {{Parker}},
  \citenamefont {{Saunders}},\ and\ \citenamefont
  {{Watson}}}]{Beutler2012MNRAS}%
  \BibitemOpen
  \bibfield  {author} {\bibinfo {author} {\bibfnamefont {F.}~\bibnamefont
  {{Beutler}}}, \bibinfo {author} {\bibfnamefont {C.}~\bibnamefont {{Blake}}},
  \bibinfo {author} {\bibfnamefont {M.}~\bibnamefont {{Colless}}}, \bibinfo
  {author} {\bibfnamefont {D.~H.}\ \bibnamefont {{Jones}}}, \bibinfo {author}
  {\bibfnamefont {L.}~\bibnamefont {{Staveley-Smith}}}, \bibinfo {author}
  {\bibfnamefont {G.~B.}\ \bibnamefont {{Poole}}}, \bibinfo {author}
  {\bibfnamefont {L.}~\bibnamefont {{Campbell}}}, \bibinfo {author}
  {\bibfnamefont {Q.}~\bibnamefont {{Parker}}}, \bibinfo {author}
  {\bibfnamefont {W.}~\bibnamefont {{Saunders}}}, \ and\ \bibinfo {author}
  {\bibfnamefont {F.}~\bibnamefont {{Watson}}},\ }\href {\doibase
  10.1111/j.1365-2966.2012.21136.x} {\bibfield  {journal} {\bibinfo  {journal}
  {\mnras}\ }\textbf {\bibinfo {volume} {423}},\ \bibinfo {pages} {3430}
  (\bibinfo {year} {2012})},\ \Eprint {http://arxiv.org/abs/1204.4725}
  {arXiv:1204.4725 [astro-ph.CO]} \BibitemShut {NoStop}%
\bibitem [{\citenamefont {{Blake}}\ \emph {et~al.}(2012)\citenamefont
  {{Blake}}, \citenamefont {{Brough}}, \citenamefont {{Colless}}, \citenamefont
  {{Contreras}}, \citenamefont {{Couch}}, \citenamefont {{Croom}},
  \citenamefont {{Croton}}, \citenamefont {{Davis}}, \citenamefont
  {{Drinkwater}}, \citenamefont {{Forster}}, \citenamefont {{Gilbank}},
  \citenamefont {{Gladders}}, \citenamefont {{Glazebrook}}, \citenamefont
  {{Jelliffe}}, \citenamefont {{Jurek}}, \citenamefont {{Li}}, \citenamefont
  {{Madore}}, \citenamefont {{Martin}}, \citenamefont {{Pimbblet}},
  \citenamefont {{Poole}}, \citenamefont {{Pracy}}, \citenamefont {{Sharp}},
  \citenamefont {{Wisnioski}}, \citenamefont {{Woods}}, \citenamefont
  {{Wyder}},\ and\ \citenamefont {{Yee}}}]{Blake2012MNRAS}%
  \BibitemOpen
  \bibfield  {author} {\bibinfo {author} {\bibfnamefont {C.}~\bibnamefont
  {{Blake}}}, \bibinfo {author} {\bibfnamefont {S.}~\bibnamefont {{Brough}}},
  \bibinfo {author} {\bibfnamefont {M.}~\bibnamefont {{Colless}}}, \bibinfo
  {author} {\bibfnamefont {C.}~\bibnamefont {{Contreras}}}, \bibinfo {author}
  {\bibfnamefont {W.}~\bibnamefont {{Couch}}}, \bibinfo {author} {\bibfnamefont
  {S.}~\bibnamefont {{Croom}}}, \bibinfo {author} {\bibfnamefont
  {D.}~\bibnamefont {{Croton}}}, \bibinfo {author} {\bibfnamefont {T.~M.}\
  \bibnamefont {{Davis}}}, \bibinfo {author} {\bibfnamefont {M.~J.}\
  \bibnamefont {{Drinkwater}}}, \bibinfo {author} {\bibfnamefont
  {K.}~\bibnamefont {{Forster}}}, \bibinfo {author} {\bibfnamefont
  {D.}~\bibnamefont {{Gilbank}}}, \bibinfo {author} {\bibfnamefont
  {M.}~\bibnamefont {{Gladders}}}, \bibinfo {author} {\bibfnamefont
  {K.}~\bibnamefont {{Glazebrook}}}, \bibinfo {author} {\bibfnamefont
  {B.}~\bibnamefont {{Jelliffe}}}, \bibinfo {author} {\bibfnamefont {R.~J.}\
  \bibnamefont {{Jurek}}}, \bibinfo {author} {\bibfnamefont {I.-h.}\
  \bibnamefont {{Li}}}, \bibinfo {author} {\bibfnamefont {B.}~\bibnamefont
  {{Madore}}}, \bibinfo {author} {\bibfnamefont {D.~C.}\ \bibnamefont
  {{Martin}}}, \bibinfo {author} {\bibfnamefont {K.}~\bibnamefont
  {{Pimbblet}}}, \bibinfo {author} {\bibfnamefont {G.~B.}\ \bibnamefont
  {{Poole}}}, \bibinfo {author} {\bibfnamefont {M.}~\bibnamefont {{Pracy}}},
  \bibinfo {author} {\bibfnamefont {R.}~\bibnamefont {{Sharp}}}, \bibinfo
  {author} {\bibfnamefont {E.}~\bibnamefont {{Wisnioski}}}, \bibinfo {author}
  {\bibfnamefont {D.}~\bibnamefont {{Woods}}}, \bibinfo {author} {\bibfnamefont
  {T.~K.}\ \bibnamefont {{Wyder}}}, \ and\ \bibinfo {author} {\bibfnamefont
  {H.~K.~C.}\ \bibnamefont {{Yee}}},\ }\href {\doibase
  10.1111/j.1365-2966.2012.21473.x} {\bibfield  {journal} {\bibinfo  {journal}
  {\mnras}\ }\textbf {\bibinfo {volume} {425}},\ \bibinfo {pages} {405}
  (\bibinfo {year} {2012})},\ \Eprint {http://arxiv.org/abs/1204.3674}
  {arXiv:1204.3674 [astro-ph.CO]} \BibitemShut {NoStop}%
\bibitem [{\citenamefont {{Chuang}}\ \emph {et~al.}(2013)\citenamefont
  {{Chuang}}, \citenamefont {{Prada}}, \citenamefont {{Beutler}}, \citenamefont
  {{Eisenstein}}, \citenamefont {{Escoffier}}, \citenamefont {{Ho}},
  \citenamefont {{Kneib}}, \citenamefont {{Manera}}, \citenamefont {{Nuza}},
  \citenamefont {{Schlegel}}, \citenamefont {{Schneider}}, \citenamefont
  {{Weaver}}, \citenamefont {{Brownstein}}, \citenamefont {{Dawson}},
  \citenamefont {{Maraston}},\ and\ \citenamefont
  {{Thomas}}}]{2013arXiv1312.4889C}%
  \BibitemOpen
  \bibfield  {author} {\bibinfo {author} {\bibfnamefont {C.-H.}\ \bibnamefont
  {{Chuang}}}, \bibinfo {author} {\bibfnamefont {F.}~\bibnamefont {{Prada}}},
  \bibinfo {author} {\bibfnamefont {F.}~\bibnamefont {{Beutler}}}, \bibinfo
  {author} {\bibfnamefont {D.~J.}\ \bibnamefont {{Eisenstein}}}, \bibinfo
  {author} {\bibfnamefont {S.}~\bibnamefont {{Escoffier}}}, \bibinfo {author}
  {\bibfnamefont {S.}~\bibnamefont {{Ho}}}, \bibinfo {author} {\bibfnamefont
  {J.-P.}\ \bibnamefont {{Kneib}}}, \bibinfo {author} {\bibfnamefont
  {M.}~\bibnamefont {{Manera}}}, \bibinfo {author} {\bibfnamefont {S.~E.}\
  \bibnamefont {{Nuza}}}, \bibinfo {author} {\bibfnamefont {D.~J.}\
  \bibnamefont {{Schlegel}}}, \bibinfo {author} {\bibfnamefont {D.~P.}\
  \bibnamefont {{Schneider}}}, \bibinfo {author} {\bibfnamefont {B.~A.}\
  \bibnamefont {{Weaver}}}, \bibinfo {author} {\bibfnamefont {J.~R.}\
  \bibnamefont {{Brownstein}}}, \bibinfo {author} {\bibfnamefont {K.~S.}\
  \bibnamefont {{Dawson}}}, \bibinfo {author} {\bibfnamefont {C.}~\bibnamefont
  {{Maraston}}}, \ and\ \bibinfo {author} {\bibfnamefont {D.}~\bibnamefont
  {{Thomas}}},\ }\href@noop {} {\bibfield  {journal} {\bibinfo  {journal}
  {ArXiv e-prints}\ } (\bibinfo {year} {2013})},\ \Eprint
  {http://arxiv.org/abs/1312.4889} {arXiv:1312.4889 [astro-ph.CO]} \BibitemShut
  {NoStop}%
\bibitem [{\citenamefont {{Samushia}}\ \emph {et~al.}(2014)\citenamefont
  {{Samushia}}, \citenamefont {{Reid}}, \citenamefont {{White}}, \citenamefont
  {{Percival}}, \citenamefont {{Cuesta}}, \citenamefont {{Zhao}}, \citenamefont
  {{Ross}}, \citenamefont {{Manera}}, \citenamefont {{Aubourg}}, \citenamefont
  {{Beutler}}, \citenamefont {{Brinkmann}}, \citenamefont {{Brownstein}},
  \citenamefont {{Dawson}}, \citenamefont {{Eisenstein}}, \citenamefont {{Ho}},
  \citenamefont {{Honscheid}}, \citenamefont {{Maraston}}, \citenamefont
  {{Montesano}}, \citenamefont {{Nichol}}, \citenamefont {{Roe}}, \citenamefont
  {{Ross}}, \citenamefont {{S{\'a}nchez}}, \citenamefont {{Schlegel}},
  \citenamefont {{Schneider}}, \citenamefont {{Streblyanska}}, \citenamefont
  {{Thomas}}, \citenamefont {{Tinker}}, \citenamefont {{Wake}}, \citenamefont
  {{Weaver}},\ and\ \citenamefont {{Zehavi}}}]{2014MNRAS.439.3504S}%
  \BibitemOpen
  \bibfield  {author} {\bibinfo {author} {\bibfnamefont {L.}~\bibnamefont
  {{Samushia}}}, \bibinfo {author} {\bibfnamefont {B.~A.}\ \bibnamefont
  {{Reid}}}, \bibinfo {author} {\bibfnamefont {M.}~\bibnamefont {{White}}},
  \bibinfo {author} {\bibfnamefont {W.~J.}\ \bibnamefont {{Percival}}},
  \bibinfo {author} {\bibfnamefont {A.~J.}\ \bibnamefont {{Cuesta}}}, \bibinfo
  {author} {\bibfnamefont {G.-B.}\ \bibnamefont {{Zhao}}}, \bibinfo {author}
  {\bibfnamefont {A.~J.}\ \bibnamefont {{Ross}}}, \bibinfo {author}
  {\bibfnamefont {M.}~\bibnamefont {{Manera}}}, \bibinfo {author}
  {\bibfnamefont {{\'E}.}~\bibnamefont {{Aubourg}}}, \bibinfo {author}
  {\bibfnamefont {F.}~\bibnamefont {{Beutler}}}, \bibinfo {author}
  {\bibfnamefont {J.}~\bibnamefont {{Brinkmann}}}, \bibinfo {author}
  {\bibfnamefont {J.~R.}\ \bibnamefont {{Brownstein}}}, \bibinfo {author}
  {\bibfnamefont {K.~S.}\ \bibnamefont {{Dawson}}}, \bibinfo {author}
  {\bibfnamefont {D.~J.}\ \bibnamefont {{Eisenstein}}}, \bibinfo {author}
  {\bibfnamefont {S.}~\bibnamefont {{Ho}}}, \bibinfo {author} {\bibfnamefont
  {K.}~\bibnamefont {{Honscheid}}}, \bibinfo {author} {\bibfnamefont
  {C.}~\bibnamefont {{Maraston}}}, \bibinfo {author} {\bibfnamefont
  {F.}~\bibnamefont {{Montesano}}}, \bibinfo {author} {\bibfnamefont {R.~C.}\
  \bibnamefont {{Nichol}}}, \bibinfo {author} {\bibfnamefont {N.~A.}\
  \bibnamefont {{Roe}}}, \bibinfo {author} {\bibfnamefont {N.~P.}\ \bibnamefont
  {{Ross}}}, \bibinfo {author} {\bibfnamefont {A.~G.}\ \bibnamefont
  {{S{\'a}nchez}}}, \bibinfo {author} {\bibfnamefont {D.~J.}\ \bibnamefont
  {{Schlegel}}}, \bibinfo {author} {\bibfnamefont {D.~P.}\ \bibnamefont
  {{Schneider}}}, \bibinfo {author} {\bibfnamefont {A.}~\bibnamefont
  {{Streblyanska}}}, \bibinfo {author} {\bibfnamefont {D.}~\bibnamefont
  {{Thomas}}}, \bibinfo {author} {\bibfnamefont {J.~L.}\ \bibnamefont
  {{Tinker}}}, \bibinfo {author} {\bibfnamefont {D.~A.}\ \bibnamefont
  {{Wake}}}, \bibinfo {author} {\bibfnamefont {B.~A.}\ \bibnamefont
  {{Weaver}}}, \ and\ \bibinfo {author} {\bibfnamefont {I.}~\bibnamefont
  {{Zehavi}}},\ }\href {\doibase 10.1093/mnras/stu197} {\bibfield  {journal}
  {\bibinfo  {journal} {\mnras}\ }\textbf {\bibinfo {volume} {439}},\ \bibinfo
  {pages} {3504} (\bibinfo {year} {2014})},\ \Eprint
  {http://arxiv.org/abs/1312.4899} {arXiv:1312.4899} \BibitemShut {NoStop}%
\bibitem [{\citenamefont {{Beutler}}\ \emph {et~al.}(2014)\citenamefont
  {{Beutler}}, \citenamefont {{Saito}}, \citenamefont {{Seo}}, \citenamefont
  {{Brinkmann}}, \citenamefont {{Dawson}}, \citenamefont {{Eisenstein}},
  \citenamefont {{Font-Ribera}}, \citenamefont {{Ho}}, \citenamefont
  {{McBride}}, \citenamefont {{Montesano}}, \citenamefont {{Percival}},
  \citenamefont {{Ross}}, \citenamefont {{Ross}}, \citenamefont {{Samushia}},
  \citenamefont {{Schlegel}}, \citenamefont {{S{\'a}nchez}}, \citenamefont
  {{Tinker}},\ and\ \citenamefont {{Weaver}}}]{2014MNRAS.443.1065B}%
  \BibitemOpen
  \bibfield  {author} {\bibinfo {author} {\bibfnamefont {F.}~\bibnamefont
  {{Beutler}}}, \bibinfo {author} {\bibfnamefont {S.}~\bibnamefont {{Saito}}},
  \bibinfo {author} {\bibfnamefont {H.-J.}\ \bibnamefont {{Seo}}}, \bibinfo
  {author} {\bibfnamefont {J.}~\bibnamefont {{Brinkmann}}}, \bibinfo {author}
  {\bibfnamefont {K.~S.}\ \bibnamefont {{Dawson}}}, \bibinfo {author}
  {\bibfnamefont {D.~J.}\ \bibnamefont {{Eisenstein}}}, \bibinfo {author}
  {\bibfnamefont {A.}~\bibnamefont {{Font-Ribera}}}, \bibinfo {author}
  {\bibfnamefont {S.}~\bibnamefont {{Ho}}}, \bibinfo {author} {\bibfnamefont
  {C.~K.}\ \bibnamefont {{McBride}}}, \bibinfo {author} {\bibfnamefont
  {F.}~\bibnamefont {{Montesano}}}, \bibinfo {author} {\bibfnamefont {W.~J.}\
  \bibnamefont {{Percival}}}, \bibinfo {author} {\bibfnamefont {A.~J.}\
  \bibnamefont {{Ross}}}, \bibinfo {author} {\bibfnamefont {N.~P.}\
  \bibnamefont {{Ross}}}, \bibinfo {author} {\bibfnamefont {L.}~\bibnamefont
  {{Samushia}}}, \bibinfo {author} {\bibfnamefont {D.~J.}\ \bibnamefont
  {{Schlegel}}}, \bibinfo {author} {\bibfnamefont {A.~G.}\ \bibnamefont
  {{S{\'a}nchez}}}, \bibinfo {author} {\bibfnamefont {J.~L.}\ \bibnamefont
  {{Tinker}}}, \ and\ \bibinfo {author} {\bibfnamefont {B.~A.}\ \bibnamefont
  {{Weaver}}},\ }\href {\doibase 10.1093/mnras/stu1051} {\bibfield  {journal}
  {\bibinfo  {journal} {\mnras}\ }\textbf {\bibinfo {volume} {443}},\ \bibinfo
  {pages} {1065} (\bibinfo {year} {2014})},\ \Eprint
  {http://arxiv.org/abs/1312.4611} {arXiv:1312.4611} \BibitemShut {NoStop}%
\bibitem [{\citenamefont {{Guzzo}}\ \emph {et~al.}(2008)\citenamefont
  {{Guzzo}}, \citenamefont {{Pierleoni}}, \citenamefont {{Meneux}},
  \citenamefont {{Branchini}}, \citenamefont {{Le F{\`e}vre}}, \citenamefont
  {{Marinoni}}, \citenamefont {{Garilli}}, \citenamefont {{Blaizot}},
  \citenamefont {{De Lucia}}, \citenamefont {{Pollo}}, \citenamefont
  {{McCracken}}, \citenamefont {{Bottini}}, \citenamefont {{Le Brun}},
  \citenamefont {{Maccagni}}, \citenamefont {{Picat}}, \citenamefont
  {{Scaramella}}, \citenamefont {{Scodeggio}}, \citenamefont {{Tresse}},
  \citenamefont {{Vettolani}}, \citenamefont {{Zanichelli}}, \citenamefont
  {{Adami}}, \citenamefont {{Arnouts}}, \citenamefont {{Bardelli}},
  \citenamefont {{Bolzonella}}, \citenamefont {{Bongiorno}}, \citenamefont
  {{Cappi}}, \citenamefont {{Charlot}}, \citenamefont {{Ciliegi}},
  \citenamefont {{Contini}}, \citenamefont {{Cucciati}}, \citenamefont {{de la
  Torre}}, \citenamefont {{Dolag}}, \citenamefont {{Foucaud}}, \citenamefont
  {{Franzetti}}, \citenamefont {{Gavignaud}}, \citenamefont {{Ilbert}},
  \citenamefont {{Iovino}}, \citenamefont {{Lamareille}}, \citenamefont
  {{Marano}}, \citenamefont {{Mazure}}, \citenamefont {{Memeo}}, \citenamefont
  {{Merighi}}, \citenamefont {{Moscardini}}, \citenamefont {{Paltani}},
  \citenamefont {{Pell{\`o}}}, \citenamefont {{Perez-Montero}}, \citenamefont
  {{Pozzetti}}, \citenamefont {{Radovich}}, \citenamefont {{Vergani}},
  \citenamefont {{Zamorani}},\ and\ \citenamefont
  {{Zucca}}}]{2008Natur.451..541G}%
  \BibitemOpen
  \bibfield  {author} {\bibinfo {author} {\bibfnamefont {L.}~\bibnamefont
  {{Guzzo}}}, \bibinfo {author} {\bibfnamefont {M.}~\bibnamefont
  {{Pierleoni}}}, \bibinfo {author} {\bibfnamefont {B.}~\bibnamefont
  {{Meneux}}}, \bibinfo {author} {\bibfnamefont {E.}~\bibnamefont
  {{Branchini}}}, \bibinfo {author} {\bibfnamefont {O.}~\bibnamefont {{Le
  F{\`e}vre}}}, \bibinfo {author} {\bibfnamefont {C.}~\bibnamefont
  {{Marinoni}}}, \bibinfo {author} {\bibfnamefont {B.}~\bibnamefont
  {{Garilli}}}, \bibinfo {author} {\bibfnamefont {J.}~\bibnamefont
  {{Blaizot}}}, \bibinfo {author} {\bibfnamefont {G.}~\bibnamefont {{De
  Lucia}}}, \bibinfo {author} {\bibfnamefont {A.}~\bibnamefont {{Pollo}}},
  \bibinfo {author} {\bibfnamefont {H.~J.}\ \bibnamefont {{McCracken}}},
  \bibinfo {author} {\bibfnamefont {D.}~\bibnamefont {{Bottini}}}, \bibinfo
  {author} {\bibfnamefont {V.}~\bibnamefont {{Le Brun}}}, \bibinfo {author}
  {\bibfnamefont {D.}~\bibnamefont {{Maccagni}}}, \bibinfo {author}
  {\bibfnamefont {J.~P.}\ \bibnamefont {{Picat}}}, \bibinfo {author}
  {\bibfnamefont {R.}~\bibnamefont {{Scaramella}}}, \bibinfo {author}
  {\bibfnamefont {M.}~\bibnamefont {{Scodeggio}}}, \bibinfo {author}
  {\bibfnamefont {L.}~\bibnamefont {{Tresse}}}, \bibinfo {author}
  {\bibfnamefont {G.}~\bibnamefont {{Vettolani}}}, \bibinfo {author}
  {\bibfnamefont {A.}~\bibnamefont {{Zanichelli}}}, \bibinfo {author}
  {\bibfnamefont {C.}~\bibnamefont {{Adami}}}, \bibinfo {author} {\bibfnamefont
  {S.}~\bibnamefont {{Arnouts}}}, \bibinfo {author} {\bibfnamefont
  {S.}~\bibnamefont {{Bardelli}}}, \bibinfo {author} {\bibfnamefont
  {M.}~\bibnamefont {{Bolzonella}}}, \bibinfo {author} {\bibfnamefont
  {A.}~\bibnamefont {{Bongiorno}}}, \bibinfo {author} {\bibfnamefont
  {A.}~\bibnamefont {{Cappi}}}, \bibinfo {author} {\bibfnamefont
  {S.}~\bibnamefont {{Charlot}}}, \bibinfo {author} {\bibfnamefont
  {P.}~\bibnamefont {{Ciliegi}}}, \bibinfo {author} {\bibfnamefont
  {T.}~\bibnamefont {{Contini}}}, \bibinfo {author} {\bibfnamefont
  {O.}~\bibnamefont {{Cucciati}}}, \bibinfo {author} {\bibfnamefont
  {S.}~\bibnamefont {{de la Torre}}}, \bibinfo {author} {\bibfnamefont
  {K.}~\bibnamefont {{Dolag}}}, \bibinfo {author} {\bibfnamefont
  {S.}~\bibnamefont {{Foucaud}}}, \bibinfo {author} {\bibfnamefont
  {P.}~\bibnamefont {{Franzetti}}}, \bibinfo {author} {\bibfnamefont
  {I.}~\bibnamefont {{Gavignaud}}}, \bibinfo {author} {\bibfnamefont
  {O.}~\bibnamefont {{Ilbert}}}, \bibinfo {author} {\bibfnamefont
  {A.}~\bibnamefont {{Iovino}}}, \bibinfo {author} {\bibfnamefont
  {F.}~\bibnamefont {{Lamareille}}}, \bibinfo {author} {\bibfnamefont
  {B.}~\bibnamefont {{Marano}}}, \bibinfo {author} {\bibfnamefont
  {A.}~\bibnamefont {{Mazure}}}, \bibinfo {author} {\bibfnamefont
  {P.}~\bibnamefont {{Memeo}}}, \bibinfo {author} {\bibfnamefont
  {R.}~\bibnamefont {{Merighi}}}, \bibinfo {author} {\bibfnamefont
  {L.}~\bibnamefont {{Moscardini}}}, \bibinfo {author} {\bibfnamefont
  {S.}~\bibnamefont {{Paltani}}}, \bibinfo {author} {\bibfnamefont
  {R.}~\bibnamefont {{Pell{\`o}}}}, \bibinfo {author} {\bibfnamefont
  {E.}~\bibnamefont {{Perez-Montero}}}, \bibinfo {author} {\bibfnamefont
  {L.}~\bibnamefont {{Pozzetti}}}, \bibinfo {author} {\bibfnamefont
  {M.}~\bibnamefont {{Radovich}}}, \bibinfo {author} {\bibfnamefont
  {D.}~\bibnamefont {{Vergani}}}, \bibinfo {author} {\bibfnamefont
  {G.}~\bibnamefont {{Zamorani}}}, \ and\ \bibinfo {author} {\bibfnamefont
  {E.}~\bibnamefont {{Zucca}}},\ }\href {\doibase 10.1038/nature06555}
  {\bibfield  {journal} {\bibinfo  {journal} {\nat}\ }\textbf {\bibinfo
  {volume} {451}},\ \bibinfo {pages} {541} (\bibinfo {year} {2008})},\ \Eprint
  {http://arxiv.org/abs/0802.1944} {arXiv:0802.1944} \BibitemShut {NoStop}%
\bibitem [{\citenamefont {{Howlett}}\ \emph {et~al.}(2015)\citenamefont
  {{Howlett}}, \citenamefont {{Ross}}, \citenamefont {{Samushia}},
  \citenamefont {{Percival}},\ and\ \citenamefont
  {{Manera}}}]{2014arXiv1409.3238H}%
  \BibitemOpen
  \bibfield  {author} {\bibinfo {author} {\bibfnamefont {C.}~\bibnamefont
  {{Howlett}}}, \bibinfo {author} {\bibfnamefont {A.~J.}\ \bibnamefont
  {{Ross}}}, \bibinfo {author} {\bibfnamefont {L.}~\bibnamefont {{Samushia}}},
  \bibinfo {author} {\bibfnamefont {W.~J.}\ \bibnamefont {{Percival}}}, \ and\
  \bibinfo {author} {\bibfnamefont {M.}~\bibnamefont {{Manera}}},\ }\href
  {\doibase 10.1093/mnras/stu2693} {\bibfield  {journal} {\bibinfo  {journal}
  {\mnras}\ }\textbf {\bibinfo {volume} {449}},\ \bibinfo {pages} {848}
  (\bibinfo {year} {2015})},\ \Eprint {http://arxiv.org/abs/1409.3238}
  {arXiv:1409.3238} \BibitemShut {NoStop}%
\bibitem [{\citenamefont {{Samushia}}\ \emph {et~al.}(2012)\citenamefont
  {{Samushia}}, \citenamefont {{Percival}},\ and\ \citenamefont
  {{Raccanelli}}}]{2012MNRAS.420.2102S}%
  \BibitemOpen
  \bibfield  {author} {\bibinfo {author} {\bibfnamefont {L.}~\bibnamefont
  {{Samushia}}}, \bibinfo {author} {\bibfnamefont {W.~J.}\ \bibnamefont
  {{Percival}}}, \ and\ \bibinfo {author} {\bibfnamefont {A.}~\bibnamefont
  {{Raccanelli}}},\ }\href {\doibase 10.1111/j.1365-2966.2011.20169.x}
  {\bibfield  {journal} {\bibinfo  {journal} {\mnras}\ }\textbf {\bibinfo
  {volume} {420}},\ \bibinfo {pages} {2102} (\bibinfo {year} {2012})},\ \Eprint
  {http://arxiv.org/abs/1102.1014} {arXiv:1102.1014 [astro-ph.CO]} \BibitemShut
  {NoStop}%
\bibitem [{\citenamefont {{de la Torre}}\ \emph {et~al.}(2013)\citenamefont
  {{de la Torre}}, \citenamefont {{Guzzo}}, \citenamefont {{Peacock}},
  \citenamefont {{Branchini}}, \citenamefont {{Iovino}}, \citenamefont
  {{Granett}}, \citenamefont {{Abbas}}, \citenamefont {{Adami}}, \citenamefont
  {{Arnouts}}, \citenamefont {{Bel}}, \citenamefont {{Bolzonella}},
  \citenamefont {{Bottini}}, \citenamefont {{Cappi}}, \citenamefont {{Coupon}},
  \citenamefont {{Cucciati}}, \citenamefont {{Davidzon}}, \citenamefont {{De
  Lucia}}, \citenamefont {{Fritz}}, \citenamefont {{Franzetti}}, \citenamefont
  {{Fumana}}, \citenamefont {{Garilli}}, \citenamefont {{Ilbert}},
  \citenamefont {{Krywult}}, \citenamefont {{Le Brun}}, \citenamefont {{Le
  F{\`e}vre}}, \citenamefont {{Maccagni}}, \citenamefont {{Ma{\l}ek}},
  \citenamefont {{Marulli}}, \citenamefont {{McCracken}}, \citenamefont
  {{Moscardini}}, \citenamefont {{Paioro}}, \citenamefont {{Percival}},
  \citenamefont {{Polletta}}, \citenamefont {{Pollo}}, \citenamefont
  {{Schlagenhaufer}}, \citenamefont {{Scodeggio}}, \citenamefont {{Tasca}},
  \citenamefont {{Tojeiro}}, \citenamefont {{Vergani}}, \citenamefont
  {{Zanichelli}}, \citenamefont {{Burden}}, \citenamefont {{Di Porto}},
  \citenamefont {{Marchetti}}, \citenamefont {{Marinoni}}, \citenamefont
  {{Mellier}}, \citenamefont {{Monaco}}, \citenamefont {{Nichol}},
  \citenamefont {{Phleps}}, \citenamefont {{Wolk}},\ and\ \citenamefont
  {{Zamorani}}}]{2013A&A...557A..54D}%
  \BibitemOpen
  \bibfield  {author} {\bibinfo {author} {\bibfnamefont {S.}~\bibnamefont {{de
  la Torre}}}, \bibinfo {author} {\bibfnamefont {L.}~\bibnamefont {{Guzzo}}},
  \bibinfo {author} {\bibfnamefont {J.~A.}\ \bibnamefont {{Peacock}}}, \bibinfo
  {author} {\bibfnamefont {E.}~\bibnamefont {{Branchini}}}, \bibinfo {author}
  {\bibfnamefont {A.}~\bibnamefont {{Iovino}}}, \bibinfo {author}
  {\bibfnamefont {B.~R.}\ \bibnamefont {{Granett}}}, \bibinfo {author}
  {\bibfnamefont {U.}~\bibnamefont {{Abbas}}}, \bibinfo {author} {\bibfnamefont
  {C.}~\bibnamefont {{Adami}}}, \bibinfo {author} {\bibfnamefont
  {S.}~\bibnamefont {{Arnouts}}}, \bibinfo {author} {\bibfnamefont
  {J.}~\bibnamefont {{Bel}}}, \bibinfo {author} {\bibfnamefont
  {M.}~\bibnamefont {{Bolzonella}}}, \bibinfo {author} {\bibfnamefont
  {D.}~\bibnamefont {{Bottini}}}, \bibinfo {author} {\bibfnamefont
  {A.}~\bibnamefont {{Cappi}}}, \bibinfo {author} {\bibfnamefont
  {J.}~\bibnamefont {{Coupon}}}, \bibinfo {author} {\bibfnamefont
  {O.}~\bibnamefont {{Cucciati}}}, \bibinfo {author} {\bibfnamefont
  {I.}~\bibnamefont {{Davidzon}}}, \bibinfo {author} {\bibfnamefont
  {G.}~\bibnamefont {{De Lucia}}}, \bibinfo {author} {\bibfnamefont
  {A.}~\bibnamefont {{Fritz}}}, \bibinfo {author} {\bibfnamefont
  {P.}~\bibnamefont {{Franzetti}}}, \bibinfo {author} {\bibfnamefont
  {M.}~\bibnamefont {{Fumana}}}, \bibinfo {author} {\bibfnamefont
  {B.}~\bibnamefont {{Garilli}}}, \bibinfo {author} {\bibfnamefont
  {O.}~\bibnamefont {{Ilbert}}}, \bibinfo {author} {\bibfnamefont
  {J.}~\bibnamefont {{Krywult}}}, \bibinfo {author} {\bibfnamefont
  {V.}~\bibnamefont {{Le Brun}}}, \bibinfo {author} {\bibfnamefont
  {O.}~\bibnamefont {{Le F{\`e}vre}}}, \bibinfo {author} {\bibfnamefont
  {D.}~\bibnamefont {{Maccagni}}}, \bibinfo {author} {\bibfnamefont
  {K.}~\bibnamefont {{Ma{\l}ek}}}, \bibinfo {author} {\bibfnamefont
  {F.}~\bibnamefont {{Marulli}}}, \bibinfo {author} {\bibfnamefont {H.~J.}\
  \bibnamefont {{McCracken}}}, \bibinfo {author} {\bibfnamefont
  {L.}~\bibnamefont {{Moscardini}}}, \bibinfo {author} {\bibfnamefont
  {L.}~\bibnamefont {{Paioro}}}, \bibinfo {author} {\bibfnamefont {W.~J.}\
  \bibnamefont {{Percival}}}, \bibinfo {author} {\bibfnamefont
  {M.}~\bibnamefont {{Polletta}}}, \bibinfo {author} {\bibfnamefont
  {A.}~\bibnamefont {{Pollo}}}, \bibinfo {author} {\bibfnamefont
  {H.}~\bibnamefont {{Schlagenhaufer}}}, \bibinfo {author} {\bibfnamefont
  {M.}~\bibnamefont {{Scodeggio}}}, \bibinfo {author} {\bibfnamefont
  {L.~A.~M.}\ \bibnamefont {{Tasca}}}, \bibinfo {author} {\bibfnamefont
  {R.}~\bibnamefont {{Tojeiro}}}, \bibinfo {author} {\bibfnamefont
  {D.}~\bibnamefont {{Vergani}}}, \bibinfo {author} {\bibfnamefont
  {A.}~\bibnamefont {{Zanichelli}}}, \bibinfo {author} {\bibfnamefont
  {A.}~\bibnamefont {{Burden}}}, \bibinfo {author} {\bibfnamefont
  {C.}~\bibnamefont {{Di Porto}}}, \bibinfo {author} {\bibfnamefont
  {A.}~\bibnamefont {{Marchetti}}}, \bibinfo {author} {\bibfnamefont
  {C.}~\bibnamefont {{Marinoni}}}, \bibinfo {author} {\bibfnamefont
  {Y.}~\bibnamefont {{Mellier}}}, \bibinfo {author} {\bibfnamefont
  {P.}~\bibnamefont {{Monaco}}}, \bibinfo {author} {\bibfnamefont {R.~C.}\
  \bibnamefont {{Nichol}}}, \bibinfo {author} {\bibfnamefont {S.}~\bibnamefont
  {{Phleps}}}, \bibinfo {author} {\bibfnamefont {M.}~\bibnamefont {{Wolk}}}, \
  and\ \bibinfo {author} {\bibfnamefont {G.}~\bibnamefont {{Zamorani}}},\
  }\href {\doibase 10.1051/0004-6361/201321463} {\bibfield  {journal} {\bibinfo
   {journal} {\aap}\ }\textbf {\bibinfo {volume} {557}},\ \bibinfo {eid} {A54}
  (\bibinfo {year} {2013})},\ \Eprint {http://arxiv.org/abs/1303.2622}
  {arXiv:1303.2622 [astro-ph.CO]} \BibitemShut {NoStop}%
\bibitem [{\citenamefont {{Seo}}\ \emph {et~al.}(2011)\citenamefont {{Seo}},
  \citenamefont {{Sato}}, \citenamefont {{Dodelson}}, \citenamefont {{Jain}},\
  and\ \citenamefont {{Takada}}}]{2011ApJ...729L..11S}%
  \BibitemOpen
  \bibfield  {author} {\bibinfo {author} {\bibfnamefont {H.-J.}\ \bibnamefont
  {{Seo}}}, \bibinfo {author} {\bibfnamefont {M.}~\bibnamefont {{Sato}}},
  \bibinfo {author} {\bibfnamefont {S.}~\bibnamefont {{Dodelson}}}, \bibinfo
  {author} {\bibfnamefont {B.}~\bibnamefont {{Jain}}}, \ and\ \bibinfo {author}
  {\bibfnamefont {M.}~\bibnamefont {{Takada}}},\ }\href {\doibase
  10.1088/2041-8205/729/1/L11} {\bibfield  {journal} {\bibinfo  {journal}
  {\apjl}\ }\textbf {\bibinfo {volume} {729}},\ \bibinfo {eid} {L11} (\bibinfo
  {year} {2011})},\ \Eprint {http://arxiv.org/abs/1008.0349} {arXiv:1008.0349
  [astro-ph.CO]} \BibitemShut {NoStop}%
\bibitem [{\citenamefont {{Joachimi}}\ \emph {et~al.}(2011)\citenamefont
  {{Joachimi}}, \citenamefont {{Taylor}},\ and\ \citenamefont
  {{Kiessling}}}]{2011MNRAS.418..145J}%
  \BibitemOpen
  \bibfield  {author} {\bibinfo {author} {\bibfnamefont {B.}~\bibnamefont
  {{Joachimi}}}, \bibinfo {author} {\bibfnamefont {A.~N.}\ \bibnamefont
  {{Taylor}}}, \ and\ \bibinfo {author} {\bibfnamefont {A.}~\bibnamefont
  {{Kiessling}}},\ }\href {\doibase 10.1111/j.1365-2966.2011.19472.x}
  {\bibfield  {journal} {\bibinfo  {journal} {\mnras}\ }\textbf {\bibinfo
  {volume} {418}},\ \bibinfo {pages} {145} (\bibinfo {year} {2011})},\ \Eprint
  {http://arxiv.org/abs/1104.1399} {arXiv:1104.1399 [astro-ph.CO]} \BibitemShut
  {NoStop}%
\bibitem [{\citenamefont {{Yu}}\ \emph {et~al.}(2011)\citenamefont {{Yu}},
  \citenamefont {{Zhang}}, \citenamefont {{Lin}}, \citenamefont {{Cui}},\ and\
  \citenamefont {{Fry}}}]{2011PhRvD..84b3523Y}%
  \BibitemOpen
  \bibfield  {author} {\bibinfo {author} {\bibfnamefont {Y.}~\bibnamefont
  {{Yu}}}, \bibinfo {author} {\bibfnamefont {P.}~\bibnamefont {{Zhang}}},
  \bibinfo {author} {\bibfnamefont {W.}~\bibnamefont {{Lin}}}, \bibinfo
  {author} {\bibfnamefont {W.}~\bibnamefont {{Cui}}}, \ and\ \bibinfo {author}
  {\bibfnamefont {J.~N.}\ \bibnamefont {{Fry}}},\ }\href {\doibase
  10.1103/PhysRevD.84.023523} {\bibfield  {journal} {\bibinfo  {journal}
  {\prd}\ }\textbf {\bibinfo {volume} {84}},\ \bibinfo {eid} {023523} (\bibinfo
  {year} {2011})},\ \Eprint {http://arxiv.org/abs/1103.2858} {arXiv:1103.2858
  [astro-ph.CO]} \BibitemShut {NoStop}%
\bibitem [{\citenamefont {{Eisenstein}}\ \emph {et~al.}(2007)\citenamefont
  {{Eisenstein}}, \citenamefont {{Seo}}, \citenamefont {{Sirko}},\ and\
  \citenamefont {{Spergel}}}]{EisSeoRecon}%
  \BibitemOpen
  \bibfield  {author} {\bibinfo {author} {\bibfnamefont {D.~J.}\ \bibnamefont
  {{Eisenstein}}}, \bibinfo {author} {\bibfnamefont {H.-J.}\ \bibnamefont
  {{Seo}}}, \bibinfo {author} {\bibfnamefont {E.}~\bibnamefont {{Sirko}}}, \
  and\ \bibinfo {author} {\bibfnamefont {D.~N.}\ \bibnamefont {{Spergel}}},\
  }\href {\doibase 10.1086/518712} {\bibfield  {journal} {\bibinfo  {journal}
  {\apj}\ }\textbf {\bibinfo {volume} {664}},\ \bibinfo {pages} {675} (\bibinfo
  {year} {2007})},\ \Eprint {http://arxiv.org/abs/arXiv:astro-ph/0604362}
  {arXiv:astro-ph/0604362} \BibitemShut {NoStop}%
\bibitem [{\citenamefont {{Padmanabhan}}\ \emph {et~al.}(2012)\citenamefont
  {{Padmanabhan}}, \citenamefont {{Xu}}, \citenamefont {{Eisenstein}},
  \citenamefont {{Scalzo}}, \citenamefont {{Cuesta}}, \citenamefont {{Mehta}},\
  and\ \citenamefont {{Kazin}}}]{Padmanabhan2012}%
  \BibitemOpen
  \bibfield  {author} {\bibinfo {author} {\bibfnamefont {N.}~\bibnamefont
  {{Padmanabhan}}}, \bibinfo {author} {\bibfnamefont {X.}~\bibnamefont {{Xu}}},
  \bibinfo {author} {\bibfnamefont {D.~J.}\ \bibnamefont {{Eisenstein}}},
  \bibinfo {author} {\bibfnamefont {R.}~\bibnamefont {{Scalzo}}}, \bibinfo
  {author} {\bibfnamefont {A.~J.}\ \bibnamefont {{Cuesta}}}, \bibinfo {author}
  {\bibfnamefont {K.~T.}\ \bibnamefont {{Mehta}}}, \ and\ \bibinfo {author}
  {\bibfnamefont {E.}~\bibnamefont {{Kazin}}},\ }\href {\doibase
  10.1111/j.1365-2966.2012.21888.x} {\bibfield  {journal} {\bibinfo  {journal}
  {\mnras}\ }\textbf {\bibinfo {volume} {427}},\ \bibinfo {pages} {2132}
  (\bibinfo {year} {2012})},\ \Eprint {http://arxiv.org/abs/1202.0090}
  {arXiv:1202.0090} \BibitemShut {NoStop}%
\bibitem [{\citenamefont {{Anderson}}\ \emph {et~al.}(2014)\citenamefont
  {{Anderson}}, \citenamefont {{Aubourg}}, \citenamefont {{Bailey}},
  \citenamefont {{Beutler}}, \citenamefont {{Bhardwaj}}, \citenamefont
  {{Blanton}}, \citenamefont {{Bolton}}, \citenamefont {{Brinkmann}},
  \citenamefont {{Brownstein}}, \citenamefont {{Burden}}, \citenamefont
  {{Chuang}}, \citenamefont {{Cuesta}}, \citenamefont {{Dawson}}, \citenamefont
  {{Eisenstein}}, \citenamefont {{Escoffier}}, \citenamefont {{Gunn}},
  \citenamefont {{Guo}}, \citenamefont {{Ho}}, \citenamefont {{Honscheid}},
  \citenamefont {{Howlett}}, \citenamefont {{Kirkby}}, \citenamefont
  {{Lupton}}, \citenamefont {{Manera}}, \citenamefont {{Maraston}},
  \citenamefont {{McBride}}, \citenamefont {{Mena}}, \citenamefont
  {{Montesano}}, \citenamefont {{Nichol}}, \citenamefont {{Nuza}},
  \citenamefont {{Olmstead}}, \citenamefont {{Padmanabhan}}, \citenamefont
  {{Palanque-Delabrouille}}, \citenamefont {{Parejko}}, \citenamefont
  {{Percival}}, \citenamefont {{Petitjean}}, \citenamefont {{Prada}},
  \citenamefont {{Price-Whelan}}, \citenamefont {{Reid}}, \citenamefont
  {{Roe}}, \citenamefont {{Ross}}, \citenamefont {{Ross}}, \citenamefont
  {{Sabiu}}, \citenamefont {{Saito}}, \citenamefont {{Samushia}}, \citenamefont
  {{S{\'a}nchez}}, \citenamefont {{Schlegel}}, \citenamefont {{Schneider}},
  \citenamefont {{Scoccola}}, \citenamefont {{Seo}}, \citenamefont {{Skibba}},
  \citenamefont {{Strauss}}, \citenamefont {{Swanson}}, \citenamefont
  {{Thomas}}, \citenamefont {{Tinker}}, \citenamefont {{Tojeiro}},
  \citenamefont {{Maga{\~n}a}}, \citenamefont {{Verde}}, \citenamefont
  {{Wake}}, \citenamefont {{Weaver}}, \citenamefont {{Weinberg}}, \citenamefont
  {{White}}, \citenamefont {{Xu}}, \citenamefont {{Y{\`e}che}}, \citenamefont
  {{Zehavi}},\ and\ \citenamefont {{Zhao}}}]{2014_BOSS_BAO}%
  \BibitemOpen
  \bibfield  {author} {\bibinfo {author} {\bibfnamefont {L.}~\bibnamefont
  {{Anderson}}}, \bibinfo {author} {\bibfnamefont {{\'E}.}~\bibnamefont
  {{Aubourg}}}, \bibinfo {author} {\bibfnamefont {S.}~\bibnamefont {{Bailey}}},
  \bibinfo {author} {\bibfnamefont {F.}~\bibnamefont {{Beutler}}}, \bibinfo
  {author} {\bibfnamefont {V.}~\bibnamefont {{Bhardwaj}}}, \bibinfo {author}
  {\bibfnamefont {M.}~\bibnamefont {{Blanton}}}, \bibinfo {author}
  {\bibfnamefont {A.~S.}\ \bibnamefont {{Bolton}}}, \bibinfo {author}
  {\bibfnamefont {J.}~\bibnamefont {{Brinkmann}}}, \bibinfo {author}
  {\bibfnamefont {J.~R.}\ \bibnamefont {{Brownstein}}}, \bibinfo {author}
  {\bibfnamefont {A.}~\bibnamefont {{Burden}}}, \bibinfo {author}
  {\bibfnamefont {C.-H.}\ \bibnamefont {{Chuang}}}, \bibinfo {author}
  {\bibfnamefont {A.~J.}\ \bibnamefont {{Cuesta}}}, \bibinfo {author}
  {\bibfnamefont {K.~S.}\ \bibnamefont {{Dawson}}}, \bibinfo {author}
  {\bibfnamefont {D.~J.}\ \bibnamefont {{Eisenstein}}}, \bibinfo {author}
  {\bibfnamefont {S.}~\bibnamefont {{Escoffier}}}, \bibinfo {author}
  {\bibfnamefont {J.~E.}\ \bibnamefont {{Gunn}}}, \bibinfo {author}
  {\bibfnamefont {H.}~\bibnamefont {{Guo}}}, \bibinfo {author} {\bibfnamefont
  {S.}~\bibnamefont {{Ho}}}, \bibinfo {author} {\bibfnamefont {K.}~\bibnamefont
  {{Honscheid}}}, \bibinfo {author} {\bibfnamefont {C.}~\bibnamefont
  {{Howlett}}}, \bibinfo {author} {\bibfnamefont {D.}~\bibnamefont {{Kirkby}}},
  \bibinfo {author} {\bibfnamefont {R.~H.}\ \bibnamefont {{Lupton}}}, \bibinfo
  {author} {\bibfnamefont {M.}~\bibnamefont {{Manera}}}, \bibinfo {author}
  {\bibfnamefont {C.}~\bibnamefont {{Maraston}}}, \bibinfo {author}
  {\bibfnamefont {C.~K.}\ \bibnamefont {{McBride}}}, \bibinfo {author}
  {\bibfnamefont {O.}~\bibnamefont {{Mena}}}, \bibinfo {author} {\bibfnamefont
  {F.}~\bibnamefont {{Montesano}}}, \bibinfo {author} {\bibfnamefont {R.~C.}\
  \bibnamefont {{Nichol}}}, \bibinfo {author} {\bibfnamefont {S.~E.}\
  \bibnamefont {{Nuza}}}, \bibinfo {author} {\bibfnamefont {M.~D.}\
  \bibnamefont {{Olmstead}}}, \bibinfo {author} {\bibfnamefont
  {N.}~\bibnamefont {{Padmanabhan}}}, \bibinfo {author} {\bibfnamefont
  {N.}~\bibnamefont {{Palanque-Delabrouille}}}, \bibinfo {author}
  {\bibfnamefont {J.}~\bibnamefont {{Parejko}}}, \bibinfo {author}
  {\bibfnamefont {W.~J.}\ \bibnamefont {{Percival}}}, \bibinfo {author}
  {\bibfnamefont {P.}~\bibnamefont {{Petitjean}}}, \bibinfo {author}
  {\bibfnamefont {F.}~\bibnamefont {{Prada}}}, \bibinfo {author} {\bibfnamefont
  {A.~M.}\ \bibnamefont {{Price-Whelan}}}, \bibinfo {author} {\bibfnamefont
  {B.}~\bibnamefont {{Reid}}}, \bibinfo {author} {\bibfnamefont {N.~A.}\
  \bibnamefont {{Roe}}}, \bibinfo {author} {\bibfnamefont {A.~J.}\ \bibnamefont
  {{Ross}}}, \bibinfo {author} {\bibfnamefont {N.~P.}\ \bibnamefont {{Ross}}},
  \bibinfo {author} {\bibfnamefont {C.~G.}\ \bibnamefont {{Sabiu}}}, \bibinfo
  {author} {\bibfnamefont {S.}~\bibnamefont {{Saito}}}, \bibinfo {author}
  {\bibfnamefont {L.}~\bibnamefont {{Samushia}}}, \bibinfo {author}
  {\bibfnamefont {A.~G.}\ \bibnamefont {{S{\'a}nchez}}}, \bibinfo {author}
  {\bibfnamefont {D.~J.}\ \bibnamefont {{Schlegel}}}, \bibinfo {author}
  {\bibfnamefont {D.~P.}\ \bibnamefont {{Schneider}}}, \bibinfo {author}
  {\bibfnamefont {C.~G.}\ \bibnamefont {{Scoccola}}}, \bibinfo {author}
  {\bibfnamefont {H.-J.}\ \bibnamefont {{Seo}}}, \bibinfo {author}
  {\bibfnamefont {R.~A.}\ \bibnamefont {{Skibba}}}, \bibinfo {author}
  {\bibfnamefont {M.~A.}\ \bibnamefont {{Strauss}}}, \bibinfo {author}
  {\bibfnamefont {M.~E.~C.}\ \bibnamefont {{Swanson}}}, \bibinfo {author}
  {\bibfnamefont {D.}~\bibnamefont {{Thomas}}}, \bibinfo {author}
  {\bibfnamefont {J.~L.}\ \bibnamefont {{Tinker}}}, \bibinfo {author}
  {\bibfnamefont {R.}~\bibnamefont {{Tojeiro}}}, \bibinfo {author}
  {\bibfnamefont {M.~V.}\ \bibnamefont {{Maga{\~n}a}}}, \bibinfo {author}
  {\bibfnamefont {L.}~\bibnamefont {{Verde}}}, \bibinfo {author} {\bibfnamefont
  {D.~A.}\ \bibnamefont {{Wake}}}, \bibinfo {author} {\bibfnamefont {B.~A.}\
  \bibnamefont {{Weaver}}}, \bibinfo {author} {\bibfnamefont {D.~H.}\
  \bibnamefont {{Weinberg}}}, \bibinfo {author} {\bibfnamefont
  {M.}~\bibnamefont {{White}}}, \bibinfo {author} {\bibfnamefont
  {X.}~\bibnamefont {{Xu}}}, \bibinfo {author} {\bibfnamefont {C.}~\bibnamefont
  {{Y{\`e}che}}}, \bibinfo {author} {\bibfnamefont {I.}~\bibnamefont
  {{Zehavi}}}, \ and\ \bibinfo {author} {\bibfnamefont {G.-B.}\ \bibnamefont
  {{Zhao}}},\ }\href {\doibase 10.1093/mnras/stu523} {\bibfield  {journal}
  {\bibinfo  {journal} {\mnras}\ }\textbf {\bibinfo {volume} {441}},\ \bibinfo
  {pages} {24} (\bibinfo {year} {2014})},\ \Eprint
  {http://arxiv.org/abs/1312.4877} {arXiv:1312.4877} \BibitemShut {NoStop}%
\end{thebibliography}%

\end{document}